\documentclass[11pt, letterpaper]{article}
\usepackage{geometry}
\usepackage{amsmath,amsfonts,amssymb,amsthm,epsfig,color, graphicx, natbib, appendix,comment,subcaption, amssymb,xcolor,geometry,bm,enumerate}

\usepackage{pgfplots}
\usepgfplotslibrary{fillbetween}
\usetikzlibrary{patterns}
\usepackage{subcaption}
\usepgfplotslibrary{groupplots}

\definecolor{bid}{RGB}{0,102,204}
\definecolor{units}{RGB}{180,40,40}
\definecolor{value}{RGB}{0,140,70}

\pgfplotsset{compat=1.18}

\definecolor{maroonshade}{RGB}{128,0,0}

\newcommand{\twopartdef}[4]
{
	\left\{
		\begin{array}{ll}
			#1 & \mbox{if } #2 \\
			#3 & \mbox{if } #4
		\end{array}
	\right.
}

\usepackage{marginnote}

\usepackage{bbm}
\usepackage{soul}

\definecolor{bid}{RGB}{0,102,204}
\definecolor{units}{RGB}{180,40,40}

\usepackage{setspace} %

\usepackage[colorlinks=true,
    linkcolor=blue,
    filecolor=magenta,      
    urlcolor=blue]{hyperref}
\usepackage{tikz}
\usetikzlibrary{fit,positioning}
\usetikzlibrary{shapes,arrows}
\usetikzlibrary{intersections}
\usetikzlibrary{external}
\usepackage[nameinlink]{cleveref}

\newtheorem{theorem}{Theorem}
\newtheorem*{theoremstar}{Theorem}
\newtheorem{lemma}{Lemma}
\newtheorem*{definition}{Definition}
\newtheorem{assumption}{Assumption}

\newtheorem{proposition}{Proposition}
\newtheorem{corollary}{Corollary}
\theoremstyle{definition}
\newtheorem{example}{Example}
\theoremstyle{remark}

\newtheorem*{example*}{Example}
\theoremstyle{remark}

\theoremstyle{remark} \newtheorem{fact}{Fact}

\usepackage{tikz}
\usetikzlibrary{positioning,quotes}

\definecolor{darkblue}{rgb}{0.0, 0.0, 0.55}
\hypersetup{colorlinks,linkcolor={darkblue},citecolor={darkblue},urlcolor={darkblue}} 
\allowdisplaybreaks
\title{Robust Analysis of First-Price Auctions with Large Supply \thanks{We thank Ben Golub, Bruno Strulovici and Jeroen Swinkels for very valuable discussions. We used LLM tools including Refine.ink and Google DeepReview to check writing and check proofs.}}

\author{Jason Hartline \thanks{Department of Computer Science, Northwestern University. Email:hartline@northwestern.edu.} \and Anant Shah \thanks{University of Cambridge. Email:anantshah200@gmail.com.} \and Matthew vonAllmen \thanks{Department of Computer Science, Northwestern University. Email:matthewvonallmen2026@u.northwestern.edu.}}

\newcommand{\numagents}[0]{n}
\newcommand{\agentset}[0]{\mathcal{N}}
\newcommand{\numunits}[0]{k}
\newcommand{\subseqind}{l}

\newcommand{\supportinfset}[0]{\mathbf{\Gamma}}

\newcommand{\supportinfcriticalset}[1]{\mathbf{\Gamma}_{-#1}}

\newcommand{\supportinfcriticalsetelement}[2]{{\Gamma^{(#2)}_{-#1}}}

\newcommand{\supportinf}[1]{\gamma_{#1}}
\newcommand{\supportsup}[1]{\lambda_{#1}}

\newcommand{\supportinforderstat}[1]{\gamma^{(#1)}}
\newcommand{\supportsuporderstat}[1]{\lambda^{(#1)}}

\newcommand{\supportinfhighest}[1]{\gamma_{(#1)}}
\newcommand{\supportsuphighest}[1]{\lambda_{(#1)}}

\newcommand{\supportinfcritical}[1]{\gamma^{(\numunits)}_{-#1}}

\newcommand{\valueCDF}[1]{\mathcal{V}_{#1}}
\newcommand{\bidCDF}[1]{\mathcal{E}_{#1}}
\newcommand{\khighbidCDF}{\mathcal{B}^{(\numunits)}}
\newcommand{\khighbidCDFsubseq}{\mathcal{B}^{(\numunits_l)}}
\newcommand{\khighbidCDFexpec}{\widehat{B}}

\newcommand{\criticalbidCDF}[1]{\mathcal{B}^{(\numunits)}_{-#1}}

\newcommand{\interimalloc}[1]{x_{#1}}

\newcommand{\lowagents}[0]{L}
\newcommand{\topagents}[0]{T}

\newcommand{\revenuecover}[0]{\mu}
\newcommand{\bidsinductive}[1]{\mathbb{B}_{#1}}
\newcommand{\highpsi}[1]{\overline{\psi}^{(\numunits)}_{#1}}
\newcommand{\highpsisubseq}[1]{\overline{\psi}^{(\numunits_{\subseqind})}_{#1}}
\newcommand{\lowpsi}[1]{\underline{\psi}^{(\numunits)}_{#1}}
\newcommand{\lowpsisubseq}[1]{\underline{\psi}^{(\numunits_{\subseqind})}_{#1}}

\newcommand{\correspondingagentset}[0]{\agentset^{(\numunits)}}

\newcommand{\supp}{\text{supp}}

\newcommand{\deterministicbidprofile}{\mathbf{b}}

\newcommand{\expectedcriticalbid}[1]{\widehat{B}_{#1}}
\newcommand{\bid}[1]{b_{#1}}
\newcommand{\auction}{\mathcal{A}}
\newcommand{\feasibleallocation}{\bm{y}}

\newcommand{\equilibrium}{\mathcal{E}}

\newcommand{\bidpairlow}[1]{\underline{b}_{#1}^{(\numunits)}}
\newcommand{\bidpairhigh}[1]{\overline{b}_{#1}^{(\numunits)}}
\newcommand{\bidpairlowsubseq}[1]{\underline{b}_{#1}^{(\numunits_{\subseqind})}}
\newcommand{\bidpairlowlimit}{\underline{b}^*}
\newcommand{\bidpairhighlimit}{\overline{b}^*}
\newcommand{\bidpairhighsubseq}[1]{\overline{b}_{#1}^{(\numunits_{\subseqind})}}

\DeclareMathOperator*{\prob}{\mathbb{P}}
\DeclareMathOperator*{\expect}{\mathbb{E}}

\begin{document}

\maketitle

\begin{abstract}



This paper performs a robust analysis of welfare (i.e., the price of anarchy) in multi-unit first-price auctions with independent private values and unit-demand bidders. We employ a smoothness framework (see \citealp{ST-13}) to analyze the price of anarchy when supply is large. 

The main barrier to analysis in this framework is a worst-case bid distribution that establishes a lower bound on the price of anarchy of $1-
\frac{1}{e}$. We show that such worst-case distributions cannot arise in equilibrium when agents' values satisfy a mild density lower bound condition. This characterization leads to novel asymptotic lower bounds. In particular, the price of anarchy for $\epsilon$-approximate equilibrium as the number of units goes to infinity is at least $1-\Theta \left(\epsilon \log \left(\frac{1}{\epsilon}\right)\right)$. 

Our analysis relies on a key structural property of equilibrium bid distributions: for a welfare analysis in any multi-unit first-price auction, it is without loss to analyze (approximate) equilibria with no atoms. In addition to this, we establish continuity and monotonicity properties of Bayes-Nash equilibria. We believe these results are of independent interest.    

\end{abstract}



\section{Introduction}
\label{s:introduction}

Many economic environments involve the sale of multiple identical goods. A leading example is online marketplaces, where firms sell products from large inventories. Policymakers and online market designers are often interested in the welfare guarantees of standard market mechanisms as the market grows large. We study this question in a canonical first-price auction with large supply. 


An agent has an independent private value for obtaining a good and submits a single bid to obtain a single unit of the good. A first-price auction allocates goods to the highest bidders, up to the total supply, and the winners pay their bids. This auction format, and variants of it, are commonly used in practice because they are simple and transparent. For instance, in 2019, Google Ad Manager transitioned to ``unified first-price auctions'' as it provided an increase in transparency \footnote{See https://blog.google/products/admanager/rolling-out-first-price-auctions-google-ad-manager-partners/.}. A key feature of this mechanism, however, is that agents strategically shade their bids below their valuations. This causes inefficiencies in allocation: a high-value agent may lose out if she bids less aggressively than a low-value competitor. What is the extent of these inefficiencies as supply becomes large?

To study this question, we analyze the Price of Anarchy, henceforth \textit{PoA}, described as the worst-case ratio of equilibrium welfare to the socially optimum welfare (see \citealp{RST-17} for a comprehensive survey of the PoA in auctions). A long line of work has studied the PoA in an environment with single-unit supply and independent private values. In pioneering work, \cite{ST-13} developed a \textit{smoothness framework} to analyze PoA in auctions and established a lower bound of $1-\frac{1}{e}$. Outside of this standard smoothness framework, \cite{HTW-18} improved the lower bound by leveraging the independence of value distributions to show that equilibrium welfare is a 0.743-fraction of socially optimum welfare. For the upper bound, \cite{HHT-14} provided a value distribution and corresponding equilibrium in which welfare is a 0.869-fraction of the optimum welfare. In a recent breakthrough result, \cite{JL-23} showed that for a single-unit first-price auction with independent private values, the PoA is $1-\frac{1}{e^2}$. Their analysis, which is outside the smoothness paradigm, characterizes the worst-case equilibrium bid distribution through a differential equation. 

The smoothness framework was developed by works such as \cite{R-12}, \cite{ST-13}, and \citet{HHT-14}, and is a central technique to bound the PoA. As a proof technique, the primary appeal of this framework is its robustness to exact details of equilibrium behavior. Characterizing equilibria in first-price auctions is, in general, analytically difficult. Complex auction environments, some examples of which include auctions with combinatorial demand constraints and simultaneous composition of auctions (see \citealp{H-16}), only further compound this issue. Remarkably, the smoothness framework establishes non-trivial bounds on the PoA in these complex environments without characterizing equilibrium behavior.

This paper builds on the concepts of \textit{individual efficiency} and \textit{competitive efficiency} --- a decomposition of the smoothness framework --- introduced by \cite{HHT-14}. These objects are the two main drivers of inefficiency in an auction. Individual efficiency measures losses in welfare due to aggregate uncertainty while competitive efficiency measures losses due to unpriced externalities. In any multi-unit first-price auction, unpriced externalities are absent and the only inefficiency is due to aggregate uncertainty. 

Mathematically, individual efficiency is defined as the following intuitive tradeoff: in equilibrium, either an agent gets large utility, or faces strong competition from other bidders in the auction. A primary barrier to a proof approach using individual efficiency is a worst-case  bid distribution which leads to a PoA lower bound of $1-\frac{1}{e}$ in any multi-unit first-price auction. The first contribution of this paper is to show that such worst-case bid distributions do not arise in equilibrium when values satisfy a density lower bound condition. This establishes novel PoA bounds in the large supply regime which we informally provide below. 

\begin{theoremstar}[Informal]
Consider the PoA of a $\numunits$-unit first-price auction with worst-case performance analyzed over independent value distributions satisfying a density lower bound condition and $\epsilon^{(\numunits)}$-approximate equilibria satisfying no over-bidding. Let $\epsilon := \lim_{\numunits \to \infty}\epsilon^{(\numunits)}$. \begin{enumerate}
    \item If $\epsilon=0$, then $\lim_{\numunits \to \infty} \text{PoA} = 1$.
    \item If $\epsilon > 0$, then $\lim_{\numunits \to \infty}\text{PoA} \geq 1-\Theta\left(\epsilon\log\left(\frac{1}{\epsilon}\right)\right)$.
\end{enumerate}    
\end{theoremstar}

The main challenge is to analyze structural properties of (approximate) equilibria in settings where it is analytically intractable to explicitly solve for such objects. With non-identical values, the optimization problem that bidders face can be very different, leading to asymmetric equilibrium behavior. Our approach combines a law of large numbers argument with (approximately) optimal best-response properties to establish worst-case bounds on the distribution function of the $\numunits$-th order statistic of equilibrium bids. This order statistic is closely related to the individual efficiency of an auction. Indeed, any uncertainty in this statistic quantifies a bound on the PoA. 

In \Cref{s:smoothnessframework}, we adapt the smoothness framework to show what feature of the limit distribution of the $\numunits$-th order statistic of equilibrium bids governs the PoA. Individual efficiency is typically defined for a fixed auction environment. Our framework adapts this definition to a sequence of auctions with increasing number of units. A key object we introduce is an \textit{extremizing sequence of bid pairs}. As the name suggests, this is a sequence of pairs of bids, with each bid-pair corresponding to a particular auction in the sequence. Briefly, it has the property that the $\numunits$-th order statistic lies in the interval defined by a bid-pair with probability tending to one. Then, the area below the distribution function when restricted to this bid-pair interval governs the PoA. When the distribution function is relatively flat over a large interval, the loss in welfare is high. When it is steep over a small interval, the loss in welfare is small. We show that when agents optimally best-respond ($\epsilon=0$), the $\numunits$-th order statistic of equilibrium bids converges to a step function. When agents approximately best-respond ($\epsilon>0$), we characterize a distribution which is stochastically dominated by the $
\numunits$-th order statistic in the limit; this leads to the PoA lower bound in Part (b) of the above result (see \Cref{s:welfareanalysisfpa}).

The second contribution of this paper is to establish key structural properties of (approximate) equilibrium in first-price auctions (see \Cref{s:eqlbstruct}). Our analysis is driven by the following result: for a welfare analysis, it is without loss to analyze equilibria with no atoms. We provide an informal statement of the result below. 

\begin{figure}[t]
    \centering

    \begin{tikzpicture}

    \def\alphaOne{1.85}
    \def\betaOne{4.25}
    \def\salphaOne{1.20}
    \def\sbetaOne{0.95}

    \def\alphalowOne{0.38}
    \def\alphahighOne{0.50}
    \def\betalowOne{0.62}
    \def\betahighOne{0.74}

    \def\middleOneA{2.25}
    \def\middleOneB{2.70}
    \def\middleOneC{3.20}

    \def\finalOneA{4.55}
    \def\finalOneB{4.90}
    \def\finalOneC{5.25}
    \def\finalOneD{5.60}
    \def\finalOneE{5.80}

    \pgfmathsetmacro{\alphaOneShift}{\alphaOne+\salphaOne}
    \pgfmathsetmacro{\betaOneShift}{\betaOne+\salphaOne}

    \pgfmathsetmacro{\middleOneAShift}{\middleOneA+\salphaOne}
    \pgfmathsetmacro{\middleOneBShift}{\middleOneB+\salphaOne}
    \pgfmathsetmacro{\middleOneCShift}{\middleOneC+\salphaOne}

    \pgfmathsetmacro{\finalOneAShift}{\finalOneA+\salphaOne}
    \pgfmathsetmacro{\finalOneBShift}{\finalOneB+\salphaOne}
    \pgfmathsetmacro{\finalOneCShift}{\finalOneC+\salphaOne}
    \pgfmathsetmacro{\finalOneDShift}{\finalOneD+\salphaOne}
    \pgfmathsetmacro{\finalOneEShift}{\finalOneE+\salphaOne}

    \pgfmathsetmacro{\betaOneSecondShift}{\betaOneShift+\sbetaOne}

    \pgfmathsetmacro{\finalOneASecondShift}{\finalOneAShift+\sbetaOne}
    \pgfmathsetmacro{\finalOneBSecondShift}{\finalOneBShift+\sbetaOne}
    \pgfmathsetmacro{\finalOneCSecondShift}{\finalOneCShift+\sbetaOne}
    \pgfmathsetmacro{\finalOneDSecondShift}{\finalOneDShift+\sbetaOne}
    \pgfmathsetmacro{\finalOneESecondShift}{\finalOneEShift+\sbetaOne}

    \def\alphaTwo{1.75}
    \def\betaTwo{4.35}
    \def\salphaTwo{1.05}
    \def\sbetaTwo{0.90}

    \def\alphaValTwo{0.36}
    \def\betaLowTwo{0.73}
    \def\betaHighTwo{0.82}

    \def\postATwoA{2.25}
    \def\postATwoB{2.70}
    \def\postATwoC{3.15}
    \def\postATwoD{3.55}
    \def\postATwoE{3.95}

    \def\postBTwoA{4.80}
    \def\postBTwoB{5.20}
    \def\postBTwoC{5.60}
    \def\postBTwoD{6.00}

    \pgfmathsetmacro{\alphaTwoShift}{\alphaTwo+\salphaTwo}
    \pgfmathsetmacro{\betaTwoShift}{\betaTwo+\salphaTwo}

    \pgfmathsetmacro{\postATwoAShift}{\postATwoA+\salphaTwo}
    \pgfmathsetmacro{\postATwoBShift}{\postATwoB+\salphaTwo}
    \pgfmathsetmacro{\postATwoCShift}{\postATwoC+\salphaTwo}
    \pgfmathsetmacro{\postATwoDShift}{\postATwoD+\salphaTwo}
    \pgfmathsetmacro{\postATwoEShift}{\postATwoE+\salphaTwo}

    \pgfmathsetmacro{\postBTwoAShift}{\postBTwoA+\salphaTwo}
    \pgfmathsetmacro{\postBTwoBShift}{\postBTwoB+\salphaTwo}
    \pgfmathsetmacro{\postBTwoCShift}{\postBTwoC+\salphaTwo}
    \pgfmathsetmacro{\postBTwoDShift}{\postBTwoD+\salphaTwo}

    \pgfmathsetmacro{\betaTwoSecondShift}{\betaTwoShift+\sbetaTwo}

    \pgfmathsetmacro{\postBTwoASecondShift}{\postBTwoAShift+\sbetaTwo}
    \pgfmathsetmacro{\postBTwoBSecondShift}{\postBTwoBShift+\sbetaTwo}
    \pgfmathsetmacro{\postBTwoCSecondShift}{\postBTwoCShift+\sbetaTwo}
    \pgfmathsetmacro{\postBTwoDSecondShift}{\postBTwoDShift+\sbetaTwo}

    \begin{groupplot}[
        group style={
            group size=3 by 2,
            horizontal sep=0.75cm,
            vertical sep=1.2cm
        },
        width=0.26\textwidth,
        height=0.20\textwidth,
        scale only axis,
        axis lines=left,
        xmin=0,
        xmax=8.4,
        ymin=0,
        ymax=1.08,
        xlabel={},
        xtick=\empty,
        ytick={1},
        yticklabels={$1$},
        xlabel style={
            at={(axis description cs:1.05,-0.05)},
            anchor=north east
        },
        axis line style={->,thick},
        clip=false,
    ]

    \nextgroupplot[
        ylabel={$\mathcal{B}_{1}(b)$},
        ylabel style={
            rotate=0,
            at={(axis description cs:-0.14,0.50)},
            anchor=south
        }
    ]

    \addplot[dashed, thick, domain=0:8.0] {1};

    \addplot[
        blue,
        thick,
        smooth
    ] coordinates {
        (0.00,0.00)
        (0.35,0.08)
        (0.70,0.16)
        (1.05,0.24)
        (1.40,0.31)
        (\alphaOne,\alphalowOne)
    };

    \addplot[
        only marks,
        mark=o,
        mark size=2.6pt,
        thick,
        blue,
        fill=white
    ] coordinates {
        (\alphaOne,\alphalowOne)
    };

    \addplot[
        only marks,
        mark=*,
        mark size=2.4pt,
        blue
    ] coordinates {
        (\alphaOne,\alphahighOne)
    };

    \addplot[
        blue,
        thick,
        smooth
    ] coordinates {
        (\alphaOne,\alphahighOne)
        (\middleOneA,0.54)
        (\middleOneB,0.58)
        (\middleOneC,0.61)
        (\betaOne,\betalowOne)
    };

    \addplot[
        only marks,
        mark=o,
        mark size=2.6pt,
        thick,
        blue,
        fill=white
    ] coordinates {
        (\betaOne,\betalowOne)
    };

    \addplot[
        only marks,
        mark=*,
        mark size=2.4pt,
        blue
    ] coordinates {
        (\betaOne,\betahighOne)
    };

    \addplot[
        blue,
        thick,
        smooth
    ] coordinates {
        (\betaOne,\betahighOne)
        (\finalOneA,0.79)
        (\finalOneB,0.86)
        (\finalOneC,0.91)
        (\finalOneD,0.95)
        (\finalOneE,0.96)
    };




    \nextgroupplot

    \addplot[dashed, thick, domain=0:8.0] {1};

    \addplot[
        blue,
        thick,
        smooth
    ] coordinates {
        (0.00,0.00)
        (0.35,0.08)
        (0.70,0.16)
        (1.05,0.24)
        (1.40,0.31)
        (\alphaOne,\alphalowOne)
    };

    \addplot[
        red,
        dashed,
        thick
    ] coordinates {
        (\alphaOne,\alphalowOne)
        (\alphaOneShift,\alphahighOne)
    };

    \addplot[
        blue,
        thick,
        smooth
    ] coordinates {
        (\alphaOneShift,\alphahighOne)
        (\middleOneAShift,0.54)
        (\middleOneBShift,0.58)
        (\middleOneCShift,0.61)
        (\betaOneShift,\betalowOne)
    };

    \addplot[
        only marks,
        mark=o,
        mark size=2.6pt,
        thick,
        blue,
        fill=white
    ] coordinates {
        (\betaOneShift,\betalowOne)
    };

    \addplot[
        only marks,
        mark=*,
        mark size=2.4pt,
        blue
    ] coordinates {
        (\betaOneShift,\betahighOne)
    };

    \addplot[
        blue,
        thick,
        smooth
    ] coordinates {
        (\betaOneShift,\betahighOne)
        (\finalOneAShift,0.79)
        (\finalOneBShift,0.86)
        (\finalOneCShift,0.91)
        (\finalOneDShift,0.95)
        (\finalOneEShift,0.96)
    };




    \nextgroupplot

    \addplot[dashed, thick, domain=0:8.0] {1};

    \addplot[
        blue,
        thick,
        smooth
    ] coordinates {
        (0.00,0.00)
        (0.35,0.08)
        (0.70,0.16)
        (1.05,0.24)
        (1.40,0.31)
        (\alphaOne,\alphalowOne)
    };

    \addplot[
        red,
        dashed,
        thick
    ] coordinates {
        (\alphaOne,\alphalowOne)
        (\alphaOneShift,\alphahighOne)
    };

    \addplot[
        blue,
        thick,
        smooth
    ] coordinates {
        (\alphaOneShift,\alphahighOne)
        (\middleOneAShift,0.54)
        (\middleOneBShift,0.58)
        (\middleOneCShift,0.61)
        (\betaOneShift,\betalowOne)
    };

    \addplot[
        red,
        dashed,
        thick
    ] coordinates {
        (\betaOneShift,\betalowOne)
        (\betaOneSecondShift,\betahighOne)
    };

    \addplot[
        blue,
        thick,
        smooth
    ] coordinates {
        (\betaOneSecondShift,\betahighOne)
        (\finalOneASecondShift,0.79)
        (\finalOneBSecondShift,0.86)
        (\finalOneCSecondShift,0.91)
        (\finalOneDSecondShift,0.95)
        (\finalOneESecondShift,0.96)
    };




    \nextgroupplot[
        ylabel={$\mathcal{B}_{2}(b)$},
        ylabel style={
            rotate=0,
            at={(axis description cs:-0.14,0.50)},
            anchor=south
        }
    ]

    \addplot[dashed, thick, domain=0:8.0] {1};

    \addplot[
        blue,
        thick,
        smooth
    ] coordinates {
        (0.00,0.00)
        (0.45,0.11)
        (0.90,0.22)
        (1.30,0.30)
        (\alphaTwo,\alphaValTwo)
        (\postATwoA,0.46)
        (\postATwoB,0.57)
        (\postATwoC,0.64)
        (\postATwoD,0.69)
        (\postATwoE,0.71)
        (\betaTwo,\betaLowTwo)
    };

    \addplot[
        only marks,
        mark=o,
        mark size=2.6pt,
        thick,
        blue,
        fill=white
    ] coordinates {
        (\betaTwo,\betaLowTwo)
    };

    \addplot[
        only marks,
        mark=*,
        mark size=2.4pt,
        blue
    ] coordinates {
        (\betaTwo,\betaHighTwo)
    };

    \addplot[
        blue,
        thick,
        smooth
    ] coordinates {
        (\betaTwo,\betaHighTwo)
        (\postBTwoA,0.85)
        (\postBTwoB,0.87)
        (\postBTwoC,0.885)
        (\postBTwoD,0.89)
    };

    \draw[thick] (axis cs:\alphaTwo,0) -- (axis cs:\alphaTwo,0.04);
    \draw[thick] (axis cs:\betaTwo,0) -- (axis cs:\betaTwo,0.04);

    \node[below] at (axis cs:\alphaTwo,0) {$\alpha$};
    \node[below] at (axis cs:\betaTwo,0) {$\beta$};

    \node[anchor=north, font=\small] at (rel axis cs:0.5,-0.25) {(a)};

    \nextgroupplot

    \addplot[dashed, thick, domain=0:8.0] {1};

    \addplot[
        blue,
        thick,
        smooth
    ] coordinates {
        (0.00,0.00)
        (0.45,0.11)
        (0.90,0.22)
        (1.30,0.30)
        (\alphaTwo,\alphaValTwo)
    };

    \addplot[
        red,
        dashed,
        thick
    ] coordinates {
        (\alphaTwo,\alphaValTwo)
        (\alphaTwoShift,\alphaValTwo)
    };

    \addplot[
        blue,
        thick,
        smooth
    ] coordinates {
        (\alphaTwoShift,\alphaValTwo)
        (\postATwoAShift,0.46)
        (\postATwoBShift,0.57)
        (\postATwoCShift,0.64)
        (\postATwoDShift,0.69)
        (\postATwoEShift,0.71)
        (\betaTwoShift,\betaLowTwo)
    };

    \addplot[
        only marks,
        mark=o,
        mark size=2.6pt,
        thick,
        blue,
        fill=white
    ] coordinates {
        (\betaTwoShift,\betaLowTwo)
    };

    \addplot[
        only marks,
        mark=*,
        mark size=2.4pt,
        blue
    ] coordinates {
        (\betaTwoShift,\betaHighTwo)
    };

    \addplot[
        blue,
        thick,
        smooth
    ] coordinates {
        (\betaTwoShift,\betaHighTwo)
        (\postBTwoAShift,0.85)
        (\postBTwoBShift,0.87)
        (\postBTwoCShift,0.885)
        (\postBTwoDShift,0.89)
    };

    \draw[thick] (axis cs:\alphaTwoShift,0) -- (axis cs:\alphaTwoShift,0.04);
    \draw[thick] (axis cs:\betaTwoShift,0) -- (axis cs:\betaTwoShift,0.04);

    \node[below, yshift=-2pt, font=\scriptsize] at (axis cs:\alphaTwoShift,0) {$\alpha+s_{\alpha}$};
    \node[below, font=\scriptsize] at (axis cs:\betaTwoShift,0) {$\beta+s_{\alpha}$};

    \node[anchor=north, font=\small] at (rel axis cs:0.5,-0.25) {(b)};

    \nextgroupplot

    \addplot[dashed, thick, domain=0:8.0] {1};

    \addplot[
        blue,
        thick,
        smooth
    ] coordinates {
        (0.00,0.00)
        (0.45,0.11)
        (0.90,0.22)
        (1.30,0.30)
        (\alphaTwo,\alphaValTwo)
    };

    \addplot[
        red,
        dashed,
        thick
    ] coordinates {
        (\alphaTwo,\alphaValTwo)
        (\alphaTwoShift,\alphaValTwo)
    };

    \addplot[
        blue,
        thick,
        smooth
    ] coordinates {
        (\alphaTwoShift,\alphaValTwo)
        (\postATwoAShift,0.46)
        (\postATwoBShift,0.57)
        (\postATwoCShift,0.64)
        (\postATwoDShift,0.69)
        (\postATwoEShift,0.71)
        (\betaTwoShift,\betaLowTwo)
    };

    \addplot[
        red,
        dashed,
        thick
    ] coordinates {
        (\betaTwoShift,\betaLowTwo)
        (\betaTwoSecondShift,\betaHighTwo)
    };

    \addplot[
        blue,
        thick,
        smooth
    ] coordinates {
        (\betaTwoSecondShift,\betaHighTwo)
        (\postBTwoASecondShift,0.85)
        (\postBTwoBSecondShift,0.87)
        (\postBTwoCSecondShift,0.885)
        (\postBTwoDSecondShift,0.89)
    };

    \draw[thick] (axis cs:\alphaTwoShift,0) -- (axis cs:\alphaTwoShift,0.04);
    \draw[thick] (axis cs:\betaTwoSecondShift,0) -- (axis cs:\betaTwoSecondShift,0.04);

    \node[below, yshift=-2pt, font=\scriptsize] at (axis cs:\alphaTwoShift,0) {$\alpha+s_{\alpha}$};
    \node[below, font=\scriptsize] at (axis cs:\betaTwoSecondShift,0) {$\beta+s_{\alpha}+s_{\beta}$};

    \node[anchor=north, font=\small] at (rel axis cs:0.5,-0.25) {(c)};

    \end{groupplot}
    \end{tikzpicture}

    \caption{This figure illustrates the construction of an equilibrium with no atoms $\widetilde{\equilibrium}$ from one with atoms $\equilibrium$. Column (a) depicts the distribution function of equilibrium bids for each agent: $\mathcal{B}_{1}(\cdot)$ for agent $1$ and $\mathcal{B}_{2}(\cdot)$ for agent $2$. Agent $1$ has atoms at $\alpha$ and $\beta$ while agent $2$ only has an atom at $\beta$. Column (b) depicts the shift in bids to the right of atom $\alpha$ by $s_{\alpha}$. The red dashed line shows the uniform randomization of the bid at $\alpha$. Column (c) depicts the shift in bids to the right of the atom at $\beta + s_{\alpha}$ by $s_{\beta}$.  
    }
    \label{fig:no-atoms-constructionintro}
\end{figure}
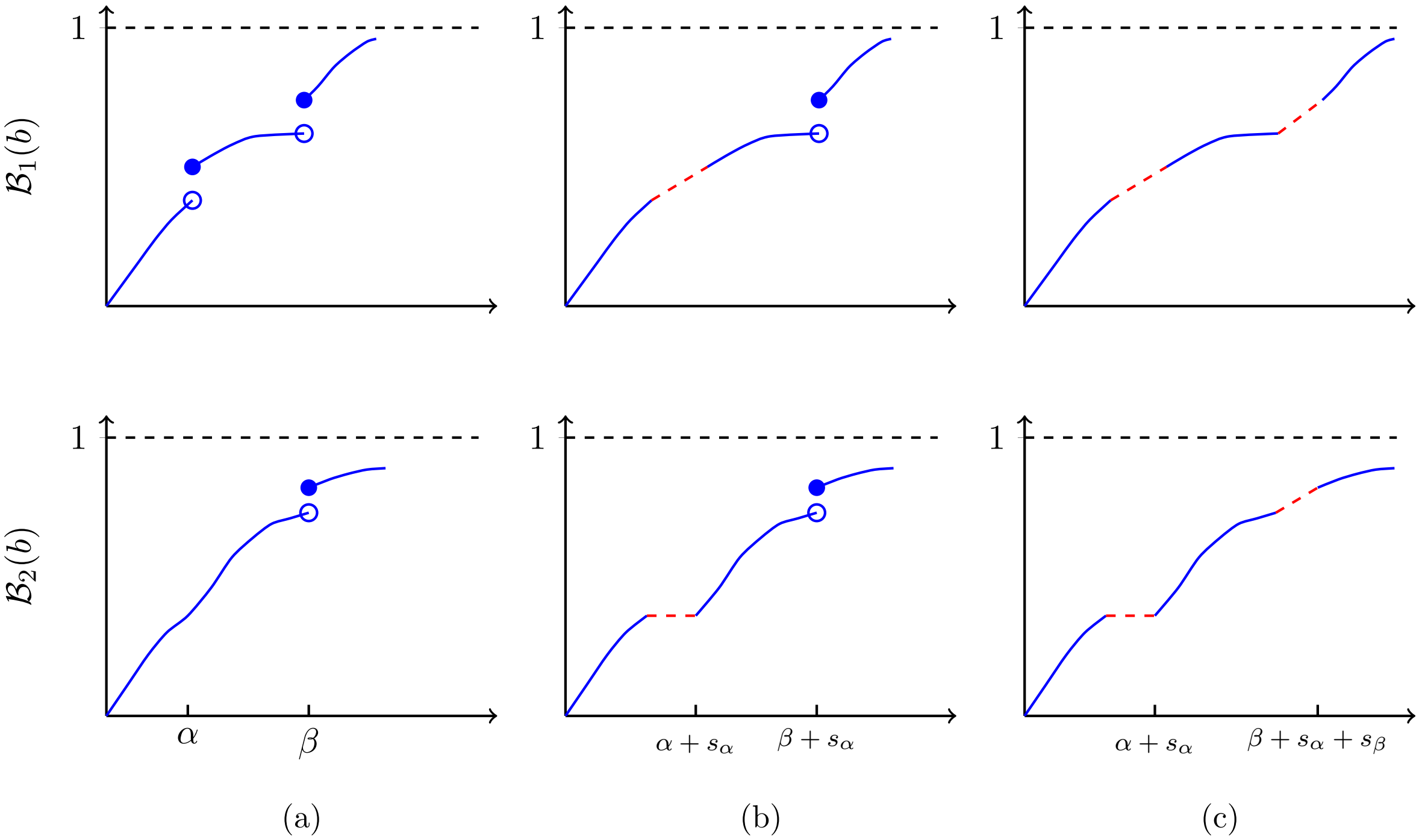 

\begin{theoremstar}[Informal]
      Consider any $\epsilon$-approximate equilibrium $\equilibrium$. For every $\omega>0$, there exists a profile of bid distributions $\widetilde{\equilibrium}$ satisfying the following properties: \begin{enumerate}
        \item $\widetilde{\equilibrium}$ is an $(\epsilon+\omega)$-approximate equilibrium.
        \item For every agent $i$, the bid distribution $\widetilde{\equilibrium}_{i}$ has no atoms.
        \item $\text{welfare}(\equilibrium) = \text{welfare}(\widetilde{\equilibrium}).$
    \end{enumerate}
\end{theoremstar}

Approximate equilibrium bids are typically characterized by atoms, which makes them challenging to analyze. The above result reduces the problem to working with distributions without atoms. This is useful in our large market analysis and greatly simplifies agents' optimal bidding problem. To give a quick sketch of the construction of the atom-less distribution, fix any $\epsilon$-approximate equilibrium and the locations of atoms in the agents' bid distributions. Cumulatively shift the atoms to the right and add a small uniform random bid at each atom. The total cumulative shift can be arbitrarily small, so the resulting distribution is still an approximate equilibrium. \Cref{fig:no-atoms-constructionintro} provides a pictorial description of this construction. Surprisingly, the allocations under the constructed strategy and the original strategy have identical welfare.  

Finally, we establish monotonicity and continuity properties of Bayes-Nash equilibrium in any multi-unit first-price auction. This generalizes analogous structural results in the single-unit setting \citep[see][]{L99,MR00,MR03,WSZ20}. The fundamental challenge in our setting is that an agent's bid distribution need not necessarily impact the competing bid distribution of a different agent. This is because an agent's allocation is determined by the $\numunits$-th highest bid among the remaining agents. Addressing this issue requires a different set of arguments. We believe that our structural results may be of independent interest.

\subsection*{Related Work}

There is a rich body of work on robust analysis in auctions; see \citet{RST-17} for a comprehensive survey. A central technique in this literature is the smoothness framework, introduced by \citet{R-12}, \citet{ST-13}, and \citet{R-15}, which has proven effective for analyzing efficiency guarantees across a range of mechanisms. \citet{HHT-14} further refine this approach by decomposing smoothness into two key components—individual efficiency and competitive efficiency—which highlight structural properties of auctions that ensure near-optimal outcomes. 

We view our analysis of the $\numunits$-th order statistic as a contribution to this literature. The $1-1/e$ bound in \cite{HHT-14} was a barrier to the proof approach of separating individual efficiency and competitive efficiency. The challenge in overcoming this barrier is proving that the worst-case allocation rules for individual efficiency cannot arise in equilibrium. This paper characterizes bounds on the worst-case bid distributions allowing improvements over the $1-1/e$ bound for both Bayes-Nash and approximate equilibria. 

A broader generalization of the smoothness framework, developed by \citet{FILRS-16}, extends the analysis of inefficiency to large games and has been applied to mechanisms such as the simultaneous uniform-price auction and greedy combinatorial auction. Their results are complementary to ours, as both address efficiency guarantees in large-scale market environments.

In the economics literature, there is a long tradition of studying large-market limits of auctions and games \citep[see][]{W-77,RSW-94,S-99,S-01}. The study most closely related to ours is \citet{S-99}, which analyzes large-market limits of Bayes-Nash equilibria in multi-unit first-price auctions. This work directly characterizes the limit PoA bound of $1$ when $\epsilon^{(\numunits)}=0$ for every $\numunits \in \mathbb{N}$. Their analysis is outside the smoothness paradigm of individual efficiency in the sense that it establishes welfare guarantees without characterizing the behavior of the $\numunits$-th order statistic. We have a detailed discussion of how our analysis relates to this work when we provide our PoA bounds in \Cref{s:welfareanalysisfpa}. 

The inefficiency of finite auction environments has also received significant attention. Notable contributions include \citet{FFGL-13}, \citet{BR-11}, and \citet{MT-15}, who analyze the efficiency of uniform-price and simultaneous-item auctions. The single-unit first-price auction, in particular, has been studied extensively (see \citealp{KS-12}; \citealp{HTW-18}). A central challenge across this literature lies in the characterization of Bayes-Nash equilibria, which often admit complex or non-explicit forms. In a recent breakthrough, \citet{JL-23} show that the price of anarchy for the single-unit first-price auction is $1 - 1/e^2$. Their analysis characterizes the worst-case equilibrium bid distribution by solving a differential equation. Extending this type of characterization to the multi-unit setting presents substantial analytical challenges. 

In orthogonal work on multi-unit mechanisms that are not winner-pays-bid, a growing body of evidence suggests that such mechanisms tend to achieve stronger welfare performance than their single-unit counterparts. For example, \citet{Y-11} uses a correlation gap argument to show that, in posted-price mechanisms with independent values, the price of anarchy converges asymptotically to $1$ at a rate of $1 - O(1 / \sqrt{k})$. This bound is tight when agent values are identically distributed. Similarly, \citet{CDL-23} study anonymous posted-price mechanisms and demonstrate that, under independent values, the worst-case distribution of agent demand is Poisson, allowing welfare approximation ratios to be computed for any finite number of units. Building on this line of work, \citet{JJM-23} establish that the exact rate of convergence of the welfare approximation ratio is $1 - \Theta(\sqrt{\ln(k) / k})$.

\section{Model}
\label{s:model}

There are $\numunits \in \mathbb{N}$ identical units and there are $n^{(\numunits)}$ single-dimensional single-unit demand agents, indexed by $\agentset^{(\numunits)} = \left\{1,2,\dots,n^{(\numunits)}\right \}$. We assume $\numunits < n^{(\numunits)}$ because there is no loss in efficiency when $k \geq n^{(\numunits)}$ --- all agents are allocated a unit. Each agent $i$ has a private valuation drawn independently from a distribution $\mathcal{V}_{i}$. We denote this auction environment by $\auction^{(\numunits)}$. Consider a sequence of auctions $\left(\auction^{(\numunits)}\right)_{\numunits=1}^{\infty}$ with value distributions satisfying the following conditions: 

\begin{assumption}\label{as:lowerboundexpectation}
    \begin{enumerate}[(a)]
        \item (Bounded Support) There exists a constant $H < \infty$ such that for any $\numunits$ and any agent $i \in \agentset^{(\numunits)}$, their value $v_{i}$ is at most $H$ almost surely.
        \item (Density Lower Bound)
        There exists a constant $\kappa>0$ such that for any $\numunits$ and any agent $i \in \mathcal{N}^{(\numunits)}$, for any $\Delta>0$ and any value $v \in [0,H-\Delta]$, the measure of values \begin{align*}
           \Pr\left[V_{i} \in (v,v+\Delta)\right] > \kappa \Delta.
        \end{align*}
    \end{enumerate}
\end{assumption}

Fix any $\numunits \in \mathbb{N}$. We suppress the dependence on $\numunits$ in our notation when describing the auction rules. Each agent submits a single non-negative bid $\bid{i}$. Units are sold using a first-price auction. The $\numunits$ agents with the highest bids each receive one unit and pay their submitted bid, with ties broken uniformly at random. Agents have a quasi-linear utility. 

A strategy for agent $i$ is a mapping $\equilibrium_{i} : \mathcal{V}_{i} \to \Delta \left( \mathbb{R}_{\geq 0} \right)$ from her valuation to a distribution over bids. We will denote the distribution over bids for a value $v_{i}$ by $\equilibrium_{i,v_{i}}$. Sometimes we will overload notation and denote the marginal distribution over bids by $\equilibrium_{i}$. We will clarify when we do so. We denote a strategy profile by $\equilibrium := \left(\equilibrium_{i}\right)_{i \in \agentset}$. 

Given a strategy profile, distribution over bids determines interim outcomes. The interim allocation rule for agent $i$ is a function $x_{i} : \mathbb{R}_{\geq 0} \to [0,1]$ that specifies the probability that agent $i$ receives a unit when submitting bid $b_{i}$. Similarly, the interim payment rule $p_{i}: \mathbb{R}_{\geq 0} \to \mathbb{R}_{\geq 0}$ gives agent $i$'s expected payment conditional on bidding $b_{i}$. The interim utility of agent $i$ with valuation $v_{i}$ and bid $b_{i}$ is
\begin{align*}
    u_{i}\left(v_{i},\bid{i}\right) = v_{i}x_{i}(\bid{i}) - p_{i}(\bid{i}) \text{.}
\end{align*} The optimal interim utility of agent $i$ with valuation $v_{i}$ is $u_{i}\left(v_{i}\right)$. 

Our solution concept is an $\epsilon$-approximate Bayes-Nash equilibrium. We provide a definition below for completeness.

\begin{definition}[$\epsilon$-approximate Bayes Nash Equilibrium]
    A strategy profile $\equilibrium$ is said to be an $\epsilon$-approximate Bayes-Nash equilibrium if for every agent $i \in \agentset$, we have \begin{align*}
        \mathbb{E}_{b \sim \equilibrium_{i,v_{i}}} \left[ u_{i}\left(v_{i},b\right) \right] \geq \sup_{b_{i} \in \mathbb{R}_{\geq 0}}u_{i}(v_{i},b_{i}) - \epsilon \quad \text{ for every } v_{i} \in \text{supp}(\valueCDF{i}). 
    \end{align*} 
\end{definition}

We impose a mild no over-bidding assumption on equilibrium bidding behavior. Our results for exact Bayes-Nash equilibrium continue to hold without this assumption. 

\begin{assumption}[No Over-Bidding]
    \label{as:nooverbidding}
    Consider any auction $\auction$ and equilibrium $\equilibrium$. For any agent $i \in \agentset$ and value $v$, the equilibrium bid $\equilibrium_{i,v}$ is at most $v$ almost surely. 
\end{assumption}


Our goal is to perform a robust analysis of equilibrium welfare as the number of units $\numunits$ becomes large. To that end, consider any $\numunits$-unit first-price auction $\auction^{(\numunits)}$ and any corresponding equilibrium $\equilibrium^{(\numunits)}$. Let expected equilibrium welfare be $\mathcal{W}\left(\equilibrium^{(\numunits)},\auction^{(\numunits)}\right)$ and expected optimal welfare be $\mathcal{O}(\auction^{(\numunits)})$. The rest of the paper focuses on analyzing asymptotic inefficiencies in equilibria, measured by the following quantity:  \begin{align*}
   \inf_{\left(\left\{\auction^{(\numunits)},\equilibrium^{(\numunits)}\right\}\right)_{\numunits=1}^{\infty}} \liminf_{\numunits \to \infty} \frac{\mathcal{W}\left(\equilibrium^{(\numunits)},\auction^{(\numunits)}\right)}{\mathcal{O}\left(\auction^{(\numunits)}\right)}.
\end{align*}

In the objective above, the value distribution parameters $H>0$, $\kappa>0$ and the best-response approximation $\lim_{\numunits \to \infty} \epsilon^{(\numunits)} = \epsilon$ are fixed and define the family of auction-equilibrium sequences over which the infimum is evaluated.


\section{Robust Framework For Equilibrium Analysis}
\label{s:smoothnessframework}

This section introduces a framework that facilitates a robust analysis of equilibrium welfare in an auction. In general, an auction admits multiple equilibria with very different welfare. Furthermore, their structure is extremely sensitive to the agents' value distribution. This makes a direct analysis of all possible equilibria intractable. The smoothness framework provides a robust approach by establishing welfare guarantees that hold uniformly across value distributions and their corresponding equilibria. 

\subsection{Framework Parameters} \label{ss:fixauctionefficiency} We begin by defining two properties attributed to any auction: individual efficiency and competitive efficiency  (see \citealp{HHT-23} and \citealp{ST-13} for closely related definitions). Together, they determine the worst-case welfare loss in every equilibrium.  

  We begin with the \textit{individual efficiency} of an auction. To introduce some notation, consider an auction $\auction$ and any corresponding equilibrium $\equilibrium$. The individual efficiency of an auction is defined in terms of the utility of an entrant agent who competes against the incumbents playing according to equilibrium $\equilibrium$. Let $\khighbidCDF$ denote the entrant's critical bid, that is, the minimum bid required to be allocated, and let $\khighbidCDFexpec$ denote its expectation. Furthermore, let $\khighbidCDF(\cdot)$ denote its distribution function. For example, in a multi-unit first-price auction with $\numunits$ units, $\khighbidCDF$ is the $\numunits$-th highest equilibrium bid. The entrant has value $v$ and best responds to obtain utility $u(v)$. 

\begin{definition}[Individual Efficiency] \label{def:individualefficiency} An auction $\auction$ is said to be $\eta$-individually efficient if for any $\epsilon$-approximate Bayes-Nash equilibrium $\equilibrium$, we have \begin{align*}
    u(v) + \widehat{B} \geq \eta v \quad \text{ for every value } v.
\end{align*} The individual efficiency of auction $\auction$ is the largest $\eta$ such that the auction is $\eta$-individually efficient.    
\end{definition}

\begin{figure}[t]
    \centering
    \begin{tikzpicture}
\begin{axis}[
    width=0.6\linewidth,
    height=0.4\linewidth,
    axis lines=left,
    xmin=0, xmax=6.2,
    ymin=0, ymax=1.25,
    xlabel={},
    ylabel={$\khighbidCDF(b)$},
    xtick={3.5},
    xticklabels={$v$},
    ytick={1},
    yticklabels={$1$},
    axis line style={->, thick},
    clip=false,
]

\def\bid{1.2}
\def\vval{3.5}
\def\Bbid{0.58}

\path[
    draw=none,
    fill=green!10,
    pattern=vertical lines,
    pattern color=green!40
]
(axis cs:0,0) rectangle (axis cs:\vval,1);

\addplot[
    name path=topline,
    draw=none,
    domain=0:4.45
] {1};

\addplot[
    name path=curvefill,
    draw=none,
    smooth
] coordinates {
    (0,0)
    (0.15,0.08)
    (0.35,0.32)
    (0.70,0.48)
    (1.20,0.58)
    (1.80,0.64)
    (2.50,0.68)
    (3.20,0.75)
    (3.80,0.84)
    (4.20,0.95)
    (4.45,1.00)
};

\addplot[
    draw=none,
    pattern=north east lines,
    pattern color=red!150
] fill between[of=topline and curvefill];

\node[
    text=red!80!black,
    font=\large
] at (axis cs:1.8,0.88) {$\widehat{B}$};

\addplot[
    dashed,
    thick,
    domain=0:6
] {1};

\path[
    draw=black,
    fill=blue!10,
    pattern=crosshatch,
    pattern color=blue!50
]
(axis cs:\bid,0) rectangle (axis cs:\vval,\Bbid);

\node[
    text=blue!80!black,
    font=\large
] at (axis cs:2.3,0.29) {$u(v)$};

\addplot[
    blue,
    thick,
    smooth
] coordinates {
    (0,0)
    (0.15,0.08)
    (0.35,0.32)
    (0.70,0.48)
    (1.20,0.58)
    (1.80,0.64)
    (2.50,0.68)
    (3.20,0.75)
    (3.80,0.84)
    (4.20,0.95)
    (4.45,1.00)
    (5.80,1.00)
};

\end{axis}
\end{tikzpicture}
    \caption{The solid blue curve plots the distribution function of the $\numunits$-th highest bid $\khighbidCDF(\cdot)$ as a function of bids. The blue cross-hatched area is the optimal utility $u(v)$. The red area above the curve (slanting lines) is the expected critical bid $\widehat{B}$. The green region (vertical lines) has area $v$. Individual efficiency measures what fraction of the green area does the total of blue and red area cover.}
    \label{fig:etadef}
\end{figure}
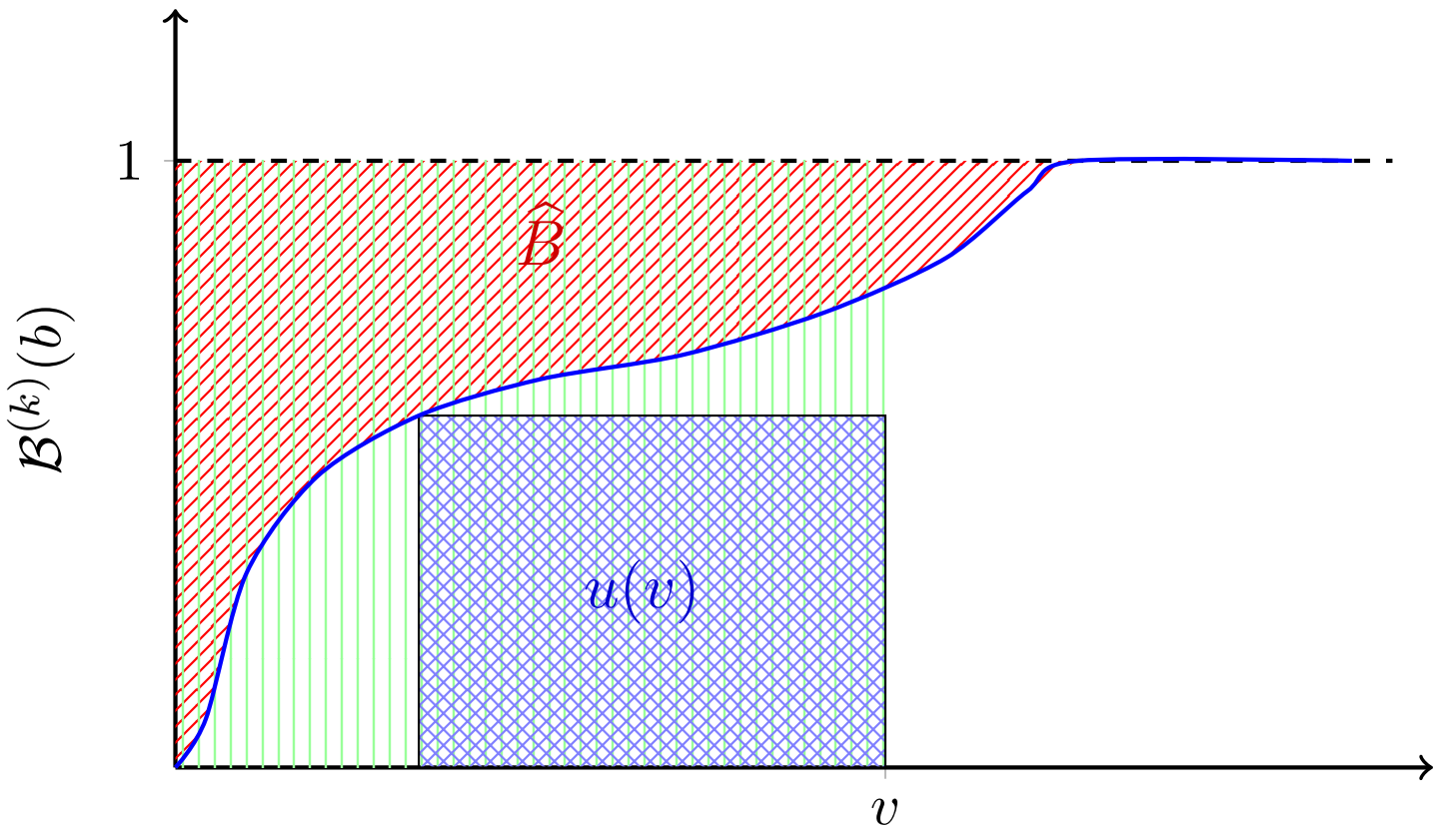

We provide some intuition for this definition. The individual efficiency of an auction captures an inherent tradeoff: either the optimal utility is large ($u(v)$ is large) or the agent faces strong competition in the auction (large expected critical bid $\khighbidCDFexpec$). Then, the sum of utility and expected critical bid recovers a large fraction of available surplus. \Cref{fig:etadef} provides a geometric interpretation. 

To further illustrate this definition, we present the individual efficiency guarantee for any multi-unit first-price auction together with a proof (see \citealp{HHT-23} for individual efficiency guarantees in other auction environments). 

\begin{example}
    \label{ex:singleuniteta}
     The individual efficiency of any multi-unit first-price auction is at least $1-1/e \approx 0.63$. The proof constructs the following randomized bidding strategy for an entrant with value $v$. The entrant draws a bid according to the density $s(v,b) = 1/(v-b)$ supported on $b \in [0,(1-1/e)v]$. The incumbents' bids induce a distribution over the entrant's critical bid, denoted by $\khighbidCDF$. Under this strategy, the entrant's expected utility is $$\mathbb{E}_{\hat{b} \sim \khighbidCDF} \left[\int_{\hat{b}}^{(1-1/e)v}(v-b) \cdot \frac{1}{v-b} db \right]  \geq \left(1-\frac{1}{e}\right)v - \khighbidCDFexpec,$$ where the expectation is over the critical bid. The optimal utility $u(v)$, then, satisfies $$u(v) \geq \left(1-\frac{1}{e}\right)v - \khighbidCDFexpec.$$ Rearranging the equation establishes the bound on individual efficiency. Notably, the argument does not rely on the critical bid $\khighbidCDF$ being induced by a fixed equilibrium. As a result, the individual efficiency bound holds uniformly across all equilibria.
\end{example}

Next, the framework defines the \textit{competitive efficiency} of an auction. This concept measures the extent to which competition among bidders translates into revenue for the seller. Let $\deterministicbidprofile = \left(\bid{1},\bid{2},\dots,\bid{\numagents}\right)$ be a deterministic bid profile and $\hat{b}$ be the critical bid of an entrant agent given this bid profile. Furthermore, let $\mathrm{Revenue}(\deterministicbidprofile)$ denote the revenue of the auction under bid profile $\deterministicbidprofile$.

\begin{definition}[Competitive Efficiency] \label{def:competitiveefficiency}
    An auction $\auction$ is $\revenuecover$-competitive efficient if, for every bid profile $\deterministicbidprofile$ and every feasible allocation $\bm{y} \in \mathcal{X}$, where $\mathcal{X}$ is the set of feasible allocations, we have \begin{align*}
    \revenuecover \mathrm{Revenue}(\deterministicbidprofile) \geq \hat{b} \left( \sum_{i \in \agentset} y_{i} \right).
\end{align*} The competitive efficiency of an auction is the smallest $\revenuecover$ such that the auction is $\revenuecover$-competitive efficient.
\end{definition}

Here, competition among bidders is measured by the total critical bid of allocated agents, given by $$\hat{b}\sum_{i \in \agentset}y_{i},$$ where $\bm{y}$ is a feasible allocation. Then, a small competitive efficiency (small $\revenuecover$) translates to high revenue for the seller, whereas a large competitive efficiency (large $\revenuecover$) translates to low revenue. A straightforward argument establishes the competitive efficiency for any multi-unit first-price auction. 

\begingroup
\addtocounter{example}{-1}
\begin{example}[continued]
Any multi-unit first-price auction has competitive efficiency $1$.
\end{example}
\endgroup

Together, individual and competitive efficiency and the extent to which agents optimize their bids determine the PoA of an auction. We now provide a brief discussion of how these quantities interact. Welfare can be decomposed as the total sum of agents' utility and revenue to the seller. The PoA, then, is the fraction of available surplus that equilibrium welfare can recover. 


Individual efficiency measures the fraction of available surplus that the auction recovers from an entrant agent. The quantity of interest, however, is the fraction of surplus recovered from a participating agent. How does individual efficiency relate to this? Suppose that a participating agent has identical value to the entrant. By marginally increasing the entrant player's bid, this participating agent can obtain utility that is weakly greater than the entrant. Because the participant is $\epsilon$-approximate best responding, the equilibrium utility is at least the utility of entrant minus an additive $\epsilon$ factor. By definition of individual efficiency, then, the sum of the participating agent's utility and expected critical bid recovers an identical fraction of surplus with an additive $\epsilon$ loss. 

Competitive efficiency measures the fraction of the critical bid that converts into revenue for the seller. To analyze the PoA, then, the fraction of surplus recovered by the agent's utility and critical bid needs to be scaled by the competitive efficiency. Our focus is on multi-unit first-price auctions which have competitive efficiency $1$. Therefore the PoA depends only on individual efficiency and the best response approximation factor. We formalize the discussion above in the result below.  

\begin{proposition}
    \label{p:poafixauction}
    Suppose the $\numunits$-unit first-price auction is $\eta$-individually efficient. For every $\epsilon$-approximate equilibrium $\equilibrium$ satisfying \Cref{as:nooverbidding}, the ratio of equilibrium welfare to optimal welfare satisfies \begin{align*}
\frac{\mathcal{W}\left(\equilibrium,\auction\right)}{\mathcal{O}\left(\mathcal{A}\right)} \geq \eta -  \epsilon \cdot \frac{\numunits}{\mathcal{O}(\auction)}
    \end{align*}
\end{proposition}

\begin{proof}
    By definition of $\eta$-individual efficiency, the utility of the entrant agent with value $v$ satisfies \begin{align} \label{eq:etadef}
         u(v)  + \widehat{B} \geq \eta v.
    \end{align} We denote the entrant agent's strategy by $\equilibrium_{*}$. Consider any participating agent $i \in \agentset$. Fix a $\Delta>0$ and suppose agent $i$ plays the bid $b+\Delta$ where $b \sim \equilibrium_{*}$. For any value $v_{i}$, the agent's equilibrium utility satisfies \begin{align*}
        u_{i}(v_{i}) &\geq \mathbb{E}_{b \sim \equilibrium_{*}}\left[u_{i}(v_{i},b+\Delta)\right] - \epsilon, \\
        & \geq u(v_{i}) - \Delta - \epsilon,
    \end{align*} where the second inequality follows from \Cref{fact:lballoc}. Substituting (\ref{eq:etadef}) in the inequality above, we get \begin{align} \label{eq:saucutil}
        u_{i}(v_{i}) + \widehat{B} \geq \eta v_{i} - \Delta - \epsilon. 
    \end{align} Consider a valuation profile $\bm{v} = (v_{1},v_{2},\dots,v_{\numagents^{(\numunits)}})$ and agent $i$. Suppose $\feasibleallocation^*(\bm{v})$ is a welfare maximizing allocation. Since $y_{i}^*(\bm{v}) \in [0,1]$, we can multiply (\ref{eq:saucutil}) by $y_{i}^*(\bm{v})$ to obtain \begin{align*}
        u_{i}(v_{i}) + \widehat{B}y_{i}^{*}(\bm{v}) \geq \eta v_{i} y_{i}^*(\bm{v}) - \Delta y_{i}^*(\bm{v}) - \epsilon y_{i}^*(\bm{v}).
    \end{align*} The feasibility constraint of the auction ensures that the total allocation does not exceed the supply, that is, $\sum_{i \in \agentset}y_{i}^*(\bm{v}) = k$. Substituting in the equation above gives \begin{align} \label{eq:sauctotalutil}
       \sum_{i \in \agentset}u_{i}(v_{i}) + \widehat{B} \numunits \geq \eta \sum_{i \in \agentset}v_{i}y_{i}^*(\bm{v}) - \numunits \left(\Delta + \epsilon\right).
    \end{align} Now, the competitive efficiency of a multi-unit winner-pays-bid auction is $1$, that is, the expected revenue $\expect \left[\mathrm{Revenue}(\deterministicbidprofile)\right] \geq \expect \left[\expectedcriticalbid{}\sum_{i \in \agentset}y_{i}^*(\bm{v})\right]$. Substituting this in (\ref{eq:sauctotalutil}) gives \begin{align*}
        \sum_{i \in \agentset}u_{i}(v_{i}) + \expect \left[\mathrm{Revenue}(\deterministicbidprofile)\right] \geq \eta \sum_{i \in \agentset}v_{i}y_{i}^*(\bm{v}) - \numunits \left(\Delta + \epsilon\right).
    \end{align*}  Taking an expectation with respect to the value distributions $\bm{\mathcal{V}} = \left(\valueCDF{i}\right)_{i \in \agentset^{(\numunits)}}$ gives \begin{align*}
\frac{\mathcal{W}\left(\equilibrium,\auction\right)}{\mathcal{O}\left(\mathcal{A}\right)} \geq \eta -  \left(\epsilon + \Delta \right) \cdot \frac{\numunits}{\mathcal{O}(\auction)}.
    \end{align*} Taking the limit $\Delta \to 0$ establishes the statement of the result. 
\end{proof}

\begin{figure}[t]
    \centering
    \begin{subfigure}[t]{0.48\textwidth}
        \centering

        \begin{tikzpicture}
        \begin{axis}[
            width=\linewidth,
            height=0.83\linewidth,
            axis lines=left,
            xmin=0, xmax=6.2,
            ymin=0, ymax=1.25,
            xlabel={},
            ylabel={$\khighbidCDF(b)$},
            xtick={3.5},
            xticklabels={$v$},
            ytick={1},
            yticklabels={$1$},
            axis line style={->, thick},
            clip=false,
        ]

        \def\bid{1.2}
        \def\vval{3.5}
        \def\Bbid{0.58}

        \path[
            draw=none,
            fill=green!10,
            pattern=vertical lines,
            pattern color=green!40
        ]
        (axis cs:0,0) rectangle (axis cs:\vval,1);

        \addplot[
            name path=toplineA,
            draw=none,
            domain=0:4.45
        ] {1};

        \addplot[
            name path=curvefillA,
            draw=none,
            smooth
        ] coordinates {
            (0,0)
            (0.15,0.08)
            (0.35,0.32)
            (0.70,0.48)
            (1.20,0.58)
            (1.80,0.64)
            (2.50,0.68)
            (3.20,0.75)
            (3.80,0.84)
            (4.20,0.95)
            (4.45,1.00)
        };

        \addplot[
            draw=none,
            pattern=north east lines,
            pattern color=red!80
        ]
        fill between[
            of=toplineA and curvefillA
        ];

        \node[
            text=red!80!black,
            font=\large
        ] at (axis cs:1.8,0.88) {$\widehat{B}$};

        \addplot[
            dashed,
            thick,
            domain=0:6
        ] {1};

        \path[
            draw=black,
            fill=blue!10,
            pattern=crosshatch,
            pattern color=blue!50
        ]
        (axis cs:\bid,0)
        rectangle
        (axis cs:\vval,\Bbid);

        \node[
            text=blue!80!black,
            font=\large
        ] at (axis cs:2.3,0.29) {$u(v)$};

        \addplot[
            blue,
            thick,
            smooth
        ] coordinates {
            (0,0)
            (0.15,0.08)
            (0.35,0.32)
            (0.70,0.48)
            (1.20,0.58)
            (1.80,0.64)
            (2.50,0.68)
            (3.20,0.75)
            (3.80,0.84)
            (4.20,0.95)
            (4.45,1.00)
            (5.80,1.00)
        };

        \end{axis}
        \end{tikzpicture}

        \caption{$\eta<1$}
        \label{fig:eta-general}
    \end{subfigure}
    \hfill
    \begin{subfigure}[t]{0.48\textwidth}
        \centering

        \begin{tikzpicture}
        \begin{axis}[
            width=\linewidth,
            height=0.83\linewidth,
            axis lines=left,
            xmin=0, xmax=6.2,
            ymin=0, ymax=1.25,
            xlabel={},
            ylabel={
                $\khighbidCDF(b)$
            },
            xtick={4.0},
            xticklabels={$v$},
            ytick={1},
            yticklabels={$1$},
            axis line style={->, thick},
            clip=false,
        ]

        \def\bid{2.4}
        \def\vval{4.0}
        \def\Bbid{1.0}

        \path[
            draw=none,
            fill=green!10,
            pattern=vertical lines,
            pattern color=green!40
        ]
        (axis cs:0,0)
        rectangle
        (axis cs:\vval,\Bbid);

        \addplot[
            draw=none,
            pattern=north east lines,
            pattern color=red!80
        ] coordinates {
            (0,0)
            (\bid,0)
            (\bid,\Bbid)
            (0,\Bbid)
            (0,0)
        };

        \node[
            text=red!80!black,
            font=\large
        ] at (axis cs:1.2,0.50) {$\widehat{B}$};

        \path[
            draw=black,
            fill=blue!10,
            pattern=crosshatch,
            pattern color=blue!50
        ]
        (axis cs:\bid,0)
        rectangle
        (axis cs:\vval,\Bbid);

        \node[
            text=blue!80!black,
            font=\large
        ] at (axis cs:3.2,0.50) {$u(v)$};

        \addplot[
            dashed,
            thick,
            domain=0:\bid
        ] {1};

        \addplot[
            blue,
            thick,
            domain=0:\bid
        ] {0};

        \addplot[
            blue,
            thick,
            domain=\bid:5.8
        ] {1};

        \addplot[
            only marks,
            mark=o,
            mark size=2.2pt,
            thick,
            blue
        ] coordinates {
            (\bid,0)
        };

        \addplot[
            only marks,
            mark=*,
            mark size=2.2pt,
            blue
        ] coordinates {
            (\bid,1)
        };
        \end{axis}
        \end{tikzpicture}

        \caption{$\eta=1$}
        \label{fig:eta-deterministic}
    \end{subfigure}

    \caption{
        The solid blue curves plot the distribution function of the
        $\numunits$-th highest bid. The blue cross-hatched area represents
        the optimal utility $u(v)$, the red area represents the expected
        critical bid $\widehat{B}$, and the green vertically hatched region
        has area $v$.
    }
    \label{fig:etacomparison}
\end{figure}
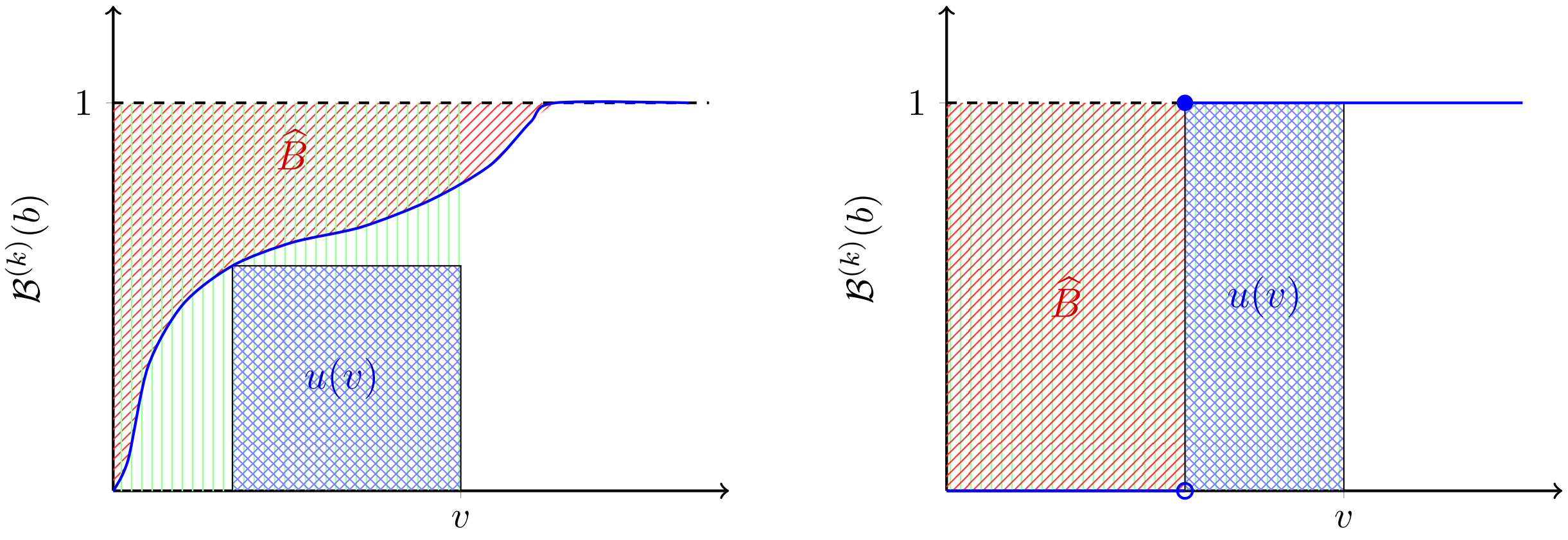

The result provides a useful interpretation of how individual efficiency determines welfare loss which we discuss below. This will play an important role in the asymptotic analysis of \Cref{ss:generalframework}. Inefficiency is primarily driven by uncertainty in the critical bid $\khighbidCDF$. To see how this relates to individual efficiency $\eta$, compare the two panels in \Cref{fig:etacomparison}. When there is uncertainty in the critical bid, as depicted in \Cref{fig:eta-general}, we have $\eta<1$. When the critical bid is deterministic, as depicted in \Cref{fig:eta-deterministic}, we have $\eta=1$. 

In view of what we have just said, it is important to understand properties of the critical bid and their effect on inefficiency. This is the primary focus of the rest of the paper. In particular, we analyze the structure of $\khighbidCDF$ as the number of units $\numunits$ goes to infinity and how this governs PoA in the limit. Furthermore, we study how this structure depends on agents' approximate best response behavior. 

\subsection{Framework for Asymptotic Efficiency} \label{ss:generalframework}

The framework and corresponding objects defined thus far have analyzed an auction with a fixed environment. This section analyzes the role these objects play in determining PoA in a sequence of auctions with an increasing number of units. 

We begin with some notation and then introduce an object --- extremizing sequence of bid pairs --- that plays a central role in our asymptotic analysis. Consider a sequence of $\numunits$-unit first-price auctions $\left(\auction^{(\numunits)}\right)_{\numunits \in \mathbb{N}}$ and a corresponding sequence of equilibria $\left(\equilibrium^{(\numunits)}\right)_{\numunits \in \mathbb{N}}$. The critical bid is denoted by $\khighbidCDF$ and we overload notation to denote its cumulative distribution by the function $\khighbidCDF(\cdot)$.




A bid pair $ \left \{\bidpairlow{},\bidpairhigh{}\right \}$, where the bids $\bidpairlow{} \leq \bidpairhigh{}$, is defined for every auction $\auction^{(\numunits)}$. We now define what it means for a sequence of bid pairs to be \textit{extremizing}.

\begin{definition}[Extremizing Sequence Of Bid Pairs] A sequence of bid pairs $$\left(\left\{\bidpairlow{},\bidpairhigh{}\right\}\right)_{\numunits \in \mathbb{N}},$$ is said to be \textit{extremizing} along subsequence $\left(\auction^{\left(\numunits_{l}\right)}\right)_{l=1}^{\infty}$ if 
    \begin{enumerate}[(a)]
        \item The limit of the distribution function $\khighbidCDFsubseq$ evaluated at $\bidpairlowsubseq{}$ satisfies 
       \begin{align*}
            \lim_{l \to \infty} \khighbidCDFsubseq\left(\bidpairlowsubseq{}\right) = 0 \text{.}
        \end{align*}
        \item The limit of the distribution function $\khighbidCDFsubseq$ evaluated at $\bidpairhighsubseq{}$ satisfies
        \begin{align*}
            \lim_{l \to \infty} \khighbidCDFsubseq\left(\bidpairhighsubseq{}\right) = 1 \text{.}
        \end{align*}
    \end{enumerate}
\end{definition}

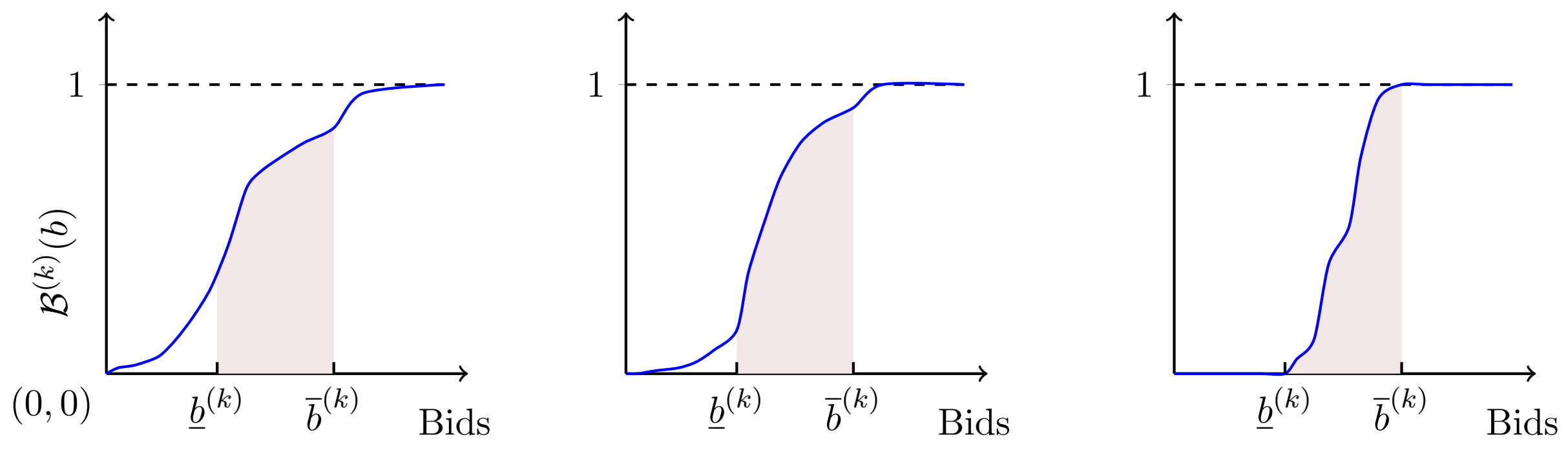
\begin{figure}[!t]
    \centering

    \begin{subfigure}[t]{0.32\textwidth}
        \centering
        \begin{tikzpicture}
        \begin{axis}[
            width=\linewidth,
            height=\linewidth,
            axis lines=left,
            xmin=0, xmax=6.2,
            ymin=0, ymax=1.25,
            xlabel={Bids},
            ylabel={$\khighbidCDF(b)$},
            xtick=\empty,
            ytick={1},
            yticklabels={$1$},
            xlabel style={at={(axis description cs:1.1,-0.06)}, anchor=north east},
            ylabel style={rotate=0, left},
            axis line style={->, thick},
            clip=false,
        ]

        \def\psilow{1.9}
        \def\psihigh{3.90}

        \addplot[
            dashed,
            thick,
            domain=0:5.8
        ] {1};

        \addplot[
            name path=baselineA,
            draw=none,
            domain=0:5.8
        ] {0};

        \addplot[
            name path=curvefillA,
            draw=none,
            smooth
        ] coordinates {
            (0,0)
            (0.20,0.02)
            (0.50,0.03)
            (0.90,0.06)
            (1.20,0.12)
            (1.50,0.20)
            (1.80,0.30)
            (2.10,0.45)
            (2.40,0.64)
            (2.65,0.70)
            (3.00,0.75)
            (3.40,0.80)
            (3.90,0.85)
            (4.40,0.97)
            (5.80,1.00)
        };

        \addplot[
            draw=none,
            pattern=crosshatch,
            pattern color=maroonshade,
            fill=maroonshade!10
        ] fill between[
            of=curvefillA and baselineA,
            soft clip={domain=\psilow:\psihigh}
        ];

        \addplot[
            blue,
            thick,
            smooth
        ] coordinates {
            (0,0)
            (0.20,0.02)
            (0.50,0.03)
            (0.90,0.06)
            (1.20,0.12)
            (1.50,0.20)
            (1.80,0.30)
            (2.10,0.45)
            (2.40,0.64)
            (2.65,0.70)
            (3.00,0.75)
            (3.40,0.80)
            (3.90,0.85)
            (4.40,0.97)
            (5.80,1.00)
        };

        \draw[thick] (axis cs:\psilow,0) -- (axis cs:\psilow,0.04);
        \draw[thick] (axis cs:\psihigh,0) -- (axis cs:\psihigh,0.04);

        \node[below] at (axis cs:\psilow,0) {$\bidpairlow{}$};
        \node[below] at (axis cs:\psihigh,0) {$\bidpairhigh{}$};
\node[below left] at (axis cs:0,0) {$(0,0)$};
        \end{axis}
        \end{tikzpicture}
        \caption{Low $\numunits$}
    \end{subfigure}
    \hfill
    \begin{subfigure}[t]{0.32\textwidth}
        \centering
        \begin{tikzpicture}
        \begin{axis}[
            width=\linewidth,
            height=\linewidth,
            axis lines=left,
            xmin=0, xmax=6.2,
            ymin=0, ymax=1.25,
            xlabel={Bids},
            ylabel={},
            xtick=\empty,
            ytick={1},
            yticklabels={$1$},
            xlabel style={at={(axis description cs:1.1,-0.06)}, anchor=north east},
            axis line style={->, thick},
            clip=false,
        ]

        \def\psilow{1.9}
        \def\psihigh{3.90}

        \addplot[
            dashed,
            thick,
            domain=0:5.8
        ] {1};

        \addplot[
            name path=baselineB,
            draw=none,
            domain=0:5.8
        ] {0};

        \addplot[
            name path=curvefillB,
            draw=none,
            smooth
        ] coordinates {
            (0,0)
            (0.20,0.00)
            (0.50,0.01)
            (0.90,0.02)
            (1.20,0.04)
            (1.50,0.08)
            (1.90,0.15)
            (2.10,0.35)
            (2.40,0.54)
            (2.65,0.68)
            (3.00,0.80)
            (3.40,0.87)
            (3.90,0.92)
            (4.40,1.00)
            (5.80,1.00)
        };

        \addplot[
            draw=none,
            pattern=crosshatch,
            pattern color=maroonshade,
            fill=maroonshade!10
        ] fill between[
            of=curvefillB and baselineB,
            soft clip={domain=\psilow:\psihigh}
        ];

        \addplot[
            blue,
            thick,
            smooth
        ] coordinates {
            (0,0)
            (0.20,0.00)
            (0.50,0.01)
            (0.90,0.02)
            (1.20,0.04)
            (1.50,0.08)
            (1.90,0.15)
            (2.10,0.35)
            (2.40,0.54)
            (2.65,0.68)
            (3.00,0.80)
            (3.40,0.87)
            (3.90,0.92)
            (4.40,1.00)
            (5.80,1.00)
        };

        \draw[thick] (axis cs:\psilow,0) -- (axis cs:\psilow,0.04);
        \draw[thick] (axis cs:\psihigh,0) -- (axis cs:\psihigh,0.04);

        \node[below] at (axis cs:\psilow,0) {$\bidpairlow{}$};
        \node[below] at (axis cs:\psihigh,0) {$\bidpairhigh{}$};

        \end{axis}
        \end{tikzpicture}
        \caption{Intermediate $\numunits$}
    \end{subfigure}
    \hfill
    \begin{subfigure}[t]{0.32\textwidth}
        \centering
        \begin{tikzpicture}
        \begin{axis}[
            width=\linewidth,
            height=\linewidth,
            axis lines=left,
            xmin=0, xmax=6.2,
            ymin=0, ymax=1.25,
            xlabel={Bids},
            ylabel={},
            xtick=\empty,
            ytick={1},
            yticklabels={$1$},
            xlabel style={at={(axis description cs:1.1,-0.06)}, anchor=north east},
            axis line style={->, thick},
            clip=false,
        ]

        \def\psilow{1.9}
        \def\psihigh{3.90}

        \addplot[
            dashed,
            thick,
            domain=0:5.8
        ] {1};

        \addplot[
            name path=baselineC,
            draw=none,
            domain=0:5.8
        ] {0};

        \addplot[
            name path=curvefillC,
            draw=none,
            smooth
        ] coordinates {
            (0,0)
            (0.20,0.00)
            (0.50,0.00)
            (0.90,0.00)
            (1.20,0.00)
            (1.50,0.00)
            (1.90,0.00)
            (2.10,0.05)
            (2.40,0.12)
            (2.65,0.38)
            (3.00,0.51)
            (3.20,0.75)
            (3.50,0.95)
            (3.90,1.00)
            (4.40,1.00)
            (5.80,1.00)
        };

        \addplot[
            draw=none,
            pattern=crosshatch,
            pattern color=maroonshade,
            fill=maroonshade!10
        ] fill between[
            of=curvefillC and baselineC,
            soft clip={domain=\psilow:\psihigh}
        ];

        \addplot[
            blue,
            thick,
            smooth
        ] coordinates {
            (0,0)
            (0.20,0.00)
            (0.50,0.00)
            (0.90,0.00)
            (1.20,0.00)
            (1.50,0.00)
            (1.90,0.00)
            (2.10,0.05)
            (2.40,0.12)
            (2.65,0.38)
            (3.00,0.51)
            (3.20,0.75)
            (3.50,0.95)
            (3.90,1.00)
            (4.40,1.00)
            (5.80,1.00)
        };

        \draw[thick] (axis cs:\psilow,0) -- (axis cs:\psilow,0.04);
        \draw[thick] (axis cs:\psihigh,0) -- (axis cs:\psihigh,0.04);

        \node[below] at (axis cs:\psilow,0) {$\bidpairlow{}$};
        \node[below] at (axis cs:\psihigh,0) {$\bidpairhigh{}$};

        \end{axis}
        \end{tikzpicture}
        \caption{Large $\numunits$}
    \end{subfigure}

    \caption{An illustration of an extremizing sequence of bid pairs. In each figure, the solid blue line plots the distribution function of the critical bid $\khighbidCDF$. The maroon cross-hatched region is the area below the curve between $\bidpairlow{}$ and $\bidpairhigh{}$. Panel (a) depicts the distribution for a low number of units, Panel (b) for an intermediate number of units, and Panel (c) for a large number of units.}
    \label{fig:extremizingsequence}
\end{figure}

We will soon analyze the interval $\left(\bidpairlow{},\bidpairhigh{}\right]$, defined above.
To see why this is important for our analysis, the critical bid lies within this interval with probability tending to one. There is no uncertainty in the critical bid outside it. Any loss in efficiency due to uncertainty is determined by the behavior of the critical bid within such an interval. \Cref{fig:extremizingsequence} shows an illustration of an extremizing sequence of bid pairs. When bid distributions have no atoms, a trivial extremizing bid pair sequence is one where the lower bid is zero and the higher bid is the maximum equilibrium bid. We provide a non-trivial construction in \Cref{s:welfareanalysisfpa}. 

Before we present the main result of this section, we introduce an extension of the standard no over-bidding assumption in  \Cref{as:nooverbidding}. 



\begin{assumption} \label{as:generalnooverbidding}
  The sequence of equilibrium bid distribution profiles $\left(\equilibrium^{(\numunits)}\right)_{\numunits \in \mathbb{N}}$ are such that there exists a sequence of positive parameters $\left(\omega^{(\numunits)}\right)_{\numunits=1}^{\infty}$ where $\omega^{(\numunits)} = o(k/n^{(\numunits)})$ so that for every $\numunits$, every agent $i \in \agentset^{(\numunits)}$ and every value $v_{i}$ in the support of $\mathcal{V}_{i}$, the equilibrium bid $\equilibrium^{(\numunits)}_{i,v_{i}}$ is almost surely at most $v_{i} + \omega^{(\numunits)}$.
\end{assumption}

We state our result under this more general condition so that we can apply it to establish PoA bounds in the next section. The crux of our approach is to reduce to equilibria with no atoms, which involves perturbations that might lead to over-bidding. It turns out that it suffices to establish results under this general setting. We now present the main result of this section. 

\begin{theorem}
    \label{lemma:generalposepslim}
  Consider any sequence of $\numunits$-unit first-price auctions $\left(\auction^{(\numunits)}\right)_{\numunits \in \mathbb{N}}$ satisfying \Cref{as:lowerboundexpectation} and any corresponding sequence of $\epsilon^{(\numunits)}$-approximate Bayes-Nash equilibrium $\left(\equilibrium^{(\numunits)}\right)_{\numunits \in \mathbb{N}}$ satisfying \Cref{as:generalnooverbidding}. Let $\epsilon := \lim_{\numunits \to \infty}\epsilon^{(\numunits)}$. 
  
  Suppose there exists a $c \in \mathbb{R}_{\geq 0}$ and an extremizing sequence of bid pairs $\left( \left\{\bidpairlow{},\bidpairhigh{}\right\} \right)_{\numunits \in \mathbb{N}}$ along subsequence $\left(\auction^{\left(\numunits_{\subseqind}\right)}\right)_{\subseqind=1}^{\infty}$ satisfying \begin{align*}
        \lim_{\subseqind \to \infty} \int_{\bidpairlowsubseq{}}^{\bidpairhighsubseq{}}\khighbidCDFsubseq(b)db \leq c. \end{align*} Then, we have \begin{align*}
        \liminf_{l \to \infty} \frac{\mathcal{W}\left(\equilibrium^{\left(\numunits_{\subseqind}\right)},\auction^{\left(\numunits_{\subseqind}\right)}\right)}{\mathcal{O}\left(\auction^{\left(\numunits_{\subseqind}\right)}\right)} \geq 1- \frac{c}{\zeta}-\frac{\epsilon}{\zeta},
    \end{align*} where $\zeta>0$ is a constant that depends on the parameters of the environment. 
\end{theorem}

\begin{proof}
Fix any $c\geq0$. By assumption, let $\left\{ \left(\bidpairlow{},\bidpairhigh{}\right) \right \}_{\numunits=1}^{\infty}$ be the extremizing sequence of bid pairs and $\left( \mathcal{A}^{(\numunits_{\subseqind})} \right)_{\subseqind = 1}^{\infty}$ the corresponding subsequence of auctions satisfying \begin{align*}
        \lim_{\subseqind \to \infty} \int_{\bidpairlowsubseq{}}^{\bidpairhighsubseq{}}\khighbidCDFsubseq(b)db \leq c.
    \end{align*} For the rest of the proof, we suppress the subsequence index to simplify notation. Thus, whenever we take limits, they are to be understood as limits along the subsequence of auctions $\left( \mathcal{A}^{(\numunits_{\subseqind})} \right)_{\subseqind = 1}^{\infty}$. 
    
    Consider any agent $i \in \agentset^{(\numunits)}$ and value $v_{i}$. Furthermore, consider a sequence of perturbation parameters $\left(\Delta^{(\numunits)}\right)_{\numunits=1}^{\infty}$ such that $\Delta^{(\numunits)} > 0$ and $\lim_{\numunits \to \infty} \Delta^{(\numunits)} = 0$. Suppose the agent bids $\bidpairhigh{}+\Delta^{(\numunits)}$ and the value $v_{i} \geq \bidpairhigh{}+\Delta^{(\numunits)}$. Let $u_{i}(v_{i})$ denote the agent's equilibrium utility and $u_{i}\left(v_{i},\bid{}\right)$ denote the utility obtained from bidding $b$. Since the strategy profile $\equilibrium^{(\numunits)}$ is an $\epsilon^{(\numunits)}$-approximate equilibrium, we have
    \begin{align}
        u_{i}(v_{i}) & \geq u_{i}\left(v_{i},\bidpairhigh{} + \Delta^{(\numunits)}\right) - \epsilon^{(\numunits)} \text{,}
        \\ &= \left(v_{i} - \bidpairhigh{}-\Delta^{(\numunits)}\right)x_{i}\left(\bidpairhigh{}+\Delta^{(\numunits)}\right) - \epsilon^{(\numunits)} \text{,}  \\
        & \geq \left(v_{i} - \bidpairhigh{}-\Delta^{(\numunits)}\right) \khighbidCDF\left(\bidpairhigh{}\right) - \epsilon^{(\numunits)} \text{.} \label{eq:lowerboundeqlbutil}
    \end{align}
    Here, the second inequality follows from Fact \ref{fact:lballoc}. By assumption, $\omega^{(\numunits)} > 0$. Substituting in (\ref{eq:lowerboundeqlbutil}) gives \begin{align}
       u_{i}\left(v_{i}\right) + \omega^{(\numunits)} \geq \left(v_{i} - \bidpairhigh{}-\Delta^{(\numunits)}\right) \khighbidCDF\left(\bidpairhigh{}\right) - \epsilon^{(\numunits)} \text{.} \label{eq:lowerboundeqlbutilnew}
    \end{align} Next, suppose $v_{i} < \bidpairhigh{}+\Delta^{(\numunits)}$. By assumption, the equilibrium bid distributions allows over-bidding. Thus, the interim utility $u_{i}(v_{i})$ can be negative. However, by assumption, it must satisfy $u_{i}\left(v_{i}\right) \geq -\omega^{(\numunits)}$. It follows that \begin{align}
        u_{i}\left(v_{i}\right) + \omega^{(\numunits)} \geq \left(v_{i} - \bidpairhigh{}-\Delta^{(\numunits)}\right) \khighbidCDF\left(\bidpairhigh{}\right) - \epsilon^{(\numunits)} \text{.} \label{eq:lowerboundeqlbutilnewoc}
    \end{align}

    We now solve for the expectation of $\expectedcriticalbid{}$. By the tail sum formula for expectations,
    \begin{align*}
        \expectedcriticalbid{} = \int_{0}^{\infty}\left(1-\khighbidCDF(b)\right) d\bid{} \text{.}
    \end{align*}
    By monotonicity of the distribution function $\khighbidCDF(\cdot)$, for any bid $b \leq \bidpairlow{}$ we have $\khighbidCDF(b) \leq \khighbidCDF\left(\bidpairlow{}\right)$. Then, a lower bound on the expectation is
    \begin{align}\label{eq:expectedcriticalbidlbgen}
        \expectedcriticalbid{} \geq \left(1-\khighbidCDF\left(\bidpairlow{}\right)\right) \bidpairlow{} + \left(\bidpairhigh{}-\bidpairlow{}\right) - \int_{\bidpairlow{}}^{\bidpairhigh{}}\khighbidCDF(b)db \text{.}
    \end{align} Combining (\ref{eq:lowerboundeqlbutilnew}), (\ref{eq:lowerboundeqlbutilnewoc}) and (\ref{eq:expectedcriticalbidlbgen}), we have \begin{align}
        u_{i}(v_{i}) + \omega^{(\numunits)} + \expectedcriticalbid{}
        & \geq v_{i}\khighbidCDF\left(\bidpairhigh{}\right) - \bidpairlow{} \khighbidCDF\left(\bidpairlow{}\right) - \epsilon^{(\numunits)} - \Delta^{(\numunits)} \khighbidCDF\left(\bidpairhigh{}\right) -\int_{\bidpairlow{}}^{\bidpairhigh{}}\khighbidCDF(b)db \text{.} \label{eq:individualefficiencyagentgen}
    \end{align} 
    
    Consider a valuation profile $\bm{v} = (v_{1},v_{2},\dots,v_{\numagents^{(\numunits)}})$ and agent $i$. Suppose $\feasibleallocation^*(\bm{v})$ is a welfare maximizing allocation. Since $y_{i}^*(\bm{v}) \in [0,1]$, we can multiply (\ref{eq:individualefficiencyagentgen}) by $y_{i}^*(\bm{v})$ to obtain
    \begin{multline*}
        u_{i}(v_{i}) + \omega^{(\numunits)} + \expectedcriticalbid{}y_{i}^*(\bm{v}) \geq \khighbidCDF\left(\bidpairhigh{}\right) v_{i}y_{i}^*(\bm{v}) - y_{i}^*(\bm{v}) \int_{\bidpairlow{}}^{\bidpairhigh{}}\khighbidCDF(b)db - \Delta^{(\numunits)} \khighbidCDF\left(\bidpairhigh{}\right) y_{i}^*(\bm{v}) \\ - \bidpairlow{} \khighbidCDF\left(\bidpairlow{}\right)y_{i}^*(\bm{v}) - y_{i}^*(\bm{v}) \epsilon^{(\numunits)} \text{.}
    \end{multline*}
    The feasibility constraint of the auction ensures that the total allocation does not exceed the supply, i.e., $\sum_{i \in \agentset^{(\numunits)}} y_i^*(\bm{v}) = k$. By definition of an extremizing bid-pair sequence, along subsequence $\left(\auction^{\left(\numunits_{\subseqind}\right)}\right)_{\subseqind=1}^{\infty}$ we have that $\lim_{l \to \infty} \khighbidCDFsubseq \left(\bidpairlowsubseq{}\right) = 0$. By the uniform bounds on values in \Cref{as:lowerboundexpectation} and \Cref{as:generalnooverbidding}, it must be that $\bidpairlowsubseq{} \leq 2H$ for $l$ large enough: every equilibrium bid is at most $2H$ almost surely for large $l$. The rest of the proof focuses on this large $\subseqind$ along this subsequence. Substituting this upper bound on the bid in the equation above gives \begin{multline} 
        \sum_{i \in \agentset^{(\numunits_{\subseqind})}}u_{i}(v_{i}) + n^{(\numunits_{\subseqind})}\omega^{(\numunits_{\subseqind})} + \expectedcriticalbid{} \sum_{i \in \agentset^{(\numunits_{\subseqind})}}y_{i}^*(\bm{v}) \geq  \khighbidCDFsubseq\left(\bidpairhighsubseq{}\right) \left(  \sum_{i \in \agentset^{(\numunits_{\subseqind})}}v_{i}y_{i}^*(\bm{v}) \right) - \numunits_{\subseqind} \int_{\bidpairlowsubseq{}{}}^{\bidpairhighsubseq{}}\khighbidCDFsubseq(b)db \\ - 2H\numunits_{\subseqind} \khighbidCDFsubseq\left(\bidpairlowsubseq{}\right) - k_{\subseqind} \epsilon^{(\numunits_{\subseqind})} - \numunits_{\subseqind} \Delta^{(\numunits_{\subseqind})} \khighbidCDFsubseq\left(\bidpairhighsubseq{}\right) \text{.} \label{eq:utilcriticomb}
    \end{multline} Now, the competitive efficiency of a multi-unit winner-pays-bid auction is one, that is, the expected revenue $\expect \left[\mathrm{Revenue}(\deterministicbidprofile)\right] \geq \expect \left[\expectedcriticalbid{}\sum_{i \in \agentset^{(\numunits)}}y_{i}^*(\bm{v})\right]$. Substituting this in (\ref{eq:utilcriticomb}) gives
    \begin{multline}
        \sum_{i \in \agentset^{(\numunits_{\subseqind})}}u_{i}(v_{i}) + n^{(\numunits_{\subseqind})}\omega^{(\numunits_{\subseqind})} + \expect \left[\mathrm{Revenue}(\deterministicbidprofile)\right] \geq \khighbidCDFsubseq\left(\bidpairhighsubseq{}\right) \left(  \sum_{i \in \agentset^{(\numunits_{\subseqind})}}v_{i}y_{i}^*(\bm{v}) \right) - \numunits_{\subseqind} \int_{\bidpairlowsubseq{}}^{\bidpairhighsubseq{}}\khighbidCDFsubseq(b)db \\ - 2H\numunits_{\subseqind} \khighbidCDFsubseq\left(\bidpairlowsubseq{}\right) - k_{\subseqind} \epsilon^{(\numunits_{\subseqind})} - \numunits_{\subseqind} \Delta^{(\numunits_{\subseqind})} \text{.}
    \end{multline}
    Taking an expectation with respect to the value distributions $\bm{\mathcal{V}} = \left(\valueCDF{i}\right)_{i \in \agentset^{(\numunits_{\subseqind})}}$ gives
    \begin{align*} \frac{\mathcal{W}\left(\equilibrium^{\left(\numunits_{\subseqind}\right)},\auction^{\left(\numunits_{\subseqind}\right)}\right)}{\mathcal{O}\left(\auction^{\left(\numunits_{\subseqind}\right)}\right)} \geq \khighbidCDFsubseq\left(\bidpairhighsubseq{}\right) - \frac{k_{\subseqind} \cdot \int_{\bidpairlowsubseq{}}^{\bidpairhighsubseq{}}\khighbidCDFsubseq(b)db + 2H\numunits_{\subseqind} \khighbidCDFsubseq\left(\bidpairlowsubseq{}\right) + k_{\subseqind} \epsilon^{(\numunits_{\subseqind})} + \numunits_{\subseqind} \Delta^{(\numunits_{\subseqind})} + n^{(\numunits_\subseqind)}\omega^{(\numunits_{\subseqind})}}{\mathcal{O}\left(\auction^{\left(\numunits_{\subseqind}\right)}\right)} \text{.}
    \end{align*}
    By \Cref{as:lowerboundexpectation}, there exists a constant $\zeta > 0$ such that $\mathcal{O}\left(\auction^{\left(\numunits_{\subseqind}\right)}\right) \geq k_{\subseqind} \zeta$. It follows that \begin{align} \label{eq:poaconvergencegen}
    \frac{\mathcal{W}\left(\equilibrium^{\left(\numunits_{\subseqind}\right)},\auction^{\left(\numunits_{\subseqind}\right)}\right)}{\mathcal{O}\left(\auction^{\left(\numunits_{\subseqind}\right)}\right)} \geq \khighbidCDFsubseq\left(\bidpairhighsubseq{}\right) - \frac{1}{\zeta} \int_{\bidpairlowsubseq{}}^{\bidpairhighsubseq{}}\khighbidCDFsubseq(b)db - \frac{2H}{\zeta} \khighbidCDFsubseq\left(\bidpairlowsubseq{}\right) -\frac{\epsilon^{(\numunits_{\subseqind})}}{\zeta} - \frac{\Delta^{(\numunits_{\subseqind})}}{\zeta} - \frac{n^{(\numunits_{\subseqind})} \omega^{(\numunits_{\subseqind})}}{k_{\subseqind}\zeta}. 
    \end{align} By assumption, the sequence of bid pairs is extremizing. Next, along subsequence $\left(\auction^{(\numunits_\subseqind)}\right)_{\subseqind=1}^{\infty}$, the limit of the integral is $$\lim_{\subseqind \to \infty} \int_{\bidpairlowsubseq{}}^{\bidpairhighsubseq{}}\khighbidCDFsubseq(b)db < c.$$ Finally, the perturbation $\Delta^{(\numunits)}$ converges to $0$ as $\numunits \to \infty$ and by assumption $\omega^{(\numunits)} = o(\numunits/n^{(\numunits)})$. Thus, substituting in (\ref{eq:poaconvergencegen}), the limit of the ratio is \begin{align*}
        \liminf_{\subseqind \to \infty} \frac{\mathcal{W}\left(\equilibrium^{\left(\numunits_{\subseqind}\right)},\auction^{\left(\numunits_{\subseqind}\right)}\right)}{\mathcal{O}\left(\auction^{\left(\numunits_{\subseqind}\right)}\right)} \geq 1 - \frac{c}{\zeta} - \frac{\epsilon}{\zeta}.
    \end{align*} The statement of the result is proved.
\end{proof}

This result says that the PoA in the limit is determined by two factors. First, the best-response approximation $\epsilon$. And second, the limit area below the distribution function of the critical bid restricted within an extremizing bid pair interval. The latter is depicted by the shaded maroon area in each panel of \Cref{fig:extremizingsequence}. This area measures the uncertainty in the critical bid relevant to limit inefficiency. The exact bound in \Cref{lemma:generalposepslim} follows from the geometry of individual efficiency described in \Cref{ss:fixauctionefficiency}. 

So far, we have described some key objects --- critical bids and extremizing bid pair sequences --- and their role in determining the PoA in the limit. In the next section, we derive explicit bounds by characterizing the structure of these objects. 

\section{Asymptotic Welfare Analysis}
\label{s:welfareanalysisfpa}

We provide the main result of this section below.

\begin{theorem}
    \label{t:efficiencyapproxeqlb}
    Consider any sequence of $\numunits$-unit first-price auctions $\left(\auction^{(\numunits)}\right)_{\numunits \in \mathbb{N}}$ satisfying \Cref{as:lowerboundexpectation} and any corresponding sequence of $\epsilon^{(\numunits)}$-approximate Bayes-Nash equilibrium $\left(\equilibrium^{(\numunits)}\right)_{\numunits \in \mathbb{N}}$ satisfying \Cref{as:nooverbidding}. Let $\epsilon := \lim_{\numunits \to \infty}\epsilon^{(\numunits)}$. 
    \begin{enumerate}
        \item If $\epsilon=0$, then $$\liminf_{\numunits \to \infty} \frac{\mathcal{W}\left(\equilibrium^{\left(\numunits\right)},\auction^{\left(\numunits\right)}\right)}{\mathcal{O}\left(\auction^{\left(\numunits\right)}\right)} = 1.$$
        \item If $\epsilon \in (0,H)$, then $$\liminf_{\numunits \to \infty} \frac{\mathcal{W}\left(\equilibrium^{\left(\numunits\right)},\auction^{\left(\numunits\right)}\right)}{\mathcal{O}\left(\auction^{\left(\numunits\right)}\right)} \geq 1 - \zeta \epsilon \left(1+\log \left(\frac{H}{\epsilon}\right)\right),$$ where $\zeta>0$ is a constant that depends on the parameters of the environment. 
    \end{enumerate} 
\end{theorem}

When agents best-respond optimally (or best-respond optimally in the limit), Part (a) of the result says that the PoA converges to $1$ as the number of units goes to infinity. When agents are $\epsilon$-approximate best-responding, the PoA in the limit is at least $1-\Theta\left(\epsilon \log \left(\frac{1}{\epsilon}\right)\right)$. 

In the special case where $\epsilon^{(\numunits)}=0$ for every $\numunits \in \mathbb{N}$, our result recovers the welfare result of \cite{S-99}. Moreover, a Chernoff bound can be applied to that analysis to establish a limit PoA of at least $1-\Theta\left(\epsilon\right)$ for $\epsilon$-approximate equilibrium. \footnote{We thank Jeroen Swinkels for pointing out this extension to the analysis in \cite{S-99}.} Their analysis, however, does not characterize the distribution of the $\numunits$-th order statistic of equilibrium bids which is crucial to applying the smoothness framework.  As described in \Cref{s:introduction}, the main contribution of our paper is to the smoothness framework literature. A key barrier to the proof approach is a worst-case bid distribution which leads to a PoA lower bound of $1-\frac{1}{e}-\epsilon$. The results in this section overcome this existing barrier by showing that such worst-case distributions cannot arise in equilibrium and use this characterization to establish the bounds in \Cref{t:efficiencyapproxeqlb}. Finally, a strength of the smoothness framework is its ability to establish PoA bounds in complex auction environments. Our asymptotic framework in \Cref{s:smoothnessframework} and improved individual efficiency analysis may apply to such settings. We leave these questions for future work. 


Before providing details of the proof, we briefly mention a key technical contribution that facilitates our analysis. In \Cref{s:eqlbstruct}, we show that it is without loss to analyze welfare assuming equilibrium bid distributions have no atoms (see \Cref{t:epsapxnoatom}). This result is surprising. Equilibrium strategies, in particular approximate equilibria, typically have atoms and this poses several analytical challenges. For the rest of this section, we provide results under the following condition. 

\begin{assumption}
    \label{a:noatomsequilibrium}
    Consider any sequence of auctions $\left(\auction^{(\numunits)}\right)_{\numunits \in \mathbb{N}}$. The corresponding sequence of equilibria $\left(\equilibrium^{(\numunits)}\right)_{\numunits \in \mathbb{N}}$ are such that the bid distribution $\equilibrium_{i}^{(\numunits)}$ for every agent $i \in \agentset^{(\numunits)}$ and $\numunits \in \mathbb{N}$ has no atoms. 
\end{assumption}

\subsection{Constructing an extremizing sequence of bid pairs}
\label{ss:extremizingconstruction}

Establishing the PoA using the robust framework described in \Cref{s:smoothnessframework} involves analyzing properties of the critical bid. As discussed previously, the limit behavior of this bid is governed by an extremizing sequence of bid pairs. We define two threshold bids that play a crucial role in constructing such a sequence. 

Consider any auction $\auction^{(\numunits)}$ and equilibrium $\equilibrium^{(\numunits)}$. For every agent $i$, let $\mathcal{T}_{i}(b)$ denote the indicator random variable that takes value $1$ if agent $i$'s equilibrium bid $\equilibrium^{(\numunits)}_{i}$ is strictly above $b$ and $0$ otherwise. The total number of equilibrium bids exceeding $b$ is then given by
\begin{align*}
    \mathcal{T}^{(\numunits)}(b) = \sum_{i \in \correspondingagentset} \mathcal{T}^{(\numunits)}_{i}(b) \text{.}
\end{align*} 


\begin{definition} Fix any constant $\delta \in (0,0.5)$. Consider the following threshold bids:
    \begin{enumerate}[(i)]
        \item The high threshold bid $\highpsi{*}$ is 
\begin{align} 
    \highpsi{*} := \min \left \{b \geq 0 \bigg | \expect[\mathcal{T}^{(\numunits)}(b)] \leq \numunits - \numunits^{0.5+\delta}-1\right \} \text{.} \label{eq:highpsi}
\end{align}
\item  The low threshold bid $\lowpsi{*}$ is \begin{align} 
    \lowpsi{*} := \min \left\{ b \geq 0 \bigg| \expect[\mathcal{T}^{(\numunits)}(b)] \leq \numunits + \numunits^{0.5+\delta} - 1 \right\} \text{.} \label{eq:lowpsi}
\end{align} 
    \end{enumerate}
\end{definition} Both of these quantities exist for $\numunits$ large enough. We briefly argue that the bid pair sequence $$\left(\left\{\lowpsi{*},\highpsi{*}\right\}\right)_{\numunits \in \mathbb{N}}$$ is extremizing and formalize this in \Cref{lemma:extremizingconstructeps} below. Because bid distributions do not have atoms, the inequality in (\ref{eq:highpsi}) holds with equality and (\ref{eq:lowpsi}) holds with equality when $\lowpsi{*} > 0$. That is, the expected total bids above $\highpsi{*}$ equals $\numunits - \numunits^{0.5+\delta} - 1$. Similarly, when $\lowpsi{*}>0$, the expected total bids above $\lowpsi{*}$ equals $\numunits + \numunits^{0.5+\delta}-1$. A standard law of large numbers argument (see \Cref{ss:chernoffbounds} for standard Chernoff bounds) argues that the $\numunits$-th highest bid is within the interval $\left(\lowpsi{*},\highpsi{*}\right)$ with high probability for large $\numunits$. The proof also accounts for the case $\lowpsi{*}=0$.

\begin{lemma}\label{lemma:extremizingconstructeps} Suppose \Cref{a:noatomsequilibrium} is satisfied. Consider any sequence of bid pairs $\left \{ \left(\bidpairlow{},\bidpairhigh{}\right) \right \}_{\numunits=1}^{\infty}$ where $\bidpairlow{} \in \left[0,\lowpsi{*}\right]$ and $\bidpairhigh{} \in \left[\highpsi{*},\infty\right)$. The sequence is extremizing over the sequence of auctions $\left(\auction^{(\numunits)}\right)_{\numunits=1}^{\infty}$. 
\end{lemma}

\begin{proof}
     Consider any $k \in \mathbb{N}$. We begin by analyzing the distribution of the $\numunits^{\text{th}}$ highest bid $\khighbidCDF$ at the higher bid $\bidpairhigh{}$. This involves solving for the probability that at most $\numunits - 1$ equilibrium bids are strictly above $\highpsi{*}$. Consider the following sequence of inequalities:
    \begin{align*}
        \prob\left[ \mathcal{T}\left(\highpsi{*}\right) \leq \numunits - 1 \right] & = \prob \left[\mathcal{T}\left(\highpsi{*}\right) - \expect \left[\mathcal{T}\left(\highpsi{*}\right) \right] \leq  \numunits - 1 - \expect \left[\mathcal{T}\left(\highpsi{*}\right) \right] \right] \text{,} \\
        & \geq \prob\left[  \mathcal{T}\left(\highpsi{*}\right) - \expect\left[ \mathcal{T}\left(\highpsi{*}\right) \right] \leq \numunits^{0.5+\delta} \right] \text{,} \\
        & \geq 1 - 2e^{-\frac{1}{3} \cdot \frac{\left(\numunits\right)^{2\delta}}{1 - \frac{1}{\numunits^{0.5 - \delta}} - \frac{1}{\numunits}}} \text{.}
    \end{align*}
    The first inequality follows when we substitute $ \expect\left[\mathcal{T}\left(\highpsi{*}\right)\right] = \numunits - \numunits^{0.5+\delta} - 1 \label{eq:psihigheq}$, which holds by definition of $\highpsi{*}$ and by assumption that no bid distribution has atoms. The final inequality follows from applying the upper tail Chernoff bound in \Cref{fact:chernoffbound} to $\mathcal{T}\left(\highpsi{*}\right)$ --- a sum of independent indicator random variables.

    We begin by analyzing the probability of the following event: the $\numunits^{\text{th}}$ highest equilibrium bid is at most $\bidpairhigh{}$. This event occurs if the total equilibrium bids strictly above $\highpsi{*}$ is at most $(\numunits-1)$. Thus, the distribution function at $\bidpairhigh{}$ is at least
    \begin{align*}
        \prob \left[\khighbidCDF \leq \bidpairhigh{}\right] \geq 1 - 2e^{-\frac{1}{3} \cdot \frac{\left(\numunits\right)^{2\delta}}{1 - \frac{1}{\numunits^{0.5-\delta}} - \frac{1}{\numunits}}} \text{,}
    \end{align*} and taking the limit $\numunits \to \infty$ implies that $\lim_{\numunits \to \infty} \prob \left[\khighbidCDF \leq \bidpairhigh{}\right] = 1 \text{.}$

    Next, we analyze the distribution $\khighbidCDF$ at the lower bid $\bidpairlow{}$. Suppose $\lowpsi{*} > 0$. By definition of $\lowpsi{*}$ and by assumption that equilibrium bids have no atoms, the expected number of equilibrium bids above $\lowpsi{*}$ satisfies $
        \expect \left[\mathcal{T}\left(\lowpsi{*}\right)\right] = \numunits + \numunits^{0.5+\delta} - 1$. We will use this to establish a concentration inequality on the distribution of the $\numunits$-th highest bid $\khighbidCDF$.  
    
   By monotonicity of the distribution function, we have $\prob  \left[\khighbidCDF \leq \bidpairlow{}\right] \leq \prob \left[\khighbidCDF \leq \lowpsi{*}\right]$. The $\numunits^{\text{th}}$ highest equilibrium bid is at most $\lowpsi{*}$, then at most $(\numunits-1)$ bids are strictly above $\lowpsi{*}$. Consider the following sequence of inequalities:
    \begin{align}
        \prob\left[\mathcal{T}\left(\lowpsi{*}\right) \leq \numunits - 1\right] & = \prob\left[ \mathcal{T}\left(\lowpsi{*}\right) - \expect \left[\mathcal{T}\left(\lowpsi{*}\right)\right] \leq \numunits-1- \expect \left[\mathcal{T}\left(\lowpsi{*}\right)\right] \right] \text{,} \\ \label{eq:lowbidfineq}
        & = \prob\left[  \mathcal{T}\left(\lowpsi{*}\right) - \expect\left[ \mathcal{T}\left(\lowpsi{*}\right) \right] \leq - \numunits^{0.5+\delta} \right] \text{,} \\ \label{eq:lowbidsineq}
        & \leq e^{-\frac{1}{2} \cdot \frac{\left(\numunits\right)^{2\delta}}{1 + \frac{1}{\numunits^{0.5 - \delta}} - \frac{1}{\numunits}}} \text{.}
    \end{align} The inequality follows from applying the lower tail Chernoff bound in \Cref{fact:chernoffbound}. Thus, the distribution function at $\bidpairlow{}$ is at most \begin{align*}
        \prob \left[\khighbidCDF \leq \bidpairlow{}\right] \leq  e^{-\frac{1}{2} \cdot \frac{\left(\numunits\right)^{2\delta}}{1 + \frac{1}{\numunits^{0.5 - \delta}} - \frac{1}{\numunits}}}.
    \end{align*} Now, suppose $\lowpsi{*}=0$. Since no distribution has an atom, $\prob \left[\khighbidCDF \leq 0\right] = 0$. Taking the limit $\numunits \to \infty$ gives $
       \lim_{\numunits \to \infty} \prob \left[\khighbidCDF \leq \bidpairlow{}\right]  = 0
    $ and the statement of the result follows. 
\end{proof} 

The loss in efficiency is determined by the behavior of the $\numunits$-th highest bid within the interval constructed above. We will show in the next section that this crucially depends on the agents' approximate best response factor $\epsilon$. Before that, we prove a useful result which establishes that the allocation probability for any agent at any bid strictly smaller than $\lowpsi{*}$ vanishes. This conclusion is not immediate from \Cref{lemma:extremizingconstructeps} because the $\numunits$-th order statistic is only a lower bound on the allocation probability (see Fact \ref{fact:lballoc}). 

\begin{lemma} \label{lemma:vanishingallocprob}
   Suppose equilibrium bid distributions $\equilibrium^{(\numunits)}$ satisfy \Cref{a:noatomsequilibrium}. Consider any sequence of bids $\left(\bidpairlow{}\right)_{\numunits=1}^{\infty}$ satisfying $\bidpairlow{} \in \left[0, \lowpsi{*}\right]$. The allocation probability of any agent $i \in \agentset^{(\numunits)}$ at bid $\bidpairlow{}$ converges to $0$ as $\numunits \to \infty$, that is,
    \begin{align*}
        \lim_{\numunits \to \infty} \max_{i \in \agentset^{(\numunits)}}x_{i}^{(\numunits)}\left(\bidpairlow{}\right) = 0.
    \end{align*}
\end{lemma}

\begin{proof} We begin by analyzing the case $\lowpsi{*} > 0$. By assumption that equilibrium bid distributions have a continuous distribution function, it must be $
\expect\left[\mathcal{T}\left(\lowpsi{*}\right)\right] = \numunits + \numunits^{0.5+\delta} - 1$. Consequently, for every bid $\bidpairlow{} \in \left[0,\lowpsi{*}\right]$, the expected number of equilibrium bids above $\bidpairlow{}$ is \begin{align} \label{eq:totaleqlbbidlb}
\expect\left[\mathcal{T}\left(\bidpairlow{}\right)\right] \geq \numunits + \numunits^{0.5+\delta} - 1. 
\end{align}

We now analyze the allocation probability of an individual $i$ at bid $\bidpairlow{}$. This involves analyzing the distribution function of their critical bid. To that end, we define
\begin{align*}
    \mathcal{T}_{-i}(b) := \sum_{l \in \correspondingagentset \setminus \{i\}} \mathcal{T}_{l}(b) \text{,}
\end{align*} as the total competing equilibrium bids to agent $i$ that are strictly above $b$. Since $\mathcal{T}_{i}\left(\bidpairlow{}\right)$ is an indicator random variable, it follows from (\ref{eq:totaleqlbbidlb}) that \begin{align} \label{eq:lbexpectlowbidalloc}
\expect\left[\mathcal{T}_{-i}\left(\bidpairlow{}\right)\right] \geq \numunits + \numunits^{0.5+\delta} -2.
\end{align} 

We solve for the allocation probability $x_{i}^{(\numunits)}\left(\bidpairlow{}\right)$. Individual $i$ is allocated at bid $\bidpairlow{}$ only if the total number of competing equilibrium bids above $\bidpairlow{}$ is at most $(\numunits-1)$. This gives the following bound on the allocation probability: \begin{align*}
x_{i}^{(\numunits)}\left(\bidpairlow{}\right) \leq \prob \left[\mathcal{T}_{-i}\left(\bidpairlow{}\right) \leq \numunits -1 \right].
\end{align*} Consider the following chain of inequalities that derive a bound for the right-hand side in the expression above: \begin{align}
   x_{i}^{(\numunits)}\left(\bidpairlow{}\right) & \leq \prob \left[\mathcal{T}_{-i}\left(\bidpairlow{}\right) \leq \numunits -1 \right] \text{,} \\ & \leq \prob \left[\mathcal{T}_{-i}\left(\bidpairlow{}\right) - \expect \left[\mathcal{T}_{-i}\left(\bidpairlow{}\right)\right] \leq \numunits - 1 - \expect \left[\mathcal{T}_{-i}\left(\bidpairlow{}\right)\right]\right] \text{,} \\
    & \leq e^{-\frac{1}{2} \cdot \frac{\left(\expect \left[\mathcal{T}_{-i}\left(\bidpairlow{}\right)\right] - \numunits +1 \right)^{2}}{\expect \left[\mathcal{T}_{-i}\left(\bidpairlow{}\right)\right]}} \text{,} \label{eq:lowbidallocub}  \\
    & \leq e^{-\frac{1}{2} \cdot \frac{\numunits^{2\delta}-\frac{2}{\numunits^{0.5-\delta}}+\frac{1}{\numunits}}{1+\frac{1}{\numunits^{0.5-\delta}}-\frac{2}{\numunits}}} \text{.} \label{eq:interiorallocprobub}
\end{align} The second inequality follows from applying the lower tail Chernoff bound in \Cref{fact:chernoffbound}. To derive an upper bound on (\ref{eq:lowbidallocub}), it suffices to analyze the function in the exponent, rewritten below for convenience: \begin{align*}
        \frac{\left(\expect \left[\mathcal{T}_{-i}\left(\bidpairlow{}\right)\right]-\numunits+1\right)^2}{\expect \left[\mathcal{T}_{-i}\left(\bidpairlow{}\right)\right]}.
    \end{align*} For a fixed $\numunits$, the term above is strictly increasing in the region $\expect \left[\mathcal{T}_{-i}\left(\bidpairlow{}\right)\right] > \numunits-1$. Utilizing the inequality in (\ref{eq:lbexpectlowbidalloc}), an upper bound to (\ref{eq:lowbidallocub}) is obtained by substituting $\expect \left[\mathcal{T}_{-i}\left(\bidpairlow{}\right)\right] = \numunits+\numunits^{0.5+\delta}-2$.

Next, suppose $\lowpsi{*}=0$. By assumption that no distributions have an atom, the allocation probability at a zero bid is zero. Combining this observation with (\ref{eq:interiorallocprobub}), we get \begin{align*}
x_{i}^{(\numunits)}\left(\bidpairlow{}\right) \leq e^{-\frac{1}{2} \cdot \frac{\numunits^{2\delta}-\frac{2}{\numunits^{0.5-\delta}}+\frac{1}{\numunits}}{1+\frac{1}{\numunits^{0.5-\delta}}-\frac{2}{\numunits}}}. 
\end{align*} Moreover, this is satisfied for every agent $i \in \agentset^{(\numunits)}$. Thus $
        \lim_{\numunits \to \infty} \max_{i \in \agentset^{(\numunits)}}x_{i}^{(\numunits)}\left(\bidpairlow{}\right) = 0$ and the statement of the result is established. 
\end{proof}  


Finally, we present a lemma that establishes an upper bound on the threshold bid $\highpsi{*}$ under the general no over-bidding assumption of \Cref{as:generalnooverbidding} and the no-atoms condition of \Cref{a:noatomsequilibrium}. We will apply this while analyzing the equilibrium behavior of agents. 

\begin{lemma} \label{l:interiorpsi}
   Suppose the sequence of bid profiles $\left(\equilibrium^{(\numunits)}\right)_{\numunits \in \mathbb{N}}$ satisfy \Cref{as:generalnooverbidding} and \Cref{a:noatomsequilibrium}. Then, we have $$\limsup_{\numunits \to \infty}\highpsi{*} \leq H,$$ where $H$ is the uniform upper bound on agents' value distribution in \Cref{as:lowerboundexpectation}. 
\end{lemma}

\begin{proof}
    Let $b_{*}^{(\numunits)}$ be the maximum of the supremum of the support of agents' bid distribution: \begin{align*}
        b_{*}^{(\numunits)} := \max_{i \in \agentset^{(\numunits)}} \sup \text{supp} \left(\equilibrium_{i}^{(\numunits)}\right).
    \end{align*} By \Cref{as:generalnooverbidding}, we have $
        \limsup_{\numunits \to \infty} b_{*}^{(\numunits)} \leq H$. By definition of $\highpsi{*}$ and \Cref{a:noatomsequilibrium}, we have $\highpsi{*} \leq b_{*}^{(\numunits)}$ for every $\numunits \in \mathbb{N}$. It follows that $$\limsup_{\numunits \to \infty}\highpsi{*} \leq H.$$ This establishes the statement of the result. 
\end{proof}


The rest of this section is organized as follows. \Cref{ss:partapfsketch} establishes the limiting behavior of the $\numunits$-th order statistic of (asymptotic) Bayes-Nash equilibrium and derives a PoA bound under \Cref{as:generalnooverbidding} and \Cref{a:noatomsequilibrium}. \Cref{ss:partbpfsketch} performs an analogous analysis for $\epsilon$-approximate equilibria.  \Cref{ss:proofofpoathm} provides a proof for \Cref{t:efficiencyapproxeqlb}. 

\subsection{Asymptotic $\numunits$-th order statistic of Bayes-Nash equilibria} \label{ss:partapfsketch} 

When agents best respond optimally (or best respond optimally asymptotically), the critical bid is deterministic in the limit. In particular, the distribution function of the critical bid converges to a step function along every subsequence of auctions with a convergent limit. The welfare guarantee follows from the geometry of individual efficiency in \Cref{ss:generalframework}. A formal result is provided below.  

\begin{proposition}
    \label{p:bnelimitstep} Consider any sequence of $\numunits$-unit first-price auctions $\left(\auction^{(\numunits)}\right)_{\numunits \in \mathbb{N}}$ satisfying \Cref{as:lowerboundexpectation} and any corresponding sequence of $\epsilon^{(\numunits)}$-approximate Bayes-Nash equilibria $\left(\equilibrium^{(\numunits)}\right)_{\numunits \in \mathbb{N}}$ satisfying \Cref{as:nooverbidding} and \Cref{a:noatomsequilibrium}. Let $\epsilon := \lim_{\numunits \to \infty}\epsilon^{(\numunits)}$ and suppose $\epsilon=0$. Along any subsequence of auctions which admits pointwise convergence of the $\numunits$-th order statistic of $\equilibrium^{(\numunits)}$, the limit satisfies \begin{align*}
   \lim_{\numunits \to \infty}\khighbidCDF(b) = \twopartdef{1}{b > \psi^*}{0}{b < \psi^*},
                \end{align*} for some $\psi^* \geq 0$.
\end{proposition}

\begin{figure}[t]
    \centering
    \begin{tikzpicture}
    \begin{axis}[
        width=0.65\linewidth,
        height=0.5\linewidth,
        axis lines=left,
        xmin=0, xmax=6.2,
        ymin=0, ymax=1.25,
        xlabel={Bids},
        ylabel={$\mathcal{B}^*(b)$},
        xtick=\empty,
        ytick={1},
        yticklabels={$1$},
        xlabel style={below},
        ylabel style={rotate=0, left},
        xlabel style={at={(axis description cs:1.1,-0.06)}, anchor=north east},
        axis line style={->, thick},
        clip=false,
    ]

    \def\psilow{1.9}
    \def\vval{3.2}
    \def\psihigh{3.50}

    \def\rectx{2.5}
    \def\recty{0.39}

    \addplot[
        dashed,
        thick,
        domain=0:5.8
    ] {1};


    \addplot[
        blue,
        thick,
        smooth
    ] coordinates {
        (0,0)
        (0.20,0.00)
        (0.50,0.00)
        (0.90,0.00)
        (1.20,0.00)
        (1.50,0.00)
        (1.90,0.00)
        (2.10,0.05)
        (2.40,0.25)
        (2.65,0.60)
        (3.00,0.80)
        (3.20,0.95)
        (3.50,1.00)
        (3.90,1.00)
        (4.40,1.00)
        (5.80,1.00)
    };

    \draw[thick] (axis cs:\psilow,0) -- (axis cs:\psilow,0.04);
    \draw[thick] (axis cs:\psihigh,0) -- (axis cs:\psihigh,0.04);
    \draw[thick] (axis cs:\vval,0) -- (axis cs:\vval,0.04);

    \path[
    draw=black,
    fill=blue!10,
    pattern=crosshatch,
    pattern color=blue!50
    ]
    (axis cs:\vval,0) rectangle (axis cs:\rectx,\recty);

    \node[below] at (axis cs:\psilow,0) {$\underline{\psi}_{*}$};
    \node[below] at (axis cs:\psihigh,0) {$\overline{\psi}_{*}$};
    \node[below] at (axis cs:\vval,0) {$v$};

    \end{axis}
    \end{tikzpicture}
    \caption{The solid curve plots the limit of the distribution function of the critical bid $\khighbidCDF$ as $\numunits \to \infty$. On the x-axis, $\underline{\psi}_{*}$ and $\overline{\psi}_{*}$ denote the limit values of $\lowpsi{*}$ and $\highpsi{*}$ respectively. The blue cross hatched area is a lower bound on the utility of any agent with value $v$.}
    \label{fig:partapf}
\end{figure}
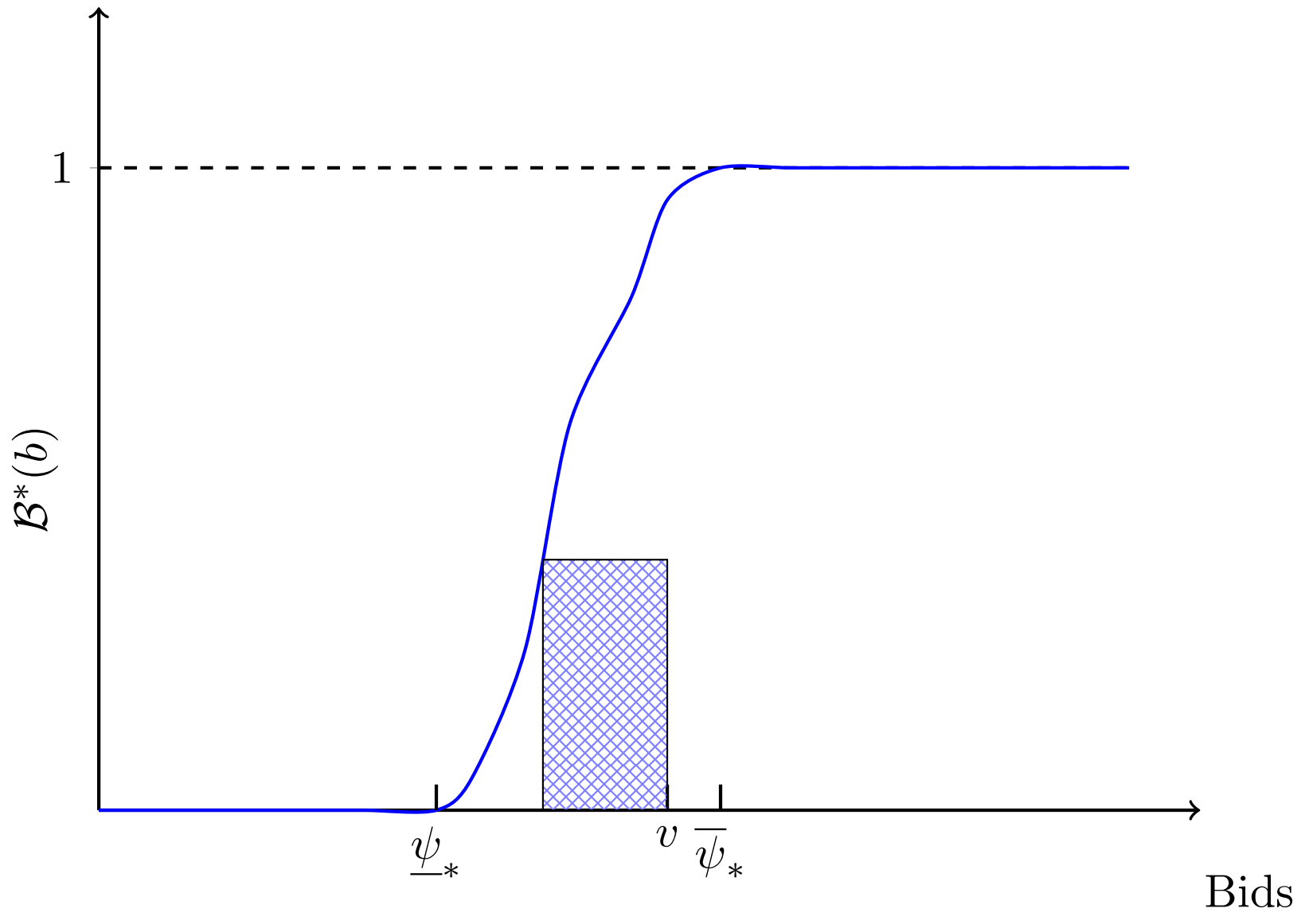

We establish from first principles that the no-overbidding condition in \Cref{as:nooverbidding} is without loss (see \Cref{lma:bidding-dichotomy}) for the special case of $\epsilon^{(\numunits)}=0$ for every $\numunits 
\in \mathbb{N}$. We now provide the high-level approach of our argument. Suppose that the threshold bids $\lowpsi{*}$ and $\highpsi{*}$ do not converge to each other and the area below the distribution function of the $\numunits$-th order statistic in the interval $\left(\lowpsi{*},\highpsi{*}\right)$ is positive in the limit (see \Cref{fig:partapf} for an illustration). Then, every agent has a constant fraction of values that obtain strictly positive utility by bidding in that interval. Furthermore, because the sequence is extremizing, these values have zero utility by bidding outside it. By \Cref{as:lowerboundexpectation}, the total expected bids in $\left(\lowpsi{*},\highpsi{*}\right)$ is at least $\kappa n$ for some constant $\kappa > 0$. However, by definition of $\lowpsi{*}$ and $\highpsi{*}$, the total expected equilibrium bids in the interval is at most $2\numunits^{0.5+\delta}$. This establishes a contradiction. In the remainder of the section, we formalize each step of this argument. 

Suppose that the following is satisfied as $\numunits \to \infty$ : the area below the distribution function of the critical bid $\khighbidCDF$ and within the interval $$\underbrace{\left(\lowpsi{*},\highpsi{*}\right)}_{\mathcal{I}^{(\numunits)}},$$ is strictly positive in the limit. By \Cref{lemma:extremizingconstructeps}, it is necessary that the threshold bid $\lowpsi{*}$ does not converge to $\highpsi{*}$. Now, there could exist a different extremizing sequence of bid pairs such that the bids lie strictly in $\mathcal{I}^{(\numunits)}$. However, when the area is strictly positive, no pair of bids in any such extremizing sequence can converge to each another. 

Any agent with value $v$ close to $\highpsi{*}$ obtains strictly positive utility by bidding in $\mathcal{I}^{(\numunits)}$. This is presented in \Cref{fig:partapf} by a strictly positive shaded area below the curve of the distribution function. In the result below, we establish that every agent has a constant length interval of values in $\mathcal{I}^{(\numunits)}$ that obtains strictly positive equilibrium utility in the limit. 

\begin{lemma} \label{lemma:posutilinterval}
   Consider any sequence of $\numunits$-unit first-price auctions $\left(\auction^{(\numunits)}\right)_{\numunits \in \mathbb{N}}$ satisfying \Cref{as:lowerboundexpectation} and a corresponding sequence of $\epsilon$-approximate Bayes-Nash equilibrium $\left(\equilibrium^{(\numunits)}\right)_{\numunits \in \mathbb{N}}$ satisfying \Cref{as:generalnooverbidding} and \Cref{a:noatomsequilibrium}. Suppose there exists a $c>0$ such that for any extremizing sequence of bid pairs $\left\{\left(\bidpairlow{},\bidpairhigh{}\right)\right\}_{\numunits=1}^{\infty}$ and any corresponding subsequence of auctions $\left(\auction^{(\numunits_{\subseqind})}\right)_{\subseqind=1}^{\infty}$ along which the bid difference converges, we have \begin{align*}
        \lim_{\subseqind \to \infty} \left(\bidpairhighsubseq{}-\bidpairlowsubseq{}\right) > c.
    \end{align*} Then, there exists a constant $\Bar{u}>0$, a subsequence $\left(\auction^{\left(\numunits_{m}\right)}\right)_{m=1}^{\infty}$ and $m^* \in \mathbb{N}$ such that for all $m > m^*$, there exist an interval $$\mathcal{W}^{(\numunits_{m})} \subset \left(\underline{\psi}_{*}^{(\numunits_{m})},\overline{\psi}_{*}^{(\numunits_{m})}\right)$$ such that \begin{enumerate}
        \item the length of the interval denoted by $\text{len}\left(\mathcal{W}^{(\numunits_{m})}\right) \geq c/8$, and, 
        \item for every agent $i \in \agentset^{\left(\numunits_m\right)}$, every value $v \in \mathcal{W}^{(\numunits_{m})}$ yields optimal utility at least $\Bar{u}$.
    \end{enumerate} 
\end{lemma}

\begin{proof}
By \Cref{lemma:extremizingconstructeps}, the bid pair sequence $\left\{\left(\lowpsi{*},\highpsi{*}\right)\right \}_{\numunits=1}^{\infty}$ is extremizing along the sequence of auctions $\left(\auction^{(\numunits)}\right)_{\numunits=1}^{\infty}$. Furthermore, by \Cref{l:interiorpsi}, for $\numunits$ large enough, $\lowpsi{*}$ and $\highpsi{*}$ lie in a compact interval. Thus, there must exist a subsequence of auctions along which the difference $\highpsi{*}-\lowpsi{*}$ converges. By assumption in the statement of the result, it must be  \begin{align} \label{eq:psidiffnoconverge}
        \lim_{\subseqind \to \infty} \left(\highpsisubseq{*}-\lowpsisubseq{*}\right) > c,
    \end{align} where $\subseqind$ is the index of any subsequence of auctions along which the bid difference converges. We will fix one such subsequence and analyze this for the rest of the proof.

We now establish that a constant length subset of values in the interval $\left(\lowpsisubseq{*},\highpsisubseq{*}\right)$ obtain strictly positive optimal utility for $\numunits_{\subseqind}$ large enough. We prove this by constructing a sequence of bids within the interval having a strictly positive distribution function in the limit. To that end, consider a sequence of bids $\left(b^{(\numunits_{\subseqind})}\right)_{\subseqind=1}^{\infty}$ with $b^{(\numunits_{\subseqind})}$ defined as \begin{align} \label{eq:posalloc}
    b^{(\numunits_{\subseqind})} := \max \left \{\highpsisubseq{*} - \frac{c}{2},0 \right \}. 
\end{align} Note that for such a subsequence, $b^{(\numunits_{\subseqind})}$ is positive for $\subseqind$ large enough. 

To establish a contradiction, suppose there exists a further subsequence of $\left(\auction^{(\numunits_{\subseqind})}\right)_{\subseqind=1}^{\infty}$, which we denote by $\left(\auction^{(\numunits_{m})}\right)_{m=1}^{\infty}$, such that the following is satisfied: \begin{enumerate}
    \item The limit of the bid difference $\overline{\psi}_{*}^{(\numunits_m)}-b^{(\numunits_m)}$ converges.
    \item The limit of the distribution function at $b^{(\numunits_m)}$ satisfies $\lim_{m \to \infty}\mathcal{B}^{(\numunits_m)}\left(b^{(\numunits_{m})}\right) = 0.$
\end{enumerate} The existence of a subsequence along which the bid difference converges follows because $\overline{\psi}_{*}^{(\numunits_{\subseqind})}$ and $b^{(\numunits_{\subseqind})}$ lie in a compact interval. Next, since $\mathcal{B}^{(\numunits_{\subseqind})}(b^{(\numunits_{\subseqind})})$ lies in the compact interval $[0,1]$ for every $\subseqind \in \mathbb{N}$, there exists a further subsequence along which the limit of the distribution function is well defined. 

As established in the proof of \Cref{lemma:extremizingconstructeps}, $\lim_{m \to \infty} \mathcal{B}^{(\numunits_{m})}\left(\overline{\psi}_{*}^{(\numunits_m)}\right)=1$. By definition, the sequence of bid pairs $\left \{ \left(b^{(\numunits_{m})},\overline{\psi}_{*}^{(\numunits_m)}\right) \right \}$ is extremizing along the subsequence of auctions $\left(\auction^{(\numunits_{m})}\right)_{m=1}^{\infty}$. However, the limit of the bid difference is $\lim_{m \to \infty} \left(\overline{\psi}_{*}^{(\numunits_m)}-b^{(\numunits_m)}\right) = c/2$. This is a contradiction to the assumption that for any extremizing sequence of bid pairs and any subsequence along which the bid difference converges, the limit of the bid difference is at least $c$. Thus, it must be that \begin{align} \label{eq:posalloccdf}
  \lim_{m \to \infty}  \mathcal{B}^{(\numunits_m)}\left(b^{(\numunits_{m})}\right) > 0.
\end{align} 

Consider any agent $i \in \mathcal{N}^{(\numunits)}$ and value $v \in \left[\highpsi{*}-\frac{c}{4},\highpsi{*}-\frac{c}{8}\right]$. Along the subsequence of auctions $\left(\auction^{(\numunits_{m})}\right)_{m=1}^{\infty}$, for $m$ large enough, the optimal interim utility at value $v$ is at least \begin{align*}
    u_{i}^{(\numunits_m)}(v) \geq \left(v-b^{(\numunits_m)}\right) \cdot x_{i}^{(\numunits_m)}\left(b^{(\numunits_m)}\right) \geq \frac{c}{4} \cdot \mathcal{B}^{(\numunits_m)}(b^{(\numunits_m)}).
\end{align*} Here, the second inequality follows from the definition of $b^{(\numunits_m)}$ in (\ref{eq:posalloc}) and observing that for large enough $m$, the bid $b^{(\numunits_m)} = \overline{\psi}_{*}^{(\numunits_m)} - c/2$. Furthermore, from Fact \ref{fact:lballoc}, we substitute $x_{i}^{(\numunits_m)}\left(b^{(\numunits_m)}\right) \geq \mathcal{B}^{(\numunits_m)}(b^{(\numunits_m)})$. Using (\ref{eq:posalloccdf}) and taking the limit of $m \to \infty$, the optimal utility for any agent is strictly positive for $m$ large enough. The statement of the result follows.  
\end{proof}

Next, we establish that when agents (asymptotically) best respond optimally, bids in extremizing sequence of bid pairs must get arbitrarily close to each other.  

\begin{lemma}
\label{lemma:continuousdistapxpoazerolim}
    Consider any sequence of $\numunits$-unit first-price auctions $\left(\auction^{(\numunits)}\right)_{\numunits \in \mathbb{N}}$ satisfying \Cref{as:lowerboundexpectation} and a corresponding sequence of $\epsilon^{(\numunits)}$-approximate Bayes-Nash equilibrium $\left(\equilibrium^{(\numunits)}\right)_{\numunits \in \mathbb{N}}$ satisfying \Cref{as:generalnooverbidding} and \Cref{a:noatomsequilibrium}. Let $\epsilon := \lim_{\numunits \to \infty}\epsilon^{(\numunits)}$ and suppose $\epsilon=0$. Then, for every $c>0$, there exists an extremizing sequence of bid pairs $\left\{\left(\bidpairlow{},\bidpairhigh{}\right)\right\}_{\numunits=1}^{\infty}$ and a corresponding subsequence of auctions $\left(\auction^{(\numunits_{\subseqind})}\right)_{\subseqind=1}^{\infty}$ along which \begin{align*}
        \lim_{\subseqind \to \infty} \left(\bidpairhighsubseq{}-\bidpairlowsubseq{}\right) < c.
    \end{align*}
\end{lemma}

\begin{proof}

Suppose this is not the case. That is, there exists a constant $c^*>0$ such that for any extremizing sequence of bid pairs and any corresponding subsequence of auctions along which the bid difference converges, we have $
    \lim_{\subseqind \to \infty} \left(\bidpairhighsubseq{}-\bidpairlowsubseq{}\right) \geq c^*$. By \Cref{lemma:posutilinterval}, there exists a subsequence denoted by $\left(\auction^{(\numunits_m)}\right)_{m=1}^{\infty}$ such that for every $m$ large enough, there is a compact interval $\mathcal{W}^{(\numunits_m)} \subset \left(\underline{\psi}_{*}^{(\numunits_m)},\overline{\psi}_{*}^{(\numunits_m)}\right)$ and a constant $\Bar{u}>0$ such that \begin{enumerate}
    \item $\text{len}\left(\mathcal{W}^{(\numunits_m)}\right) \geq c/8$, and,
    \item for every agent $i \in \agentset^{(\numunits_m)}$, every value $v \in \mathcal{W}^{(\numunits_m)}$ yields optimal utility at least $\Bar{u}$.
\end{enumerate} 

For $m$ large enough, every agent with a value in this interval obtains equilibrium utility arbitrarily close to their optimal utility which, by part (b), is at least $\Bar{u}$. Here, $\Bar{u}$ is positive. By \Cref{lemma:vanishingallocprob}, every bid in the interval $\left[0,\underline{\psi}_{*}^{(\numunits_m)}\right]$ is allocated with vanishing probability. 

We will use objects and results established in the proof of \Cref{lemma:continuousdistapxpoa}. Recall, $\mathcal{R}_{*}^{(\numunits_m)}$ (see (\ref{eq:rstar})) is the maximum measure of equilibrium bids an agent and a particular value in $\mathcal{W}^{(\numunits_m)}$ has in the set $\left[0,\underline{\psi}_{*}^{(\numunits_m)}\right] \cup \left(\overline{\psi}_{*}^{(\numunits_m)},\infty\right)$. Furthermore, in \Cref{lemma:continuousdistapxpoa}, we establish that any subsequence along which the limit of $\mathcal{R}_{*}^{(\numunits_m)}$ exists satisfies $\lim_{p \to \infty}\mathcal{R}_{*}^{(\numunits_p)} = 1$. Here, we index the subsequence with $\left(\numunits_p\right)_{p=1}^{\infty}$. 

Thus, there is a sequence of agent-value pairs $\left(\left \{i^{(\numunits_p)},v_{i^{(\numunits_p)}} \right\}\right)_{p=1}^{\infty}$ with the following properties: \begin{enumerate}
    \item for every $p \in \mathbb{N}$, the value $v_{i^{(\numunits_p)}} \in \mathcal{W}^{(\numunits_p)}$, and,
    \item the limit of the measure of equilibrium bids in $\mathcal{S}_{*}^{(\numunits_p)}$ satisfies $$
    \lim_{p \to \infty} \mathcal{R}_{i^{(\numunits_p)},v_{i^{(\numunits_p)}}}^{(\numunits_p)} = 1 .$$
\end{enumerate} For the rest of the argument, for notational convenience, we will avoid the dependence on the auction $\numunits_p$. Note that for any value $v_{i} \in \mathcal{W}$, any bid in the interval $(\overline{\psi}_{*},\infty)$ gives the agent strictly negative utility. Thus, the equilibrium utility for agent $i$ at value $v_{i}$ is \begin{align*}
u_{i}\left(v_{i},\equilibrium_{i,v_{i}}\right) \leq H x_{i} \left(\underline{\psi}_{*}\right)\mathcal{R}_{i,v_{i}} + H\left(1-\mathcal{R}_{i,v_i}\right). 
\end{align*} By \Cref{lemma:vanishingallocprob}, the allocation probability for any bid in the interval $\left[0,\lowpsi{*}\right]$ converges to zero as $p \to \infty$. Combining this with property (b) above, we have that agent $i$ with value $v$ gets vanishing equilibrium utility as $p \to \infty$. But, this is a contradiction to Part (b) in \Cref{lemma:posutilinterval} which states that every value in $\mathcal{W}$ receive equilibrium utility that is arbitrarily close to a positive $\Bar{u}$ as $p \to \infty$. Thus, for every $c>0$, there exists an extremizing sequence of bid pairs $\left\{\left(\bidpairlow{},\bidpairhigh{}\right)\right\}_{\numunits=1}^{\infty}$ and a corresponding subsequence of auctions $\left(\auction^{(\numunits_{\subseqind})}\right)_{\subseqind=1}^{\infty}$ along which $
        \lim_{\subseqind \to \infty} \left(\bidpairhighsubseq{}-\bidpairlowsubseq{}\right) < c. $
\end{proof}

We now provide a proof for \Cref{p:bnelimitstep}: for any sequence of auctions and corresponding equilibria having no atoms, there exists a subsequence of auctions along which the $\numunits$-th order statistic of (asymptotic) Bayes-Nash equilibrium bid distributions is deterministic in the limit. 

\begin{proof}[Proof of \Cref{p:bnelimitstep}]
    Consider the sequence of distribution functions $\left(\khighbidCDF\left(\cdot\right)\right)_{k \in \mathbb{N}}$. By \cite{helly1912lineare} selection theorem, the sequence admits a pointwise convergent subsequence. We want to establish that along any such subsequence, the limit denoted by $\mathcal{B}^*(\cdot)$ satisfies \begin{align}
\label{eq:steplimit}\mathcal{B^*}(b) = \twopartdef{1}{b > \psi^*}{0}{b < \psi^*},
\end{align} for some $\psi^* 
\geq 0$. 

Consider any pointwise convergent subsequence that does not satisfy (\ref{eq:steplimit}). We will overload notation and denote the elements of this subsequence by the index $\numunits \in \mathbb{N}$. Because the limit is not of the form in (\ref{eq:steplimit}), there exist two bids $\bidpairlowlimit$ and $\bidpairhighlimit$ such that: \begin{enumerate}
    \item $0 < \mathcal{B}^*\left(\bidpairlowlimit\right) \leq \mathcal{B}^*\left(\bidpairhighlimit\right) < 1$, and,
    \item $\bidpairhighlimit - \bidpairlowlimit > 0$. 
\end{enumerate} Since $\mathcal{B}^*(\cdot)$ is monotonic and the distribution function $\khighbidCDF(\cdot)$ pointwise converges to $\mathcal{B}^*(\cdot)$, by property (a) and (b), there cannot exist an extremizing sequence of bid pairs such that the bid difference is at most $\bidpairhighlimit-\bidpairlowlimit$. This is a contradiction to \Cref{lemma:continuousdistapxpoazerolim} (note that \Cref{as:nooverbidding} implies \Cref{as:generalnooverbidding} and, thus, the conditions of \Cref{lemma:continuousdistapxpoazerolim} are satisfied). The statement of the result is proved.
\end{proof}

\begin{corollary}
\label{corollary:continuousdistapxpoazerolim}
    Consider any sequence of $\numunits$-unit first-price auctions $\left(\auction^{(\numunits)}\right)_{\numunits \in \mathbb{N}}$ satisfying \Cref{as:lowerboundexpectation} and a corresponding sequence of $\epsilon^{(\numunits)}$-approximate Bayes-Nash equilibrium $\left(\equilibrium^{(\numunits)}\right)_{\numunits \in \mathbb{N}}$ satisfying \Cref{as:generalnooverbidding} and \Cref{a:noatomsequilibrium}. Let $\epsilon := \lim_{\numunits \to \infty}\epsilon^{(\numunits)}$ and suppose $\epsilon=0$. Then, \begin{align*}
        \liminf_{\numunits \to \infty} \frac{\mathcal{W}\left(\equilibrium^{\left(\numunits\right)},\auction^{\left(\numunits\right)}\right)}{\mathcal{O}\left(\auction^{\left(\numunits\right)}\right)} = 1.
    \end{align*}
\end{corollary}

\begin{proof} Suppose that $\liminf_{\numunits \to \infty} \mathcal{W}\left(\equilibrium^{\left(\numunits\right)},\auction^{\left(\numunits\right)}\right)/\mathcal{O}\left(\auction^{\left(\numunits\right)}\right) < 1$. Then, there exists a subsequence of auctions $\left(\auction^{(\numunits_{\subseqind})}\right)_{\subseqind=1}^{\infty}$ along which \begin{align} \label{eq:poaineff}
        \lim_{l \to \infty} \frac{\mathcal{W}\left(\equilibrium^{\left(\numunits_{\subseqind}\right)},\auction^{\left(\numunits_{\subseqind}\right)}\right)}{\mathcal{O}\left(\auction^{\left(\numunits_{\subseqind}\right)}\right)} < 1.
    \end{align} We will now analyze a further subsequence of $\left(\auction^{(\numunits_{\subseqind})}\right)_{\subseqind=1}^{\infty}$. By \Cref{lemma:continuousdistapxpoazerolim}, for every $c>0$, there exists an extremizing sequence of bid pairs $\left\{\left(\bidpairlow{},\bidpairhigh{}\right)\right\}_{\numunits=1}^{\infty}$ and a further subsequence of auctions $\left(\auction^{(\numunits_{m})}\right)_{m=1}^{\infty}$ along which $
        \lim_{m \to \infty} \left(\overline{b}^{(k_{m})}-\underline{b}^{(k_{m})}\right) < c$. A direct application of \Cref{lemma:generalposepslim} implies that there exists a constant $\zeta>0$ for which \begin{align*}
        \lim_{m \to \infty} \frac{\mathcal{W}\left(\equilibrium^{\left(\numunits_{m}\right)},\auction^{\left(\numunits_{m}\right)}\right)}{\mathcal{O}\left(\auction^{\left(\numunits_{m}\right)}\right)} \geq 1- \frac{c}{\zeta}.
    \end{align*} Taking $c$ sufficiently small contradicts (\ref{eq:poaineff}). The statement of the result follows. 
\end{proof}

\subsection{Asymptotic $\numunits$-th order statistic of $\epsilon$-approximate equilibria}
\label{ss:partbpfsketch} When agents $\epsilon$-approximate best respond, the contradiction argument in the previous section no longer applies to show that the critical bid is deterministic in the limit. To see why, suppose $\lowpsi{*}$ and $\highpsi{*}$ do not converge to each other. As depicted in \Cref{fig:partapf}, a value in the interval $\mathcal{I}^{(\numunits)}$ obtains a strictly positive optimal utility in the limit by bidding in $\mathcal{I}^{(\numunits)}$. But agents are approximate best-responding and may bid suboptimally outside $\mathcal{I}^{(\numunits)}$. 

\begin{figure}[t]
    \centering

    \begin{subfigure}[t]{0.48\textwidth}
        \centering
        \begin{tikzpicture}
        \begin{axis}[
            width=\linewidth,
            height=0.72\linewidth,
            axis lines=left,
            xmin=0, xmax=6.2,
            ymin=0, ymax=1.25,
            xlabel={Bids},
            ylabel={},
            xtick=\empty,
            ytick={1},
            yticklabels={$1$},
            xlabel style={
                at={(axis description cs:1.05,-0.05)},
                anchor=north east
            },
            axis line style={->,thick},
            clip=false,
        ]

        \def\psilow{1.9}
        \def\psihigh{3.5}

        \addplot[
            dashed,
            thick,
            domain=0:5.8
        ] {1};

        \path[
            draw=none,
            fill=maroonshade!20
        ]
        (axis cs:\psilow,0) rectangle (axis cs:\psihigh,1);

        \addplot[
            blue,
            thick,
            domain=0:\psilow
        ] {0};

        \addplot[
            blue,
            thick,
            smooth
        ] coordinates {
            (1.90,0.00)
            (2.05,0.02)
            (2.20,0.07)
            (2.40,0.16)
            (2.60,0.30)
            (2.80,0.47)
            (3.00,0.65)
            (3.20,0.82)
            (3.32,0.90)
            (3.40,0.95)
            (3.46,0.985)
            (3.50,1.00)
        };

        \addplot[
            blue,
            thick,
            domain=\psihigh:5.8
        ] {1};

        \draw[thick] (axis cs:\psilow,0) -- (axis cs:\psilow,0.04);
        \draw[thick] (axis cs:\psihigh,0) -- (axis cs:\psihigh,0.04);

        \node[below] at (axis cs:\psilow,0) {$\underline{\psi}_{*}$};
        \node[below] at (axis cs:\psihigh,0) {$\overline{\psi}_{*}$};

        \node[right, font=\small] at (axis cs:3.05,0.56)
            {$\mathcal{B}^{*}(b)$};

        \end{axis}
        \end{tikzpicture}
        \caption{$\overline{\psi}_{*} - \underline{\psi}_{*} < \epsilon$.}
        \label{fig:diffseps}
    \end{subfigure}
    \hfill
    \begin{subfigure}[t]{0.48\textwidth}
        \centering

        \begin{tikzpicture}

        \def\psilow{1.90}
        \def\psihigh{3.50}
        \def\epsval{0.48}

        \pgfmathsetmacro{\psiminuseps}{\psihigh-\epsval}

        \pgfmathsetmacro{\orangeleft}{\epsval/(\psihigh-\psilow)}

        \begin{axis}[
            width=\linewidth,
            height=0.72\linewidth,
            axis lines=left,
            xmin=0,
            xmax=6.2,
            ymin=0,
            ymax=1.25,
            xlabel={Bids},
            ylabel={},
            xtick=\empty,
            ytick={1},
            yticklabels={$1$},
            xlabel style={
                at={(axis description cs:1.05,-0.05)},
                anchor=north east
            },
            axis line style={->,thick},
            clip=false,
        ]

\addplot[
    name path=orangeCurve,
    draw=none,
    domain=\psilow:\psiminuseps,
    samples=200
]
{\epsval/(\psihigh-x)};

\addplot[
    name path=orangeBaseline,
    draw=none,
    domain=\psilow:\psiminuseps
] {0};

\addplot[
    draw=none,
    fill=maroonshade!20
]
fill between[
    of=orangeCurve and orangeBaseline
];

\path[
    draw=none,
    fill=maroonshade!20
]
(axis cs:\psiminuseps,0)
rectangle
(axis cs:\psihigh,1);
        \addplot[
            dashed,
            thick,
            domain=0:5.8
        ] {1};

        \addplot[
            blue,
            thick,
            domain=0:\psilow
        ] {0};

        \addplot[
            blue,
            thick,
            smooth
        ] coordinates {
            (1.90,0.00)
            (2.05,0.02)
            (2.20,0.07)
            (2.40,0.16)
            (2.60,0.30)
            (2.80,0.47)
            (3.00,0.65)
            (3.20,0.82)
            (3.32,0.90)
            (3.40,0.95)
            (3.46,0.985)
            (3.50,1.00)
        };

        \addplot[
            blue,
            thick,
            domain=\psihigh:5.8
        ] {1};

        \draw[
            orange,
            dashed,
            thick
        ]
        (axis cs:\psilow,0) -- (axis cs:\psilow,\orangeleft);

        \addplot[
            orange,
            thick,
            domain=\psilow:\psiminuseps,
            samples=200
        ]
        {\epsval/(\psihigh-x)};

        \addplot[
            orange,
            thick,
            domain=\psiminuseps:\psihigh
        ] {1};

        \draw[thick]
            (axis cs:\psilow,0)
            -- (axis cs:\psilow,0.04);

        \draw[thick]
            (axis cs:\psiminuseps,0)
            -- (axis cs:\psiminuseps,0.04);

        \draw[thick]
            (axis cs:\psihigh,0)
            -- (axis cs:\psihigh,0.04);

        \node[below]
            at (axis cs:\psilow,0)
            {$\underline{\psi}_{*}$};

        \node[
            below,
            yshift=-10pt,
            font=\small
        ]
            at (axis cs:\psiminuseps,0)
            {$\overline{\psi}_{*}-\epsilon$};

        \node[below right]
            at (axis cs:\psihigh,0)
            {$\overline{\psi}_{*}$};

        \node[
            right,
            font=\small
        ]
            at (axis cs:3.02,0.56)
            {$\mathcal{B}^{*}(b)$};

        \node[
            font=\small,
            anchor=south
        ]
            at (axis cs:2.2,0.62)
            {$\dfrac{\epsilon}{\overline{\psi}_{*}-b}$};

        \end{axis}
        \end{tikzpicture}
        \caption{$\overline{\psi}_{*} - \underline{\psi}_{*} > \epsilon$.}
        \label{fig:diffbeps}
    \end{subfigure}

    \caption{
        Two illustrations involving the bid distribution $\mathcal{B}^{*}(b)$.
        Panel (a) shows the bid distribution together with the maroon shaded region between
        $\underline{\psi}$ and $\overline{\psi}$. Panel (b) overlays the curve
        $\epsilon/(\overline{\psi}-b)$ on the same bid distribution.
    }
    \label{fig:eps-apx-criticalbid}
\end{figure}
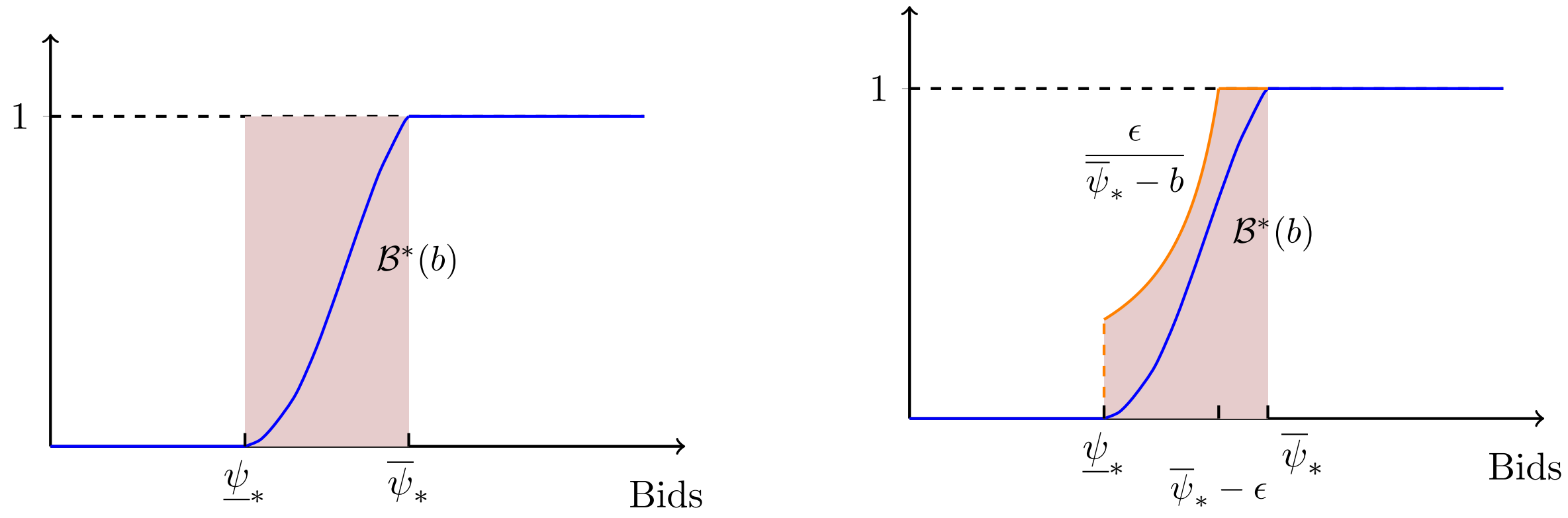

However, using the definition of $\lowpsi{*}$ and $\highpsi{*}$, we establish a bound on the distribution function of the critical bid in $\mathcal{I}^{(\numunits)}$ as a function of the approximation $\epsilon$. Let $\underline{\psi}_{*}$ and $\overline{\psi}_{*}$ be the limit of $\lowpsi{*}$ and $\highpsi{*}$ respectively. We analyze two cases. \begin{enumerate}
    \item Suppose $\overline{\psi}_{*}- \underline{\psi}_{*} < \epsilon$, that is, the critical bid is concentrated in a small region (see \Cref{fig:diffseps} for an illustration). The loss in efficiency measured as the area under the distribution function satisfies \begin{align*}
        \int_{\underline{\psi}_{*}}^{\overline{\psi}_{*}} \mathcal{B}^*(b)db < \epsilon.
    \end{align*}
    \item Suppose $\overline{\psi}_{*}- \underline{\psi}_{*} > \epsilon$, that is, the critical bid is spread across a large region (see \Cref{fig:diffbeps} for an illustration). We show that even though the region might be large, the area under the curve remains small. By construction of $\lowpsi{*}$ and $\highpsi{*}$, there must exist an agent $i$ and a corresponding value in $\left[\underline{\psi}_{*},\overline{\psi}_{*}\right]$ that almost surely bids in $\left[0,\underline{\psi}_{*}\right]$. For the purpose of illustration, suppose this value is $\overline{\psi}_{*}$. The equilibrium utility for this value is zero. By assumption that the agent is $\epsilon$-approximate best-responding, the optimal utility is at most $\epsilon$. This is only possible when the allocation probability $x_{i}(b)$ satisfies \begin{align*}
        x_{i}(b) \leq \frac{\epsilon}{\overline{\psi}_{*} - b} \quad \text{ for all } b \in \left[\underline{\psi}_{*},\overline{\psi}_{*}-\epsilon\right].
    \end{align*} The bound above is precisely the iso-utility curve depicted in \Cref{fig:diffbeps}. Under this allocation, every bid $b$ in $\left[\underline{\psi}_{*},\overline{\psi}_{*}-\epsilon\right]$ obtains utility exactly $\epsilon$. The distribution function of the $\numunits$-th highest bid stochastically dominates the allocation probability of every agent. Thus, the area under the distribution function is bounded as 
        \begin{align*}
        \int_{\underline{\psi}_{*}}^{\overline{\psi}_{*}} \mathcal{B}^*(b)db < \epsilon \log \left(\frac{H}{\epsilon}\right) + \epsilon,
    \end{align*} where $H$ is the uniform bound on values in \Cref{as:lowerboundexpectation}. 
\end{enumerate}  

We formalize this intuition in the result below. 

\begin{lemma}
    \label{lemma:continuousdistapxpoa}
    Consider any sequence of $\numunits$-unit first-price auctions $\left(\auction^{(\numunits)}\right)_{\numunits \in \mathbb{N}}$ satisfying \Cref{as:lowerboundexpectation} and a corresponding sequence of $\epsilon^{(\numunits)}$-approximate Bayes-Nash equilibrium $\left(\equilibrium^{(\numunits)}\right)_{\numunits \in \mathbb{N}}$ satisfying \Cref{as:generalnooverbidding} and \Cref{a:noatomsequilibrium}. Let $\epsilon^* := \lim_{\numunits \to \infty}\epsilon^{(\numunits)}$ and suppose $\epsilon^* \in (0,H)$. Then, \begin{align*}
        \liminf_{\numunits \to \infty} \frac{\mathcal{W}\left(\equilibrium^{\left(\numunits\right)},\auction^{\left(\numunits\right)}\right)}{\mathcal{O}\left(\auction^{\left(\numunits\right)}\right)} \geq 1 - \zeta' \epsilon^* \left(1+\log \left(\frac{H}{\epsilon^*}\right)\right),
    \end{align*} where $\zeta'>0$ is a constant that depends on the parameters of the environment.
\end{lemma}

\begin{proof} Consider any subsequence of auctions. We will overload notation and denote the index of this subsequence by $\numunits$. Suppose that there exists an extremizing sequence of bid pairs $\left\{\bidpairlow{},\bidpairhigh{}\right\}$ and a corresponding subsequence $\left(\auction_{\numunits_{\subseqind}}\right)_{\subseqind=1}^{\infty}$ along which the bid difference converges as \begin{align*}
        \lim_{\subseqind \to \infty} \left(\bidpairhighsubseq{}-\bidpairlowsubseq{}\right) \leq \epsilon^*.
    \end{align*} The integral of the allocation rule in the interval $\left(\bidpairlowsubseq{},\bidpairhighsubseq{}\right)$ is \begin{align*}\int_{\bidpairlowsubseq{}}^{\bidpairhighsubseq{}}\khighbidCDFsubseq(b)db \leq \left(\bidpairhighsubseq{}-\bidpairlowsubseq{}\right). 
    \end{align*} By \Cref{lemma:generalposepslim}, the limit of the ratio of equilibrium welfare to optimal welfare along this subsequence is 
        \begin{align} \label{eq:poaepsapxo}
        \liminf_{l \to \infty} \frac{\mathcal{W}\left(\equilibrium^{\left(\numunits_{\subseqind}\right)},\auction^{\left(\numunits_{\subseqind}\right)}\right)}{\mathcal{O}\left(\auction^{\left(\numunits_{\subseqind}\right)}\right)} \geq 1-\frac{2\epsilon^*}{\zeta}.
    \end{align} 

    Next, suppose that for every extremizing sequence of bid pairs and any corresponding subsequence of auctions along which the bid difference converges, we have \begin{align*}
    \lim_{\subseqind \to \infty} \left(\bidpairhighsubseq{}-\bidpairlowsubseq{}\right) > \epsilon^*. 
\end{align*} By \Cref{lemma:posutilinterval}, there exists a subsequence of auctions $\left(\auction^{(\numunits_m)}\right)_{m=1}^{\infty}$ such that for every $m$ large enough, there is a compact interval $\mathcal{W}^{(\numunits_m)} \subset \left(\underline{\psi}_{*}^{(\numunits_m)},\overline{\psi}_{*}^{(\numunits_m)}\right)$ and a constant $\Bar{u}>0$ such that \begin{enumerate}
    \item $\text{len}\left(\mathcal{W}^{(\numunits_m)}\right) \geq \epsilon^*/8$, and,
    \item for every agent $i \in \agentset^{(\numunits_m)}$, every value $v \in \mathcal{W}^{(\numunits_m)}$ yields optimal utility at least $\Bar{u}$.
\end{enumerate} Furthermore, as constructed in the proof of \Cref{lemma:posutilinterval}, an interval that satisfies these properties  is $\left[\overline{\psi}_{*}^{(\numunits_m)}-\epsilon^*/4, \overline{\psi}_{*}^{(\numunits_m)}-\epsilon^*/8\right]$. We will focus our attention on this particular $\mathcal{W}^{(\numunits_m)}$ for the rest of the proof.

We now introduce some notation that will be used in the rest of the proof. Let $\equilibrium_{i,v}^{(\numunits)}$ denote the equilibrium bid distribution of agent $i$ for value $v$ in auction $\auction^{(\numunits)}$. For values in the interval $\mathcal{W}^{(\numunits)}$, we will compare the measure of equilibrium bids in the following two sets:  $\underbrace{\left[0,\underline{\psi}_{*}^{(\numunits)}\right] \cup \left(\overline{\psi}_{*}^{(\numunits)},\infty\right)}_{\mathcal{S}_{*}^{(\numunits)}}$ and $\underbrace{\left(\underline{\psi}_{*}^{(\numunits)},\overline{\psi}_{*}^{(\numunits)}\right]}_{\mathcal{M}_{*}^{(\numunits)}}$. Finally, for value $v$ we will define \begin{align*}
    \mathcal{R}_{i,v}^{(\numunits)}:= \equilibrium^{(\numunits)}_{i,v}\left(\mathcal{S}_{*}^{(\numunits)}\right) \quad \text{ and } \quad 
    \mathcal{T}_{i,v}^{(\numunits)}:= \equilibrium^{(\numunits)}_{i,v}\left(\mathcal{M}_{*}^{(\numunits)}\right), 
\end{align*} as the measure of equilibrium bids in the interval $\mathcal{S}_{*}^{(\numunits)}$ and $\mathcal{M}_{*}^{(\numunits)}$ respectively. For any value $v$, we have $\mathcal{R}_{i,v}^{(\numunits)} +  \mathcal{T}_{i,v}^{(\numunits)} = 1$.  

We analyze the following two quantities: \begin{align} \label{eq:rstar}
    \mathcal{R}_{*}^{(\numunits)} := \max_{i \in \agentset^{(\numunits)}} \sup_{v \in \mathcal{W}^{(\numunits)}} \mathcal{R}_{i,v}^{(\numunits)} \quad \text{ and } \quad
    \mathcal{T}_{*}^{(\numunits)} := \min_{i \in \agentset^{(\numunits)}} \inf_{v \in \mathcal{W}^{(\numunits)}} \mathcal{T}_{i,v}^{(\numunits)}.
\end{align} $\mathcal{R}_{*}^{(\numunits)}$ is the maximum measure of equilibrium bids an agent and a particular value in $\mathcal{W}^{(\numunits)}$ have in the interval $\mathcal{S}_{*}^{(\numunits)}$. On the other hand, $\mathcal{T}_{*}^{(\numunits)}$ is the minimum measure of equilibrium bids that any agent and any value in $\mathcal{W}^{(\numunits)}$ has in the interval $\mathcal{M}_{*}^{(\numunits)}$. 

Consider any subsequence of $\left(\numunits_m\right)_{m=1}^{\infty}$ along which the limit of $\mathcal{R}_{*}^{(\numunits_m)}$ exists. We will denote such a subsequence with the index $\left(\numunits_p\right)_{p=1}^{\infty}$. We will show that $
    \lim_{p \to \infty}\mathcal{R}_{*}^{(\numunits_p)} = 1$. That is, as $p$ increases, there exists an agent and a corresponding value which places equilibrium bid measure arbitrarily close to one in $\mathcal{S}_{*}^{(\numunits_p)}$. Before we provide a proof, note that since $\mathcal{R}_{*}^{(\numunits_m)}$ lies in the compact interval $[0,1]$, there always exists a subsequence along which the limit of $\mathcal{R}_{*}^{(\numunits_m)}$ exists.

To establish a contradiction, consider any subsequence along which the limit exists and suppose $\lim_{p \to \infty} \mathcal{R}_{*}^{(\numunits_p)} < 1$. Using $\mathcal{R}_{i,v}^{(\numunits)} + \mathcal{T}_{i,v}^{(\numunits)} = 1$, we have \begin{align*}
    \lim_{p \to \infty} \mathcal{T}_{*}^{(\numunits_p)} > 0. 
\end{align*} Consequently, there exists a constant $\beta>0$ such that for $p$ large enough, every agent $i \in \agentset^{(\numunits_p)}$ and every value $v \in \mathcal{W}^{(\numunits_p)}$ has at least a $\beta$ measure of equilibrium bids in the interval $\mathcal{M}_{*}^{(\numunits_p)}$. Combining \Cref{as:lowerboundexpectation} (b), \Cref{l:interiorpsi} and observation (a) in \Cref{lemma:posutilinterval} --- that $\text{len}\left(\mathcal{W}^{(\numunits_p)}\right) \geq \epsilon^*/8$ --- there exists a constant $\zeta>0$ such that the total expected equilibrium bids in  $\mathcal{M}_{*}^{(\numunits_p)}$ is at least $\zeta n^{(\numunits_p)}$ (recall, $n^{(\numunits_p)}$ is the number of agents in $\agentset^{(\numunits_p)}$). 

We establish below that this is a contradiction to the definition of $\underline{\psi}_{*}^{(\numunits_m)}$ and $\overline{\psi}_{*}^{(\numunits_m)}$. By definition and \Cref{a:noatomsequilibrium}, \begin{align*}
    \expect \left[\mathcal{T}\left(\overline{\psi}_{*}^{(\numunits_m)}\right)\right] = \numunits_{m} - \left(\numunits_{m}\right)^{0.5+\delta}-1, \text{ and, } \expect \left[\mathcal{T}\left(\underline{\psi}_{*}^{(\numunits_m)}\right)\right] \leq \numunits_{m} + \left(\numunits_{m}\right)^{0.5+\delta}-1.
\end{align*} The total expected equilibrium bids in the interval $\left(\underline{\psi}_{*}^{(\numunits_m)},\overline{\psi}_{*}^{(\numunits_m)}\right]$ is at most $2\left(\numunits_{m}\right)^{0.5+\delta}$. But, in the paragraph above, we established that for $p$ large enough there exists a constant $\zeta>0$ such that the expected number of equilibrium bids in $\mathcal{W}^{(\numunits_p)}$ is at least $\zeta n^{(\numunits_p)}$. This is a contradiction. Thus, for any subsequence of $\left(\numunits_m\right)_{m=1}^{\infty}$ along which the limit of $\mathcal{R}_{*}^{(\numunits)}$ exists, we have $\lim_{p \to \infty}\mathcal{R}_{*}^{(\numunits_p)} = 1$.

Consequently, there is a sequence of agent-value pairs $\left(\left \{i^{(\numunits_p)},v_{i^{(\numunits_p)}} \right\}\right)_{p=1}^{\infty}$ with the following properties: \begin{enumerate}
    \item for every $p \in \mathbb{N}$, the value $v_{i^{(\numunits_p)}} \in \mathcal{W}^{(\numunits_p)}$, and,
    \item the limit of the measure of equilibrium bids in $\mathcal{S}_{*}^{(\numunits_p)}$ satisfies \begin{align*}
    \lim_{p \to \infty} \mathcal{R}_{i^{(\numunits_p)},v_{i^{(\numunits_p)}}}^{(\numunits_p)} = 1.
\end{align*} 
\end{enumerate} For the rest of the argument, for notational convenience, we will avoid the dependence on the auction $\numunits_p$. Note that for any value $v_{i} \in \mathcal{W}$, any bid in the interval $(\overline{\psi}_{*},\infty)$ gives the agent strictly negative utility. Thus, the equilibrium utility for agent $i$ at value $v_{i}$ is \begin{align*}
u_{i}\left(v_{i},\equilibrium_{i,v_{i}}\right) \leq H x_{i} \left(\underline{\psi}_{*}\right)\mathcal{R}_{i,v_{i}} + H\left(1-\mathcal{R}_{i,v_i}\right). 
\end{align*} By assumption, the equilibrium strategy $\equilibrium_{i,v_{i}}$ is an $\epsilon$-approximate equilibrium. Thus, the optimal utility at value $v_{i}$, which we denote by $u_{i}(v_{i})$, satisfies \begin{align*}
    u_{i}(v_{i}) \leq H x_{i} \left(\underline{\psi}_{*}\right)\mathcal{R}_{i,v_{i}} + H\left(1-\mathcal{R}_{i,v_i}\right) + \epsilon. 
\end{align*} Thus, the allocation probability of agent $i$ satisfies \begin{align} \label{eq:allocboundeps}
    x_{i}(b) \leq \frac{H x_{i} \left(\underline{\psi}_{*}\right)\mathcal{R}_{i,v_{i}} + H\left(1-\mathcal{R}_{i,v_i}\right) + \epsilon}{v_{i}-b} \quad \text{ for every bid } b \in \left[\underline{\psi}_{*},\overline{\psi}_{*}-\epsilon^*\right]. 
\end{align}

We now establish a bound on the integral of the distribution function of $\khighbidCDF$ in the interval $\left[\underline{\psi}_{*},\overline{\psi}_{*}\right]$: \begin{align*}
    \underbrace{{\int_{\underline{\psi}_{*}}^{\overline{\psi}_{*}}\khighbidCDF(b)db}}_{\mathcal{C}_{i}} & \leq \int_{\underline{\psi}_{*}}^{\overline{\psi}_{*}}x_{i}(b)db = \int_{\underline{\psi}_{*}}^{\overline{\psi}_{*}-\epsilon^*}x_{i}(b)db + \int_{\overline{\psi}_{*}-\epsilon^*}^{\overline{\psi}_{*}}x_{i}(b)db \text{,} \\
    & \leq \int_{\underline{\psi}_{*}}^{\overline{\psi}_{*} - \epsilon^*} \frac{H x_{i} \left(\underline{\psi}_{*}\right)\mathcal{R}_{i,v_{i}} + H\left(1-\mathcal{R}_{i,v_i}\right) + \epsilon}{v_{i}-b} db + \epsilon^* , \\
    & \leq \left(H x_{i} \left(\underline{\psi}_{*}\right)\mathcal{R}_{i,v_{i}} + H\left(1-\mathcal{R}_{i,v_i}\right) + \epsilon\right) \log \left(\frac{v_{i}-\underline{\psi}_{*}}{v_{i}-\overline{\psi}_{*} + \epsilon^*}\right) + \epsilon^*, \\
    & \leq \left(H x_{i} \left(\underline{\psi}_{*}\right)\mathcal{R}_{i,v_{i}} + H\left(1-\mathcal{R}_{i,v_i}\right) + \epsilon\right) \log \left(\frac{4H}{3\epsilon^*}\right) + \epsilon^*. \\ 
\end{align*} Here, the first inequality follows from \Cref{fact:lballoc}, the second inequality from (\ref{eq:allocboundeps}) and the final inequality from the fact that $v_{i} \in \left[\overline{\psi}_{*}-\epsilon^*/4,\overline{\psi}_{*}-\epsilon^*/8\right]$. We will overload notation and denote a subsequence along which the integral $\mathcal{C}_{i}^{(\numunits_p)}$ converges by $\left(\numunits_p\right)_{p=1}^{\infty}$. By \Cref{lemma:vanishingallocprob} and the observation that $\lim_{p \to \infty} \mathcal{R}_{i^{(\numunits_p)},v_{i^{(\numunits_p)}}}^{(\numunits_p)} = 1$, we have \begin{align*}
    \lim_{p \to \infty} \mathcal{C}_{i^{(\numunits_p)}}^{(\numunits_p)} \leq \epsilon^* \log \left(\frac{4H}{3\epsilon^*}\right) + \epsilon^*. 
\end{align*} Applying \Cref{lemma:generalposepslim}, we have that for every sequence of auctions, there exists a subsequence of auctions $\left(\auction^{(\numunits_p)}\right)_{p=1}^{\infty}$ such that \begin{align} \label{eq:poaepsapxt}
    \liminf_{p \to \infty} \frac{\mathcal{W}\left(\equilibrium^{\left(\numunits_{p}\right)},\auction^{\left(\numunits_{p}\right)}\right)}{\mathcal{O}\left(\auction^{\left(\numunits_{p}\right)}\right)} \geq 1-\frac{2\epsilon^*}{\zeta} - \frac{1}{\zeta} \cdot \epsilon^* \log \left(\frac{4H}{3\epsilon^*}\right).
\end{align} Note that the bounds established in (\ref{eq:poaepsapxo}) and (\ref{eq:poaepsapxt}) are independent of the initial subsequence chosen. A straightforward calculation then establishes that along the sequence of auctions $\left(\auction^{(\numunits)}\right)_{\numunits \in \mathbb{N}}$, we have \begin{align} \label{eq:liminfopoa}
        \liminf_{\numunits \to \infty} \frac{\mathcal{W}\left(\equilibrium^{\left(\numunits\right)},\auction^{\left(\numunits\right)}\right)}{\mathcal{O}\left(\auction^{\left(\numunits\right)}\right)} \geq \underbrace{1 - \zeta' \epsilon^* \left(1+\log \left(\frac{H}{\epsilon^*}\right)\right)}_{\mathcal{L}},
    \end{align} where $\zeta' = \left(2 + \log \left(\frac{4}{3}\right)\right) \cdot 1/\zeta$. To provide a brief argument, suppose the limit infimum is strictly smaller than $\mathcal{L}$. Then, there exists a subsequence of auctions such that the limit of the ratio of welfare to optimal welfare along this subsequence is strictly smaller than $\mathcal{L}$. This is a contradiction to what we establish in (\ref{eq:poaepsapxt}): for every sequence of auctions, there exists a subsequence of auctions along which the limit of the ratio is at least $\mathcal{L}$. The statement of the result is proved.
\end{proof}

\subsection{Proof of \Cref{t:efficiencyapproxeqlb}}
\label{ss:proofofpoathm}

Finally, we finish this section by providing a proof for \Cref{t:efficiencyapproxeqlb}. It will apply the reduction that for a welfare analysis it is without loss to analyze bid distributions without atoms, which we establish in \Cref{t:epsapxnoatom} in \Cref{s:eqlbstruct}. In the argument below, we will use some notation that is defined for the construction of distributions in \Cref{t:epsapxnoatom}. 

\begin{proof}[Proof of \Cref{t:efficiencyapproxeqlb}]
Consider any sequence of $\numunits$-unit first-price auctions $\left(\auction^{(\numunits)}\right)_{\numunits \in \mathbb{N}}$ and a corresponding sequence of $\epsilon^{(\numunits)}$-approximate Bayes-Nash equilibria $\left(\equilibrium^{(\numunits)}\right)_{\numunits \in \mathbb{N}}$. Consider a sequence of positive parameters $\left(\omega^{(\numunits)}\right)_{\numunits \in \mathbb{N}}$ such that $\omega^{(\numunits)} = o(k/n)$. For any $\numunits$, consider the construction of the $\left(\epsilon^{(\numunits)} + \omega^{(\numunits)}\right)$-approximate equilibrium in \Cref{t:epsapxnoatom} which we denote by $\widetilde{\equilibrium}^{(\numunits)}$. By the statement of \Cref{t:epsapxnoatom}, equilibrium $\widetilde{\equilibrium}^{(\numunits)}$ has no atoms and has identical welfare to $\equilibrium^{(\numunits)}$. Furthermore, by \Cref{as:nooverbidding}, we have $\widetilde{\equilibrium}^{(\numunits)}$ satisfies the general no-overbidding assumption \Cref{as:generalnooverbidding}. Let $\epsilon^* := \lim_{\numunits \to \infty} \epsilon^{(\numunits)}$.  

For Part (a), suppose $\epsilon^* = 0$. Applying \Cref{corollary:continuousdistapxpoazerolim}, we have \begin{align} \label{eq:perturbeqlbwel}
        \liminf_{\numunits \to \infty} \frac{\mathcal{W}\left(\widetilde{\equilibrium}^{\left(\numunits\right)},\auction^{\left(\numunits\right)}\right)}{\mathcal{O}\left(\auction^{\left(\numunits\right)}\right)} = 1.
    \end{align} By Part (c) of \Cref{t:epsapxnoatom}, the welfare of equilibrium $\widetilde{\equilibrium}^{(\numunits)}$ and $\equilibrium^{(\numunits)}$ are identical. Substituting in (\ref{eq:perturbeqlbwel}), we get \begin{align*}
        \liminf_{\numunits \to \infty} \frac{\mathcal{W}\left(\equilibrium^{\left(\numunits\right)},\auction^{\left(\numunits\right)}\right)}{\mathcal{O}\left(\auction^{\left(\numunits\right)}\right)} = 1.
    \end{align*}

For Part (b), suppose $\epsilon^* \in (0,H)$. Applying \Cref{lemma:continuousdistapxpoa}, we have \begin{align} \label{eq:perturbeqlbwelapx}
        \liminf_{\numunits \to \infty} \frac{\mathcal{W}\left(\widetilde{\equilibrium}^{\left(\numunits\right)},\auction^{\left(\numunits\right)}\right)}{\mathcal{O}\left(\auction^{\left(\numunits\right)}\right)} \geq 1 - \zeta' \epsilon^* \left(1+\log \left(\frac{H}{\epsilon^*}\right)\right).
    \end{align} By \Cref{t:epsapxnoatom} Part (c), the welfare of equilibrium $\widetilde{\equilibrium}^{(\numunits)}$ and $\equilibrium^{(\numunits)}$ are identical. Substituting in (\ref{eq:perturbeqlbwelapx}), we get \begin{align*}
        \liminf_{\numunits \to \infty} \frac{\mathcal{W}\left(\equilibrium^{\left(\numunits\right)},\auction^{\left(\numunits\right)}\right)}{\mathcal{O}\left(\auction^{\left(\numunits\right)}\right)} \geq 1 - \zeta' \epsilon^* \left(1+\log \left(\frac{H}{\epsilon^*}\right)\right).
    \end{align*} The statement of the result is proved. 
\end{proof}

Finally, we discuss the role of the density lower bound condition in \Cref{as:lowerboundexpectation}. The main goal of this paper is to analyze how the $\numunits$-th order statistic of equilibrium bids concentrates. Although independence of distributions and a uniformly bounded support ensures concentration of many aggregate statistics, they do not, by themselves, ensure concentration of order statistics. Under the density lower bound condition, however, a simple argument shows that order statistics of the value distributions concentrate. This does not directly establish concentration of the relevant order statistics of equilibrium bids which are characterized by strategic considerations. The results in this section analyzed the properties of the limiting distribution of this order statistic and showed how they depend on approximate bidding behavior. 

\section{Structural Properties Of Equilibria}
\label{s:eqlbstruct}

This section establishes key structural properties of equilibria that are useful not only for our analysis but, we believe, may be of independent interest. \Cref{ss:continuityapprox} focuses on approximate Bayes-Nash equilibria while \Cref{ss:continuitybne} focuses on pure Bayes-Nash equilibria. 

\subsection{Continuity of approximate equilibria}
\label{ss:continuityapprox} Approximate equilibria are typically characterized by the presence of atoms in agents' strategies. Analyzing these objects is challenging due to discontinuities in the bid distribution function introduced by such atoms. The result below shows that for a welfare analysis it is without loss to analyze equilibria with no atoms. 

\begin{theorem}
    \label{t:epsapxnoatom}
    Consider any $\epsilon$-approximate equilibrium $\equilibrium$. For every $\omega>0$, there exists a profile of bid distributions $\widetilde{\equilibrium}$ satisfying the following properties: \begin{enumerate}
        \item $\widetilde{\equilibrium}$ is an $(\epsilon+\omega)$-approximate equilibrium.
        \item For every agent $i \in \agentset$, the bid distribution $\widetilde{\equilibrium}_{i}$ has no atoms.
        \item $\text{welfare}(\equilibrium) = \text{welfare}(\widetilde{\equilibrium}).$
    \end{enumerate}
\end{theorem}

\textit{Construction of approximate equilibrium $\widetilde{\equilibrium}$:} Fix any $\epsilon$-approximate equilibrium $\equilibrium$. Let $\mathcal{P}_{i}$ denote the location of atoms in the equilibrium bid distribution of agent $i$. Define $\mathcal{P} := \cup_{i \in \mathcal{N}}\mathcal{P}_{i}$ as the set of locations of atoms across all agents. Note that the set $\mathcal{P}$ is countable. Fix a $\omega>0$. Since $\mathcal{P}$ is countable, for every atom $p \in \mathcal{P}$, we can assign a positive number $s_{p} \in \mathbb{R}_{>0}$ such that $\sum_{p \in \mathcal{P}}s_{p} < \omega$. 

The main idea behind the construction is as follows. At every atom, add a randomization to smoothen the distribution function. An issue is that the randomization may cause bids to overlap across agents: an agent who was winning earlier may now lose. This will change agents' allocation probability and consequently welfare. To avoid this, we perform cumulative shifts of \textit{all} agents' bids. A formal construction is provided below. 

\begin{figure}[t]
    \centering

    \begin{tikzpicture}

    \def\alphaOne{1.85}
    \def\betaOne{4.25}
    \def\salphaOne{1.20}
    \def\sbetaOne{0.95}

    \def\alphalowOne{0.38}
    \def\alphahighOne{0.50}
    \def\betalowOne{0.62}
    \def\betahighOne{0.74}

    \def\middleOneA{2.25}
    \def\middleOneB{2.70}
    \def\middleOneC{3.20}

    \def\finalOneA{4.55}
    \def\finalOneB{4.90}
    \def\finalOneC{5.25}
    \def\finalOneD{5.60}
    \def\finalOneE{5.80}

    \pgfmathsetmacro{\alphaOneShift}{\alphaOne+\salphaOne}
    \pgfmathsetmacro{\betaOneShift}{\betaOne+\salphaOne}

    \pgfmathsetmacro{\middleOneAShift}{\middleOneA+\salphaOne}
    \pgfmathsetmacro{\middleOneBShift}{\middleOneB+\salphaOne}
    \pgfmathsetmacro{\middleOneCShift}{\middleOneC+\salphaOne}

    \pgfmathsetmacro{\finalOneAShift}{\finalOneA+\salphaOne}
    \pgfmathsetmacro{\finalOneBShift}{\finalOneB+\salphaOne}
    \pgfmathsetmacro{\finalOneCShift}{\finalOneC+\salphaOne}
    \pgfmathsetmacro{\finalOneDShift}{\finalOneD+\salphaOne}
    \pgfmathsetmacro{\finalOneEShift}{\finalOneE+\salphaOne}

    \pgfmathsetmacro{\betaOneSecondShift}{\betaOneShift+\sbetaOne}

    \pgfmathsetmacro{\finalOneASecondShift}{\finalOneAShift+\sbetaOne}
    \pgfmathsetmacro{\finalOneBSecondShift}{\finalOneBShift+\sbetaOne}
    \pgfmathsetmacro{\finalOneCSecondShift}{\finalOneCShift+\sbetaOne}
    \pgfmathsetmacro{\finalOneDSecondShift}{\finalOneDShift+\sbetaOne}
    \pgfmathsetmacro{\finalOneESecondShift}{\finalOneEShift+\sbetaOne}

    \def\alphaTwo{1.75}
    \def\betaTwo{4.35}
    \def\salphaTwo{1.05}
    \def\sbetaTwo{0.90}

    \def\alphaValTwo{0.36}
    \def\betaLowTwo{0.73}
    \def\betaHighTwo{0.82}

    \def\postATwoA{2.25}
    \def\postATwoB{2.70}
    \def\postATwoC{3.15}
    \def\postATwoD{3.55}
    \def\postATwoE{3.95}

    \def\postBTwoA{4.80}
    \def\postBTwoB{5.20}
    \def\postBTwoC{5.60}
    \def\postBTwoD{6.00}

    \pgfmathsetmacro{\alphaTwoShift}{\alphaTwo+\salphaTwo}
    \pgfmathsetmacro{\betaTwoShift}{\betaTwo+\salphaTwo}

    \pgfmathsetmacro{\postATwoAShift}{\postATwoA+\salphaTwo}
    \pgfmathsetmacro{\postATwoBShift}{\postATwoB+\salphaTwo}
    \pgfmathsetmacro{\postATwoCShift}{\postATwoC+\salphaTwo}
    \pgfmathsetmacro{\postATwoDShift}{\postATwoD+\salphaTwo}
    \pgfmathsetmacro{\postATwoEShift}{\postATwoE+\salphaTwo}

    \pgfmathsetmacro{\postBTwoAShift}{\postBTwoA+\salphaTwo}
    \pgfmathsetmacro{\postBTwoBShift}{\postBTwoB+\salphaTwo}
    \pgfmathsetmacro{\postBTwoCShift}{\postBTwoC+\salphaTwo}
    \pgfmathsetmacro{\postBTwoDShift}{\postBTwoD+\salphaTwo}

    \pgfmathsetmacro{\betaTwoSecondShift}{\betaTwoShift+\sbetaTwo}

    \pgfmathsetmacro{\postBTwoASecondShift}{\postBTwoAShift+\sbetaTwo}
    \pgfmathsetmacro{\postBTwoBSecondShift}{\postBTwoBShift+\sbetaTwo}
    \pgfmathsetmacro{\postBTwoCSecondShift}{\postBTwoCShift+\sbetaTwo}
    \pgfmathsetmacro{\postBTwoDSecondShift}{\postBTwoDShift+\sbetaTwo}

    \begin{groupplot}[
        group style={
            group size=3 by 2,
            horizontal sep=0.75cm,
            vertical sep=1.2cm
        },
        width=0.26\textwidth,
        height=0.20\textwidth,
        scale only axis,
        axis lines=left,
        xmin=0,
        xmax=8.4,
        ymin=0,
        ymax=1.08,
        xlabel={},
        xtick=\empty,
        ytick={1},
        yticklabels={$1$},
        xlabel style={
            at={(axis description cs:1.05,-0.05)},
            anchor=north east
        },
        axis line style={->,thick},
        clip=false,
    ]

    \nextgroupplot[
        ylabel={$\mathcal{B}_{1}(b)$},
        ylabel style={
            rotate=0,
            at={(axis description cs:-0.14,0.50)},
            anchor=south
        }
    ]

    \addplot[dashed, thick, domain=0:8.0] {1};

    \addplot[
        blue,
        thick,
        smooth
    ] coordinates {
        (0.00,0.00)
        (0.35,0.08)
        (0.70,0.16)
        (1.05,0.24)
        (1.40,0.31)
        (\alphaOne,\alphalowOne)
    };

    \addplot[
        only marks,
        mark=o,
        mark size=2.6pt,
        thick,
        blue,
        fill=white
    ] coordinates {
        (\alphaOne,\alphalowOne)
    };

    \addplot[
        only marks,
        mark=*,
        mark size=2.4pt,
        blue
    ] coordinates {
        (\alphaOne,\alphahighOne)
    };

    \addplot[
        blue,
        thick,
        smooth
    ] coordinates {
        (\alphaOne,\alphahighOne)
        (\middleOneA,0.54)
        (\middleOneB,0.58)
        (\middleOneC,0.61)
        (\betaOne,\betalowOne)
    };

    \addplot[
        only marks,
        mark=o,
        mark size=2.6pt,
        thick,
        blue,
        fill=white
    ] coordinates {
        (\betaOne,\betalowOne)
    };

    \addplot[
        only marks,
        mark=*,
        mark size=2.4pt,
        blue
    ] coordinates {
        (\betaOne,\betahighOne)
    };

    \addplot[
        blue,
        thick,
        smooth
    ] coordinates {
        (\betaOne,\betahighOne)
        (\finalOneA,0.79)
        (\finalOneB,0.86)
        (\finalOneC,0.91)
        (\finalOneD,0.95)
        (\finalOneE,0.96)
    };




    \nextgroupplot

    \addplot[dashed, thick, domain=0:8.0] {1};

    \addplot[
        blue,
        thick,
        smooth
    ] coordinates {
        (0.00,0.00)
        (0.35,0.08)
        (0.70,0.16)
        (1.05,0.24)
        (1.40,0.31)
        (\alphaOne,\alphalowOne)
    };

    \addplot[
        red,
        dashed,
        thick
    ] coordinates {
        (\alphaOne,\alphalowOne)
        (\alphaOneShift,\alphahighOne)
    };

    \addplot[
        blue,
        thick,
        smooth
    ] coordinates {
        (\alphaOneShift,\alphahighOne)
        (\middleOneAShift,0.54)
        (\middleOneBShift,0.58)
        (\middleOneCShift,0.61)
        (\betaOneShift,\betalowOne)
    };

    \addplot[
        only marks,
        mark=o,
        mark size=2.6pt,
        thick,
        blue,
        fill=white
    ] coordinates {
        (\betaOneShift,\betalowOne)
    };

    \addplot[
        only marks,
        mark=*,
        mark size=2.4pt,
        blue
    ] coordinates {
        (\betaOneShift,\betahighOne)
    };

    \addplot[
        blue,
        thick,
        smooth
    ] coordinates {
        (\betaOneShift,\betahighOne)
        (\finalOneAShift,0.79)
        (\finalOneBShift,0.86)
        (\finalOneCShift,0.91)
        (\finalOneDShift,0.95)
        (\finalOneEShift,0.96)
    };




    \nextgroupplot

    \addplot[dashed, thick, domain=0:8.0] {1};

    \addplot[
        blue,
        thick,
        smooth
    ] coordinates {
        (0.00,0.00)
        (0.35,0.08)
        (0.70,0.16)
        (1.05,0.24)
        (1.40,0.31)
        (\alphaOne,\alphalowOne)
    };

    \addplot[
        red,
        dashed,
        thick
    ] coordinates {
        (\alphaOne,\alphalowOne)
        (\alphaOneShift,\alphahighOne)
    };

    \addplot[
        blue,
        thick,
        smooth
    ] coordinates {
        (\alphaOneShift,\alphahighOne)
        (\middleOneAShift,0.54)
        (\middleOneBShift,0.58)
        (\middleOneCShift,0.61)
        (\betaOneShift,\betalowOne)
    };

    \addplot[
        red,
        dashed,
        thick
    ] coordinates {
        (\betaOneShift,\betalowOne)
        (\betaOneSecondShift,\betahighOne)
    };

    \addplot[
        blue,
        thick,
        smooth
    ] coordinates {
        (\betaOneSecondShift,\betahighOne)
        (\finalOneASecondShift,0.79)
        (\finalOneBSecondShift,0.86)
        (\finalOneCSecondShift,0.91)
        (\finalOneDSecondShift,0.95)
        (\finalOneESecondShift,0.96)
    };




    \nextgroupplot[
        ylabel={$\mathcal{B}_{2}(b)$},
        ylabel style={
            rotate=0,
            at={(axis description cs:-0.14,0.50)},
            anchor=south
        }
    ]

    \addplot[dashed, thick, domain=0:8.0] {1};

    \addplot[
        blue,
        thick,
        smooth
    ] coordinates {
        (0.00,0.00)
        (0.45,0.11)
        (0.90,0.22)
        (1.30,0.30)
        (\alphaTwo,\alphaValTwo)
        (\postATwoA,0.46)
        (\postATwoB,0.57)
        (\postATwoC,0.64)
        (\postATwoD,0.69)
        (\postATwoE,0.71)
        (\betaTwo,\betaLowTwo)
    };

    \addplot[
        only marks,
        mark=o,
        mark size=2.6pt,
        thick,
        blue,
        fill=white
    ] coordinates {
        (\betaTwo,\betaLowTwo)
    };

    \addplot[
        only marks,
        mark=*,
        mark size=2.4pt,
        blue
    ] coordinates {
        (\betaTwo,\betaHighTwo)
    };

    \addplot[
        blue,
        thick,
        smooth
    ] coordinates {
        (\betaTwo,\betaHighTwo)
        (\postBTwoA,0.85)
        (\postBTwoB,0.87)
        (\postBTwoC,0.885)
        (\postBTwoD,0.89)
    };

    \draw[thick] (axis cs:\alphaTwo,0) -- (axis cs:\alphaTwo,0.04);
    \draw[thick] (axis cs:\betaTwo,0) -- (axis cs:\betaTwo,0.04);

    \node[below] at (axis cs:\alphaTwo,0) {$\alpha$};
    \node[below] at (axis cs:\betaTwo,0) {$\beta$};

    \node[anchor=north, font=\small] at (rel axis cs:0.5,-0.25) {(a)};

    \nextgroupplot

    \addplot[dashed, thick, domain=0:8.0] {1};

    \addplot[
        blue,
        thick,
        smooth
    ] coordinates {
        (0.00,0.00)
        (0.45,0.11)
        (0.90,0.22)
        (1.30,0.30)
        (\alphaTwo,\alphaValTwo)
    };

    \addplot[
        red,
        dashed,
        thick
    ] coordinates {
        (\alphaTwo,\alphaValTwo)
        (\alphaTwoShift,\alphaValTwo)
    };

    \addplot[
        blue,
        thick,
        smooth
    ] coordinates {
        (\alphaTwoShift,\alphaValTwo)
        (\postATwoAShift,0.46)
        (\postATwoBShift,0.57)
        (\postATwoCShift,0.64)
        (\postATwoDShift,0.69)
        (\postATwoEShift,0.71)
        (\betaTwoShift,\betaLowTwo)
    };

    \addplot[
        only marks,
        mark=o,
        mark size=2.6pt,
        thick,
        blue,
        fill=white
    ] coordinates {
        (\betaTwoShift,\betaLowTwo)
    };

    \addplot[
        only marks,
        mark=*,
        mark size=2.4pt,
        blue
    ] coordinates {
        (\betaTwoShift,\betaHighTwo)
    };

    \addplot[
        blue,
        thick,
        smooth
    ] coordinates {
        (\betaTwoShift,\betaHighTwo)
        (\postBTwoAShift,0.85)
        (\postBTwoBShift,0.87)
        (\postBTwoCShift,0.885)
        (\postBTwoDShift,0.89)
    };

    \draw[thick] (axis cs:\alphaTwoShift,0) -- (axis cs:\alphaTwoShift,0.04);
    \draw[thick] (axis cs:\betaTwoShift,0) -- (axis cs:\betaTwoShift,0.04);

    \node[below, yshift=-2pt, font=\scriptsize] at (axis cs:\alphaTwoShift,0) {$\alpha+s_{\alpha}$};
    \node[below, font=\scriptsize] at (axis cs:\betaTwoShift,0) {$\beta+s_{\alpha}$};

    \node[anchor=north, font=\small] at (rel axis cs:0.5,-0.25) {(b)};

    \nextgroupplot

    \addplot[dashed, thick, domain=0:8.0] {1};

    \addplot[
        blue,
        thick,
        smooth
    ] coordinates {
        (0.00,0.00)
        (0.45,0.11)
        (0.90,0.22)
        (1.30,0.30)
        (\alphaTwo,\alphaValTwo)
    };

    \addplot[
        red,
        dashed,
        thick
    ] coordinates {
        (\alphaTwo,\alphaValTwo)
        (\alphaTwoShift,\alphaValTwo)
    };

    \addplot[
        blue,
        thick,
        smooth
    ] coordinates {
        (\alphaTwoShift,\alphaValTwo)
        (\postATwoAShift,0.46)
        (\postATwoBShift,0.57)
        (\postATwoCShift,0.64)
        (\postATwoDShift,0.69)
        (\postATwoEShift,0.71)
        (\betaTwoShift,\betaLowTwo)
    };

    \addplot[
        red,
        dashed,
        thick
    ] coordinates {
        (\betaTwoShift,\betaLowTwo)
        (\betaTwoSecondShift,\betaHighTwo)
    };

    \addplot[
        blue,
        thick,
        smooth
    ] coordinates {
        (\betaTwoSecondShift,\betaHighTwo)
        (\postBTwoASecondShift,0.85)
        (\postBTwoBSecondShift,0.87)
        (\postBTwoCSecondShift,0.885)
        (\postBTwoDSecondShift,0.89)
    };

    \draw[thick] (axis cs:\alphaTwoShift,0) -- (axis cs:\alphaTwoShift,0.04);
    \draw[thick] (axis cs:\betaTwoSecondShift,0) -- (axis cs:\betaTwoSecondShift,0.04);

    \node[below, yshift=-2pt, font=\scriptsize] at (axis cs:\alphaTwoShift,0) {$\alpha+s_{\alpha}$};
    \node[below, font=\scriptsize] at (axis cs:\betaTwoSecondShift,0) {$\beta+s_{\alpha}+s_{\beta}$};

    \node[anchor=north, font=\small] at (rel axis cs:0.5,-0.25) {(c)};

    \end{groupplot}
    \end{tikzpicture}

    \caption{This figure illustrates the construction of an equilibrium with no atoms $\widetilde{\equilibrium}$ beginning from $\equilibrium$. Column (a) depicts the distribution function of equilibrium bids for each agent: $\mathcal{B}_{1}(\cdot)$ for agent $1$ and $\mathcal{B}_{2}(\cdot)$ for agent $2$. Agent $1$ has atoms at $\alpha$ and $\beta$ while agent $2$ only has an atom at $\beta$. Column (b) depicts the shift in bids to the right of atom $\alpha$ by $s_{\alpha}$. The red dashed line shows the uniform randomization of the bid at $\alpha$. Column (c) depicts the shift in bids to the right of the atom at $\beta + s_{\alpha}$ by $s_{\beta}$.  
    }
    \label{fig:no-atoms-construction}
\end{figure}
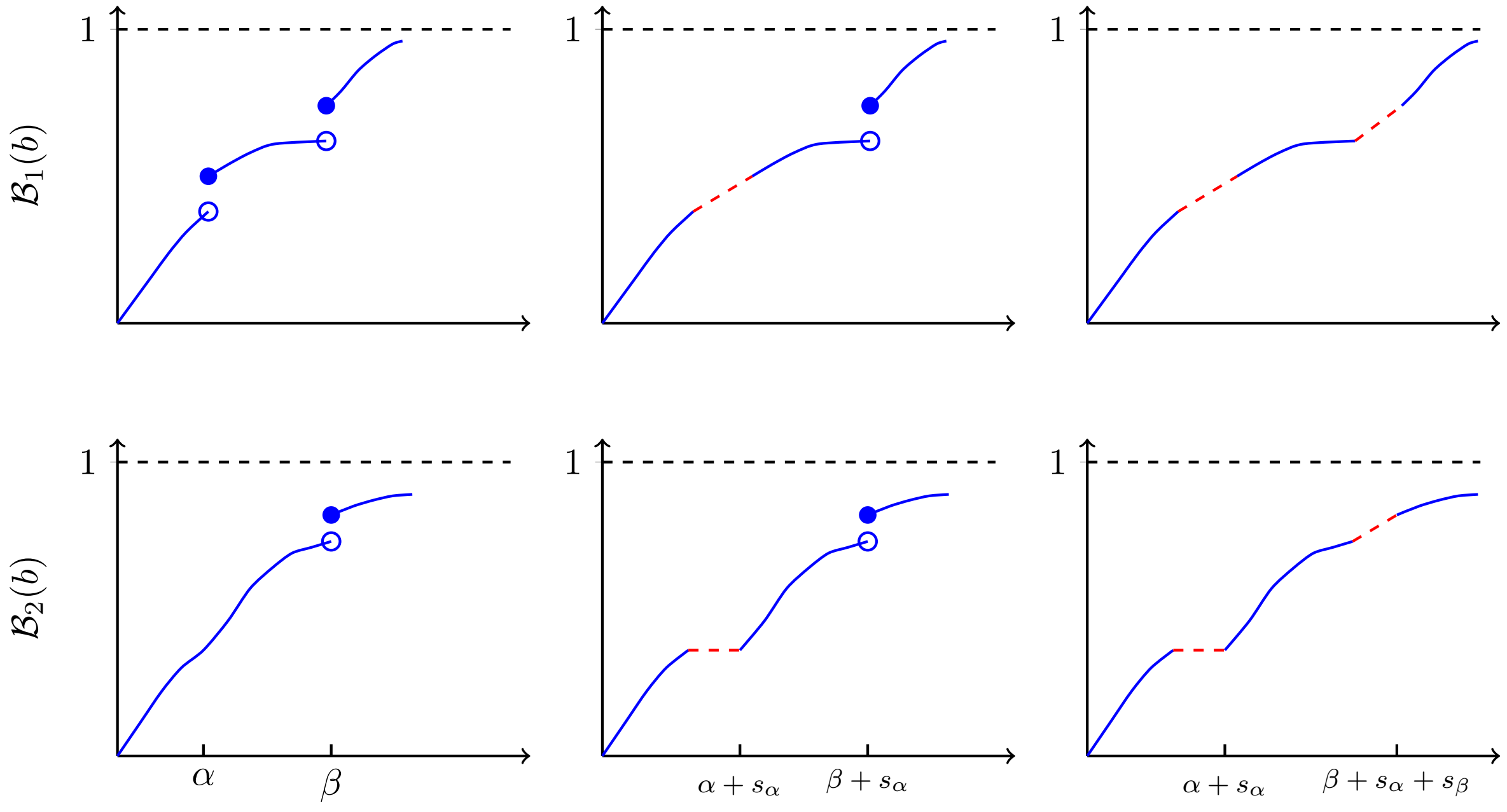

Consider the following construction of a collection of bid distributions $\widetilde{\equilibrium} = \left(\widetilde{\equilibrium}_{i}\right)_{i \in \agentset}$. For every agent $i \in \agentset$, sample the equilibrium bid $b \sim \equilibrium_{i}$. Agent $i$ plays bid $\widetilde{b}$ defined as: \begin{align}
     \text{ if } b \not \in \mathcal{P}, \quad & \text{ then } \widetilde{b} := b + \sum_{\{p \in \mathcal{P} : p < b\}}s_{p}, \label{eq:noatombid} \\
     \text{ if } b \in \mathcal{P}, \quad & \text{ then } \widetilde{b} \sim b + \sum_{\{p \in \mathcal{P} : p < b\}}s_{p} + s_{b} \cdot U[0,1]. \label{eq:atombid}  
   \end{align} Here, $U[0,1]$ is the uniform random variable in the interval $[0,1]$. Crucially, this variable is independent across atoms and across agents. \Cref{fig:no-atoms-construction} provides a pictorial construction of $\widetilde{\equilibrium}$ in a simple two agent setting. Finally, we prove Part (b), then prove Part (c) and finish by proving Part (a).

\begin{proof}[Proof of \Cref{t:epsapxnoatom} Part (b)]
    This follows by construction of bid distributions $\widetilde{\equilibrium}$. 
\end{proof}

\begin{proof}[Proof of \Cref{t:epsapxnoatom} Part (c)]
    We will compare the allocation probability $x_{i}(\cdot)$ under the bid profile $\equilibrium$ with the allocation probability $\widetilde{x}_{i}(\cdot)$ under the bid profile $\widetilde{\equilibrium}$. Consider any agent $i$ and any bid $b_{i}$. In particular, we will establish the following result.

    \begin{lemma} \label{l:idallocprob}
        Consider any agent $i \in \agentset$ and bid $\bid{i}$. The allocation probability $x_{i}(\cdot)$ and $\widetilde{x}_{i}(\cdot)$ satisfy the following: \begin{enumerate}[(i)]
            \item if $b_{i} \not \in \mathcal{P}$, then for the bid $\widetilde{b}_{i}$ defined in (\ref{eq:noatombid}) we have $x_{i}(b_{i}) = \widetilde{x}_{i}\left(\widetilde{b}_{i}\right)$.
            \item if $b_{i} \in \mathcal{P}$, then for the random bid $\widetilde{b}_{i}$ defined in (\ref{eq:atombid}) we have $x_{i}(b_{i}) = \expect \left[\widetilde{x}_{i}\left(\widetilde{b}_{i}\right)\right]$.
        \end{enumerate}
    \end{lemma}

Before we prove the lemma, we demonstrate how it implies Part (c) of \Cref{t:epsapxnoatom}, that is, $$ \textit{welfare}\left(\equilibrium\right) = \textit{welfare}\left(\widetilde{\equilibrium}\right).$$ Recall, we denote the equilibrium bid distribution of agent $i$ by $\equilibrium_{i,v}$. The welfare under equilibrium $\equilibrium$ is \begin{align} \label{eq:eqlbwelfare}
    \textit{welfare}(\equilibrium) = \mathbb{E}_{\bm{v} \sim \mathcal{V}}\left[\sum_{i \in \agentset}v_{i}\mathbb{E}_{b_{i} \sim \equilibrium_{i,v_{i}}}\left[x_{i}(b_{i})\right]\right].
\end{align} By construction of $\widetilde{\equilibrium}$ and \Cref{l:idallocprob}, we have \begin{align*}
    \mathbb{E}_{b_{i} \sim \equilibrium_{i,v_{i}}}\left[x_{i}(b_{i})\right] = \mathbb{E}_{\widetilde{b}_{i} \sim \widetilde{\equilibrium}_{i,v_{i}}}\left[\widetilde{x}_{i}\left(\widetilde{b}_{i}\right)\right].
\end{align*} Substituting in (\ref{eq:eqlbwelfare}), we have \begin{align*}
    \textit{welfare}(\equilibrium) = \mathbb{E}_{\bm{v} \sim \mathcal{V}}\left[\sum_{i \in \agentset}v_{i}\mathbb{E}_{\widetilde{b}_{i} \sim \widetilde{\equilibrium}_{i,v_{i}}}\left[\widetilde{x}_{i}\left(\widetilde{b}_{i}\right)\right]\right] = \textit{welfare}(\widetilde{\equilibrium}).
\end{align*} The rest of the proof focuses on providing arguments to establish \Cref{l:idallocprob}. We begin by proving Part (i) and then prove Part (ii).

\textit{Proof of Part (i).} Suppose $b_{i} \not \in \mathcal{P}$. We will establish that $x_{i}(b_{i}) = \widetilde{x}_{i}\left(\widetilde{b}_{i}\right)$ where $\widetilde{b}_{i}$  is defined in (\ref{eq:noatombid}). To show this, it suffices to prove that the order of bids between bid profiles $\bm{b}$ and $\widetilde{\bm{b}}$ given by (\ref{eq:noatombid}) and (\ref{eq:atombid}), is preserved as follows: for any agent $j$ and bid $b_{j}$, \begin{equation} \label{eq:orderpreserve}
       \begin{aligned}
       \text{if } & b_{i} > b_{j}, \text{ then } \widetilde{b}_{i} > \widetilde{b}_{j} \text{ almost surely and} \\
       \text{if } & b_{i} < b_{j}, \text{ then } \widetilde{b}_{i} < \widetilde{b}_{j} \text{ almost surely}.
   \end{aligned}
   \end{equation} We begin by showing that if $b_{j} < b_{i}$, the sampled bid $\widetilde{b}_{j} < \widetilde{b}_{i}$ almost surely. This is because of the following: \begin{itemize}
       \item if $b_{j} \not \in \mathcal{P}$, then because $b_{j} < b_{i}$, we have \begin{align*}
           \widetilde{b}_{j} &= b_{j} + \sum_{\{p \in \mathcal{P}:p<b_{j}\}}s_{p} < b_{i} + \sum_{\{p \in \mathcal{P}:p<b_{i}\}}s_{p} = \widetilde{b}_{i}. 
       \end{align*}
       \item if $b_{j} \in \mathcal{P}$, by definition (\ref{eq:atombid}), the sampled bid $\widetilde{b}_{j}$ is at most $b_{j} + \sum_{\{p \in \mathcal{P}:p \leq b_{j}\}}s_{p}$, almost surely. Because $b_{j} < b_{i}$, we have  \begin{align*}
           b_{j} + \sum_{\{p \in \mathcal{P}:p \leq b_{j}\}}s_{p} < b_{i} + \sum_{\{p \in \mathcal{P}:p<b_{i}\}}s_{p} = \widetilde{b}_{i}. 
       \end{align*} It follows that $\widetilde{b}_{j} < \widetilde{b}_{i}$ almost surely.
   \end{itemize} Next, for any agent $j$ and bid $b_{j}$ satisfying $b_{j} > b_{i}$, the sampled bids satisfy $\widetilde{b}_{j} > \widetilde{b}_{i}$ almost surely. This is because of the following: \begin{itemize}
       \item if $b_{j} \not \in \mathcal{P}$, then because $b_{j} > b_{i}$, we have \begin{align*}
           \widetilde{b}_{j} &= b_{j} + \sum_{\{p \in \mathcal{P}:p<b_{j}\}}s_{p} > b_{i} + \sum_{\{p \in \mathcal{P}:p<b_{i}\}}s_{p} = \widetilde{b}_{i}. 
       \end{align*}
       \item if $b_{j} \in \mathcal{P}$, by definition (\ref{eq:atombid}), the sampled bid $\widetilde{b}_{j}$ is at least $b_{j} + \sum_{\{p \in \mathcal{P}:p < b_{j}\}}s_{p}$, almost surely. Because $b_{j} > b_{i}$, we have  \begin{align*}
            b_{j} + \sum_{\{p \in \mathcal{P}:p < b_{j}\}}s_{p} > b_{i} + \sum_{\{p \in \mathcal{P}:p<b_{i}\}}s_{p} = \widetilde{b}_{i}. 
       \end{align*} It follows $\widetilde{b}_{j} > \widetilde{b}_{i}$ almost surely.
   \end{itemize} Finally, the bid $b_{i}$ is not an atom. Thus, for any agent $j$ and bid $b_{j}$, the probability that $b_{i} = b_{j}$ is zero. Combining this with the observation that the order of bids between profiles $\bm{b}$ and $\bm{\widetilde{b}}$ is preserved as defined in (\ref{eq:orderpreserve}), we have that the allocation probability $x_{i}(b_{i}) = \widetilde{x}_{i}\left(\widetilde{b}_{i}\right)$. 

   \textit{Proof of Part (ii).} Suppose $b_{i} \in \mathcal{P}$. Here, the mapped bid $\widetilde{b}_{i}$ is sampled according to \begin{align*}
       \widetilde{b}_{i} \sim b_{i} + \sum_{\{p \in \mathcal{P}:p<b_{i}\}}s_{p} + s_{b_{i}} \cdot U[0,1].
   \end{align*} We will establish that $x_{i}(b_{i}) = \expect\left[\widetilde{x}_{i}\left(\widetilde{b}_{i}\right)\right]$. Consider any profile of bids $\bm{b}$. To show this, it suffices to establish the following: \begin{itemize}
       \item any bid $b_{j}$ that does not tie with $b_{i}$ preserves the ordering as defined in (\ref{eq:orderpreserve}), and,
       \item for any set of bids $b_{j}$ that tie with $b_{i}$, the mapped random bids $\widetilde{b}_{j}$ and $\widetilde{b}_{i}$ simulate uniform tie-breaking. 
   \end{itemize} An argument identical to that presented in Case 1 establishes Part (i). To show Part (ii), let $\mathcal{J}^*$ be the set of agents who have bid $b_{i}$ in profile $\bm{b}$. By assumption that the auction breaks ties uniform randomly, agent $i$'s allocation probability at this bid profile satisfies \begin{equation}
       \label{eq:tiebreakallocapx} x_{i}(\bm{b}) =  \min \left \{ \frac{ \max \left \{\numunits - \sum_{j \in  \mathcal{N}}\mathbf{1}\left[b_{j} > b_{i}\right],0 \right \}}{|\mathcal{J}^*|} , 1 \right \},
   \end{equation} where $\bm{1}\left[\cdot\right]$ is the indicator function. Under the mapped profile $\bm{\widetilde{b}}$ and by Part (i), only agents in $\mathcal{J}^*$ compete for the remaining $\max \left \{\numunits-\sum_{j \in \mathcal{N}}\bm{1}\left[b_{j} > b_{i}\right],0 \right \}$ units. Furthermore, by (\ref{eq:atombid}), for every $j \in \mathcal{J}^*$, we have \begin{align*}
       \widetilde{b}_{j} \sim b_{i} + \sum_{p \in \mathcal{P} : p < b_{i}}s_{p} + s_{b_{i}} \cdot U[0,1],
   \end{align*} where the randomness is independent across agents. Observe that the mapped bid $\widetilde{b}_{j}$ is independent and identical across all agents in $\mathcal{J}^*$. Thus, under the random profile $\bm{\widetilde{b}}$, the expected allocation probability for agent $i$ is \begin{align} \label{eq:mappedbid}
       \expect \left[\widetilde{x}_{i}(\bm{\widetilde{b}})\right] = \min \left \{ \frac{ \max \left \{\numunits - \sum_{j \in  \mathcal{N}}\mathbf{1}\left[b_{j} > b_{i}\right],0 \right \}}{|\mathcal{J}^*|} , 1 \right \} = x_{i}(\bm{b}),
   \end{align} where the second equality follows from (\ref{eq:tiebreakallocapx}). Taking an expectation over all bid profiles $\bm{b}$ gives the result $x_{i}(b_{i}) = \expect \left[\widetilde{x}_{i}\left(\widetilde{b}_{i}\right)\right]$. Part (c) of the theorem follows.
\end{proof}

\begin{proof}[Proof of \Cref{t:epsapxnoatom} Part (a)]
    By definition of an $\epsilon$-approximate equilibrium, for every agent $i \in \agentset$ and every value $v_{i}$, the agent's expected utility is \begin{align} \label{eq:epsapxdef}
        \mathbb{E}_{b_{i} \sim \equilibrium_{i,v_{i}}} \left[\left(v_{i}-b_{i}\right)x_{i}\left(b_{i}\right)\right] \geq \sup_{\overline{b} \geq 0} \left(v_{i}-\overline{b}\right)x_{i}(\overline{b}) - \epsilon.
    \end{align} We will now establish a bound on the utility under equilibrium $\widetilde{\equilibrium}$. For any agent $i$ and value $v_{i}$, we show the expected utility satisfies \begin{align} \label{eq:lbdeltautil}
        \mathbb{E}_{\widetilde{b}_{i} \sim \widetilde{\equilibrium}_{i,v_{i}}} \left[\left(v_{i}-\widetilde{b}_{i}\right)\widetilde{x}_{i}\left(\widetilde{b}_{i}\right)\right] & \geq \mathbb{E}_{b_{i} \sim \equilibrium_{i,v_{i}}} \left[\left(v_{i}-b_{i}\right)x_{i}\left(b_{i}\right)\right] - \omega.
    \end{align} By construction of $\widetilde{\equilibrium}$, the bid $\widetilde{b}_{i}$ is mapped from a bid $b_{i}$ that is either an atom or not (see equation (\ref{eq:atombid}) and (\ref{eq:noatombid})). In both cases, we have that the difference $\widetilde{b}_{i}-b_{i} \in [0,\omega)$ almost surely. Let $\widetilde{\mathcal{P}}$ denote the event that bid $\widetilde{b}_{i}$ is mapped from an atom and $\widetilde{\mathcal{P}}^{c}$ denote its complement. We can write agent $i$'s expected utility as \begin{multline} \label{eq:utilcondexp}
        \mathbb{E}_{\widetilde{b}_{i} \sim \widetilde{\equilibrium}_{i,v_{i}}} \left[\left(v_{i}-\widetilde{b}_{i}\right)\widetilde{x}_{i}\left(\widetilde{b}_{i}\right)\right] = \mathbb{E}_{\widetilde{b}_{i} \sim \widetilde{\equilibrium}_{i,v_{i}}} \left[\left(v_{i}-\widetilde{b}_{i}\right)\widetilde{x}_{i}\left(\widetilde{b}_{i}\right)| \widetilde{\mathcal{P}}^{c}\right] \prob \left[\widetilde{\mathcal{P}}^{c}\right] \\ + \mathbb{E}_{\widetilde{b}_{i} \sim \widetilde{\equilibrium}_{i,v_{i}}} \left[\left(v_{i}-\widetilde{b}_{i}\right)\widetilde{x}_{i}\left(\widetilde{b}_{i}\right)| \widetilde{\mathcal{P}}\right] \prob \left[\widetilde{\mathcal{P}}\right].
    \end{multline} We will rewrite the right-hand side in terms of equilibrium $\equilibrium$. Consider the first term. By Part (i) of \Cref{l:idallocprob} and using $\widetilde{b}_{i} - b_{i} < \omega$, we have \begin{align} \label{eq:lbnoatom}
        \mathbb{E}_{\widetilde{b}_{i} \sim \widetilde{\equilibrium}_{i,v_{i}}} \left[\left(v_{i}-\widetilde{b}_{i}\right)\widetilde{x}_{i}\left(\widetilde{b}_{i}\right)| \widetilde{\mathcal{P}}^{c}\right] \prob \left[\widetilde{\mathcal{P}}^{c}\right] & \geq \left(\mathbb{E}_{b_{i} \sim \equilibrium_{i,v_{i}}} \left[\left(v_{i}-b_{i}\right)x_{i}(b_{i}) | b_{i} \not \in \mathcal{P}\right] - \omega \right) \prob \left[b_{i} \not \in \mathcal{P}\right],
    \end{align} where, recall, $\mathcal{P}$ is the set of atoms. For the second term, by Part (ii) of \Cref{l:idallocprob} and using $\widetilde{b}_{i}-b_{i} < \omega$ almost surely, we have \begin{align} \label{eq:lbatom}
        \mathbb{E}_{\widetilde{b}_{i} \sim \widetilde{\equilibrium}_{i,v_{i}}} \left[\left(v_{i}-\widetilde{b}_{i}\right)\widetilde{x}_{i}\left(\widetilde{b}_{i}\right)| \widetilde{\mathcal{P}}\right] \prob \left[\widetilde{\mathcal{P}}\right] \geq \left(\mathbb{E}_{b_{i} \sim \equilibrium_{i,v_{i}}} \left[\left(v_{i}-b_{i}\right)x_{i}(b_{i}) | b_{i} \in \mathcal{P}\right] - \omega \right) \prob \left[b_{i} \in \mathcal{P}\right]. 
    \end{align} Substituting (\ref{eq:lbnoatom}) and (\ref{eq:lbatom}) in (\ref{eq:utilcondexp}), we get the desired inequality in (\ref{eq:lbdeltautil}). 
    
    To establish that $\widetilde{\equilibrium}$ is an $(\epsilon+\omega)$-approximate equilibrium, it is sufficient to show \begin{align} \label{eq:optutilcomp}
        \sup_{\overline{b} \geq 0} \left(v_{i}-\overline{b}\right)x_{i}(\overline{b}) \geq \max_{\overline{b} \geq 0} \left(v_{i}-\overline{b}\right) \widetilde{x}_{i}(\overline{b}). 
    \end{align} The right-hand side in (\ref{eq:optutilcomp}) is well defined. To see why, bid distributions in $\widetilde{\equilibrium}$ have no atoms and thus $\widetilde{x}_{i}(\cdot)$ is continuous. Furthermore, every bid $\overline{b}$ in the compact interval $[0,v_{i}]$ obtains strictly greater utility than bids strictly greater than $v_{i}$. Since a continuous function over a compact interval achieves its maximum, the right-hand side is well defined. Now, our approach is to show that for any solution to the right-hand side, there is a corresponding bid which gives the agent weakly greater utility under allocation $x_{i}(\cdot)$. 

    Let $\widetilde{b}$ denote an optimal solution to the right-hand side of (\ref{eq:optutilcomp}). Suppose $\widetilde{b}$ is mapped from a bid $b$ that is not an atom (see equation (\ref{eq:noatombid})). Then by Part (i) of \Cref{l:idallocprob}, we have $x_{i}(b) = \widetilde{x}_{i}\left(\widetilde{b}\right)$. Furthermore, by construction in (\ref{eq:noatombid}), we have $\widetilde{b} \geq b$. Thus, \begin{align*}
        \sup_{\overline{b} \geq 0}\left(v_{i}-\overline{b}\right)x_{i}(\overline{b}) \geq \left(v_{i}-b\right)x_{i}(b) \geq \left(v_{i}-\widetilde{b}\right)\widetilde{x}_{i}\left(\widetilde{b}\right).
    \end{align*} Now suppose that $\widetilde{b}$ is mapped from a bid $b$ that is an atom (see equation (\ref{eq:atombid})) and $\widetilde{b} > b$. Since the number of atoms are countable, there exists a bid $b^*$ in the interval $(b,\widetilde{b})$ such that $b^*$ is not an atom. By definition of the map in (\ref{eq:noatombid}), we have \begin{align*}
        \widetilde{b^*} = b^* + \sum_{p \in \mathcal{P} : p < b^*}s_{p}
        > b + \sum_{p \in \mathcal{P}:p \leq b}s_{p} \geq \widetilde{b}. 
    \end{align*} By monotonicity of the allocation rule, we have $\widetilde{x}_{i}(\widetilde{b}^*) \geq \widetilde{x}_{i}(\widetilde{b})$. But by Part (i) of \Cref{l:idallocprob}, we have $x_{i}(b^*) = \widetilde{x}_{i}\left(\widetilde{b}^*\right)$. Combining both and using $b^* < \widetilde{b}$ gives \begin{align*}
        \sup_{\overline{b} \geq 0}\left(v_{i}-\overline{b}\right)x_{i}(\overline{b}) \geq \left(v_{i}-b^*\right)x_{i}(b^*) \geq \left(v_{i}-\widetilde{b}\right)\widetilde{x}_{i}\left(\widetilde{b}\right).
    \end{align*} This establishes (\ref{eq:optutilcomp}). Finally, suppose that $\widetilde{b}=b$. By construction, there is no atom in $\equilibrium$ that is strictly smaller than $b$. The allocation probability, then, satisfies $x_{i}\left(b\right) \geq \widetilde{x}_{i}\left(\widetilde{b}\right)$. As in the previous case, this gives \begin{align*}
        \sup_{\overline{b} \geq 0}\left(v_{i}-\overline{b}\right)x_{i}(\overline{b}) \geq \left(v_{i}-b\right)x_{i}(b) \geq \left(v_{i}-\widetilde{b}\right)\widetilde{x}_{i}\left(\widetilde{b}\right).
    \end{align*}

    Finally, consider the following chain of inequalities: \begin{align*}
        \mathbb{E}_{\widetilde{b}_{i} \sim \widetilde{\equilibrium}_{i,v_{i}}} \left[\left(v_{i}-\widetilde{b}_{i}\right)\widetilde{x}_{i}\left(\widetilde{b}_{i}\right)\right] & \geq \mathbb{E}_{b_{i} \sim \equilibrium_{i,v_{i}}} \left[\left(v_{i}-b_{i}\right)x_{i}\left(b_{i}\right)\right] - \omega \quad (\text{from } (\ref{eq:lbdeltautil})), \\
        & \geq \sup_{\overline{b} \geq 0}\left(v_{i}-\overline{b}\right)x_{i}(\overline{b}) - (\epsilon+\omega) \quad (\text{from } (\ref{eq:epsapxdef})), \\
        & \geq \max_{\overline{b} \geq 0} \left(v_{i}-\overline{b}\right) \widetilde{x}_{i}(\overline{b}) - (\epsilon+\omega) \quad (\text{from } (\ref{eq:optutilcomp})). 
    \end{align*} Since this holds for every agent $i$ and every value $v_{i}$, the bid profile $\widetilde{\equilibrium}$ is an $(\epsilon+\omega)$-approximate equilibrium. 
\end{proof}

\subsection{Continuity and Monotonicity of Bayes-Nash equilibria}
\label{ss:continuitybne} This section presents a continuity and monotonicity result of Bayes-Nash equilibrium bid distributions in multi-unit first-price auctions. In \Cref{ss:eqlbbidproperties}, we also establish some standard properties which generalize their counterparts in the single-unit setting. We will use some of those structural results in our proofs below. 

Before we present the result, we introduce some notation. Let $\supportinforderstat{\numunits}$ denote the infimum and $\supportsuporderstat{\numunits}$ denote the supremum of the support of the critical bid $\khighbidCDF$. A simple argument establishes that the infimum of the critical bid's support is $\supportinforderstat{\numunits} = 0$ under \Cref{as:lowerboundexpectation}. 

\begin{fact}
    \label{fact:gammazero}
    Suppose \Cref{as:lowerboundexpectation} is satisfied. In any Bayes-Nash equilibrium, $\supportinforderstat{\numunits} = 0$. 
\end{fact}

\begin{proof}
    Suppose that $\supportinforderstat{\numunits} > 0$. Conditional on the event that every agent $i$ has a value $v_{i} \in \left[0,\supportinforderstat{\numunits}/2\right]$, there must be an agent that is allocated a unit with strictly positive probability. Let us denote this agent by $\underline{i}$. By independence of value distributions and \Cref{as:lowerboundexpectation}, conditional on $v_{\underline{i}} \in \left[0,\supportinforderstat{\numunits}/2\right]$, the probability agent $\underline{i}$ is allocated a unit is strictly positive. Thus, by \Cref{as:lowerboundexpectation}, there must exist a value $\underline{v}$ in $\left[0,\supportinforderstat{\numunits}/2\right]$ and in the support of agent $\underline{i}$'s value distribution which has a strictly positive allocation. By definition of $\supportinforderstat{\numunits}$, the equilibrium bid for $\underline{v}$ is at least $\supportinforderstat{\numunits}$ with strictly positive probability. However, the utility for such a bid is strictly negative which is a contradiction. Thus, it must be that $\supportinforderstat{\numunits} = 0$. 
\end{proof}

We now present the result that establishes the monotonicity and continuity of equilibrium distributions. 

\begin{theorem}
   \label{t:bnecontmono}
   Consider any $\numunits$-unit first-price auction $\auction^{(\numunits)}$ and any corresponding pure Bayes-Nash equilibrium $\equilibrium^{(\numunits)}$ satisfying $\supportinforderstat{\numunits} < \supportsuporderstat{\numunits}$. The following is satisfied: \begin{enumerate}
    \item At most $(\numunits-1)$ agents have an atom at $\supportsuporderstat{\numunits}$.
     \item The distribution function of the $\numunits$-th order statistic of equilibrium bids $\khighbidCDF$ is strictly increasing in the interval $\left[\supportinforderstat{\numunits},\supportsuporderstat{\numunits}\right]$.
       \item For every agent $i$, the equilibrium bid distribution $\equilibrium_{i}^{(\numunits)}$ has no atoms in the interval $\left(\supportinforderstat{\numunits},\supportsuporderstat{\numunits}\right)$.
   \end{enumerate} 
\end{theorem}

This generalizes an analogous result in the single-unit setting by \cite{JL-23}. A key advantage of that setting is that it only requires analyzing the highest equilibrium bid, which corresponds to the first-order statistic. In the multi-unit setting, however, a fundamental challenge arises: changes in an agent's bid distribution need not necessarily impact the competing bid distribution of another agent. This is because an agent's allocation probability is determined solely by the $k^{\text{th}}$-highest bid among the remaining agents. Addressing this issue requires a different set of arguments. We use an inductive argument in conjunction with a standard neighborhood deviation argument (see \Cref{ss:eqlbbidproperties} for a statement of the result) to prove \Cref{t:bnecontmono}. 

We now provide several lemmas which build towards the final proof. A crucial result demonstrates that at most one agent can have a probability mass at any bid within the interval $\left( \supportinforderstat{\numunits}, \supportsuporderstat{\numunits} \right)$. The argument we use will serve as the foundation for several proofs later on. 

\begin{lemma}\label{lma:at-most-one-agent-with-point-mass}
    Consider any bid $b^* \in \left( \supportinforderstat{\numunits}, \supportsuporderstat{\numunits} \right)$. The bid distribution of at most $1$ agent places strictly positive probability mass at $b^*$.
\end{lemma}

\begin{proof}
    Suppose there are $2$ agents, $i$ and $j$, such that their bid distributions $\bidCDF{i}$ and $\bidCDF{j}$ have probability mass at $b$. We will show that the interim allocation for one of these agents has a discontinuity at $b$. A sufficient condition for this is if the following properties hold:
    \begin{enumerate}[(i)]
        \item With strictly positive probability, $b$ is the bid of the $\numunits^{\text{th}}$ highest agent.
        \item With strictly positive probability, exactly $\numunits - 1$ agents other than $i$ and $j$ bid strictly above $b$.
    \end{enumerate}
    The rest of the proof will focus on showing the properties above. Define $\lowagents$ as
    \begin{align*}
        \lowagents := \{ l : \bidCDF{l}(b) = 0 \} \text{.}
    \end{align*}
    The agents in $\lowagents$ bid above $b$ almost surely. It must be the case that $|\lowagents| \leq \numunits - 1$ otherwise we would have $\supportinforderstat{\numunits} \geq b$ which is not possible. Next, define $\topagents$ as
    \begin{align*}
        \topagents := \{ l : \supportsup{l} = \supportsuporderstat{\numunits} \} \text{.}
    \end{align*}
    An application of \Cref{lma:support-upper-bound-equivalence} tells us that $|\topagents| \geq \numunits + 1$. Consider a set $\topagents'$ such that
    \begin{align*}
        \topagents' \subseteq \topagents \setminus \left(\{i,j\} \cup \lowagents\right) \quad \text{ and } \quad  |\topagents'| = \numunits - 1 - |\lowagents| \text{.}
    \end{align*}
    Such a set exists because $|\topagents| \geq \numunits + 1$ and $|\lowagents| \leq \numunits - 1$. Consider the set $\mathcal{S} = \lowagents \cup \topagents'$. We will show that agents in $\mathcal{S}$ bid above $b$ with strictly positive probability.  By construction, the size of $\mathcal{S}$ is $\numunits - 1$. The agents in $\lowagents$ bid above $b$ almost surely. Any agent $l \in \topagents'$ has $\supportsup{l} > b$ and thus bids above $b$ with strictly positive probability. It follows that agents in $\mathcal{S}$ bid above $b$ with strictly positive probability.

    Now, any agent $l$ that does not belong to $\mathcal{S}$ has $\bidCDF{l}(b) > 0$. Thus, agents that do not belong to the set $\mathcal{S}$ bid at most $b$ with strictly positive probability. Since agents $i$ and $j$ have probability mass at $b$, it is the bid of the $\numunits^{\text{th}}$ highest agent with strictly positive probability. Moreover, exactly $\numunits - 1$ agents bid strictly above $b$ with positive probability. The statement of the result follows.  
\end{proof}

   \begin{lemma}\label{lma:flat-implies-no-point-mass}
        Suppose the bid distribution of the $\numunits^{\text{th}}$ highest agent $\khighbidCDF$ is flat in the interval $(\alpha, \beta) 
        \subset \left(\supportinforderstat{\numunits},\supportsuporderstat{\numunits}\right)$. Then, no agent has a point mass in their bid distribution at any bid in $(\alpha, \beta)$. 
    \end{lemma}
    
    \begin{proof}
        Suppose agent $i$ has a point mass at some bid $b$ in the interval $(\alpha, \beta)$. We will follow the approach in the proof of \Cref{lma:at-most-one-agent-with-point-mass}. Our goal is to show that with strictly positive probability there are exactly $\numunits - 1$ agents who bid strictly above $b$ and the rest bid strictly below $b$. This will imply that the bid of the $\numunits^{\text{th}}$ highest agent is $b$ with strictly positive probability which is a contradiction to the assumption that $\khighbidCDF$ is flat in $(\alpha, \beta)$.
        
        Define $\lowagents$ as
        \begin{align*}
            \lowagents := \{ l : \bidCDF{l}(b) = 0 \} \text{.}
        \end{align*}
        The agents in $\lowagents$ bid above $b$ almost surely. It must be the case that $|\lowagents| \leq \numunits - 1$ otherwise we would have $\supportinforderstat{\numunits} \geq b$ which is a contradiction to $\supportinforderstat{\numunits} < \alpha$. Next, define $\topagents$ as
        \begin{align*}
            \topagents := \{ l : \supportsup{l} = \supportsuporderstat{\numunits} \} \text{.}
        \end{align*}
        An application of \Cref{lma:support-upper-bound-equivalence} tells us that $|\topagents| \geq \numunits + 1$. Consider a set $\topagents'$ such that
        \begin{align*}
            \topagents' \subseteq \topagents \setminus \left(\{i\} \cup \lowagents\right) \quad \text{ and } \quad  |\topagents'| = \numunits - 1 - |\lowagents| \text{.}
        \end{align*}
        Such a set exists because $|\topagents| \geq \numunits + 1$ and $|\lowagents| \leq \numunits - 1$. We now consider the set $\mathcal{S} = \lowagents \cup \topagents'$. We will show that agents in $\mathcal{S}$ bid above $b$ with strictly positive probability. By construction, the size of $\mathcal{S}$ is $\numunits - 1$. The agents in $\lowagents$ bid above $b$ almost surely. Consider any agent $l$ in $\topagents'$. This agent would bid strictly above $b$ with positive probability because $\supportsup{l} > b$. It follows that agents in $\mathcal{S}$ bid above $b$ with strictly positive probability.
        
        Now, any agent $l$ that does not belong to $\mathcal{S}$ has $\bidCDF{l}(b) > 0$. Thus, agents that do not belong to the set $\mathcal{S}$ bid at most $b$ with strictly positive probability. Since agent $i$ has a probability mass at $b$, it is the bid of the $\numunits^{\text{th}}$ highest agent with strictly positive probability. This contradicts the assumption that $\khighbidCDF$ is flat in $(\alpha, \beta)$. The statement of the result follows.
    \end{proof}

We begin by providing a proof for \Cref{t:bnecontmono} Part (a) which states the following: at most $(\numunits-1)$ agents have an atom at $\supportsuporderstat{\numunits}$.

\begin{proof}[Proof of \Cref{t:bnecontmono} Part (a)]
    Finally, we will show that at most $(\numunits-1)$ agents have a point mass at $\supportsuporderstat{\numunits}$. We will first show at most $\numunits$ agents can have a point mass at $\supportsuporderstat{\numunits}$ in their equilibrium bid distributions. Suppose that at least $(\numunits + 1)$ agents have a point mass at $\supportsuporderstat{\numunits}$. We will show there must be an agent who has a discontinuity in their allocation at $\supportsuporderstat{\numunits}$. Formally, there exists an agent $i$ and a corresponding $\delta_{i} > 0$ satisfying
    \begin{align*}
        \interimalloc{i}(b) \geq \interimalloc{i}(\supportsuporderstat{\numunits}) + \delta_{i} \quad \text{for all } b > \supportsuporderstat{\numunits} \text{.}
    \end{align*}
    The allocation probability of agent $i$ for bid $b$ is
    \begin{align*}
        \interimalloc{i}(b) = \prob[\criticalbidCDF{i} < b] + \prob[\criticalbidCDF{i} = b] \prob[\text{agent } i \text{ wins in a tie-break} \mid \criticalbidCDF{i} = b] \text{.}
    \end{align*}
    Because at least $(\numunits + 1)$ agents have a point mass at $\supportsuporderstat{\numunits}$, and there are only $\numunits$ items available, there must exist an agent $i$ for which
    \begin{align*}
        \prob[\text{agent } i \text{ wins in a tie-break} \mid \criticalbidCDF{i} = \supportsuporderstat{\numunits}] < 1 \text{.}
    \end{align*}
    The allocation rule is discontinuous at $\supportsuporderstat{\numunits}$ for this agent. A neighborhood deviation argument (as stated in \Cref{lma:neighborhood-deviation-argument}) implies that $\supportsuporderstat{\numunits}$ cannot be a best response bid of agent $i$. This is a contradiction to $\supportsuporderstat{\numunits}$ being in the support of their equilibrium bid distribution. Thus, at most $\numunits$ agents can have a point mass at $\supportsuporderstat{\numunits}$ in their equilibrium bid distribution.

    Suppose exactly $\numunits$ agents have a point mass at $\supportsuporderstat{\numunits}$. By \Cref{lma:support-upper-bound-equivalence}, there exists an agent $i^{\dag}$ such that the equilibrium bid distribution $\bidCDF{i^{\dag}}$ does not have a point mass at $\supportsuporderstat{\numunits}$ but where $\supportsuporderstat{\numunits}$ belongs to its support. By the uniform tie-breaking rule, it must be that
    \begin{align}\label{eq:disconthighsupport}
        \prob[\text{agent } i^{\dag} \text{ wins in a tie-break} \mid \criticalbidCDF{i^{\dag}} = \supportsuporderstat{\numunits}] < 1 \text{.}
    \end{align} This implies that the allocation rule of $i^{\dag}$ is discontinuous at $\supportsuporderstat{\numunits}$. We will show that this is a contradiction to the fact that there are bids in the support of $\bidCDF{i^{\dag}}$ that are arbitrarily close to $\supportsuporderstat{\numunits}$. 

    Since there are bids in the support of $\bidCDF{i^{\dag}}$ that are arbitrarily close to $\supportsuporderstat{\numunits}$, it must be that for every $\delta>0$, the bid distribution $\bidCDF{i^{\dag}}$ has positive probability in $(\supportsuporderstat{\numunits}-\delta,\supportsuporderstat{\numunits}]$. Fix a $l \in \mathbb{N}$ and consider the interval $\left(\supportsuporderstat{\numunits}-\frac{1}{l},\supportsuporderstat{\numunits}\right)$. There must exist a bid-value pair $\left(b_{l},v_{l}\right)$ satisfying the following properties: $b_{l} \in \left(\supportsuporderstat{\numunits}-\frac{1}{l},\supportsuporderstat{\numunits}\right)$; bid $b_{l}$ is in the support of the equilibrium bid distribution of $v_{l}$; and $v_{l}$ is in the support of the value distribution of agent $i^{\dag}$. Furthermore, by \Cref{lma:strategy-and-utility-monotonicity}, the sequence of values $(v_{l})_{l=1}^{\infty}$ and bids $b_{l}$ can be chosen to be weakly increasing. Note that the sequence $(b_{l})_{l=1}^{\infty}$ converges to $\supportsuporderstat{\numunits}$. The utility of agent $i^{\dag}$ along this sequence is
        \begin{align*}
            (v_{l} - b_{l}) \interimalloc{i^{\dag}}(b_{l}).
        \end{align*} We will establish that the limit infimum of the sequence of values must be strictly greater than $\supportsuporderstat{\numunits}$, that is, $
            \liminf_{l \to \infty} v_{l} > \supportsuporderstat{\numunits}.$ If this was not the case, then large enough values in the sequence obtain equilibrium utility that is arbitrarily close to $0$. This is a contradiction because large enough values in the sequence can guarantee a strictly positive utility by bidding at $\left(\supportsuporderstat{\numunits}+\supportinforderstat{\numunits}\right)/2$. Thus, it must be that $\liminf_{l \to \infty} v_{l} > \supportsuporderstat{\numunits}$. 
         
         Because of the discontinuity in the allocation rule at $\supportsuporderstat{\numunits}$, it follows by \Cref{lma:neighborhood-deviation-argument} that for values large enough in the sequence above, there exists a bid $b'$ strictly greater than $\supportsuporderstat{\numunits}$ which gives the agent a strictly better utility. This is a contradiction to bids being arbitrarily close to $\supportsuporderstat{\numunits}$ belonging to the support of $\bidCDF{i^{\dag}}$. Thus, at most $\numunits - 1$ agents have a point mass at $\supportsuporderstat{\numunits}$ in their equilibrium bid distribution.
\end{proof}

We now provide a proof for \Cref{t:bnecontmono} Part (b) which states the following: the distribution function of the $\numunits$-th order statistic of equilibrium bids is strictly increasing in the interval $\left[\supportinforderstat{\numunits},\supportsuporderstat{\numunits}\right]$.

\begin{proof}[Proof of \Cref{t:bnecontmono} Part (b)]
     Suppose $\khighbidCDF$ has no bid in its support on an open interval $(\alpha, \beta)$ with $\beta < \supportsuporderstat{\numunits}$. It is without loss for the interval to be maximal in the sense that for any $\delta>0$, $\khighbidCDF$ taking realizations in $(\alpha-\delta,\alpha]$ and $[\beta,\beta+\delta)$ has positive probability. Thus, we have $\khighbidCDF(\beta) \geq \khighbidCDF(\alpha) > 0$. Consider the following cases:
    \begin{enumerate}[(i)]
        \item \textit{Case} 1: Suppose $\khighbidCDF(\beta) > \khighbidCDF(\alpha) > 0$. It must be the case that $\khighbidCDF$ has a point mass at $\beta$. And thus, there is some agent $i_{0}$ that has probability mass at $\beta$ in their equilibrium bid distribution. By \Cref{lma:at-most-one-agent-with-point-mass}, no other agent has a point mass at $\beta$ in their bid distribution. The competing bid distribution $\criticalbidCDF{i_{0}}$ is thus continuous at $\beta$. 
            
        If it is the case that agent $i_{0}$'s competing bid distribution $\criticalbidCDF{i_{0}}$ has no point in its support in the open interval $(\alpha, \beta)$, then agent $i_{0}$'s interim allocation is constant in $(\alpha,\beta]$. By \Cref{lma:neighborhood-deviation-argument}, it is suboptimal for $i_{0}$ to have probability mass at $\beta$ which is a contradiction to $\beta \in \supp(\bidCDF{i_{0}})$. 
        
        Thus, it must be that $\criticalbidCDF{i_{0}}$ has some point in its support in $(\alpha, \beta)$. But, because the distribution of the $\numunits^{\text{th}}$ highest bid was flat in $(\alpha, \beta)$ it must be the case that
        \begin{align}\label{eq:infsuplb}
            \supportinf{i_{0}} \geq \sup \left\{ (\alpha, \beta) \cap \supp\left(\criticalbidCDF{i_{0}}\right) \right\} \text{.}
        \end{align}
        This is because, if it were the case that $\supportinf{i_{0}} < \sup \left \{ (\alpha, \beta) \cap \supp\left(\criticalbidCDF{i_{0}}\right)\right \}$, then the bid of the $\numunits^{\text{th}}$ highest agent apart from $i_{0}$ becomes the bid of the $\numunits^{\text{th}}$ highest agent overall with strictly positive probability. Moreover, such bids would lie in $(\alpha, \beta)$ which is a contradiction to the assumption that the distribution of the $\numunits^{\text{th}}$ highest bid overall is flat in $(\alpha, \beta)$. 
        
        Recall, we are analyzing a scenario in which some point in $(\alpha, \beta)$ lies in the support of $\criticalbidCDF{i_{0}}$. There must be an agent $i_{1}$ (clearly not the same as $i_{0}$ because we are analyzing the competing bid distribution) and a corresponding bid $\beta_{1} \in (\alpha, \beta)$ such that $i_{1}$ has $\beta_{1}$ in its support and $\supportinf{i_{0}} \geq \beta_{1}$ (an application of \Cref{lma:flat-implies-no-point-mass} tells us that this agent will not have a point mass at $\beta_{1}$). Moreover, since no agent has a point mass at $\beta_{1}$, as shown in \Cref{lma:flat-implies-no-point-mass}, agent $i_{1}$'s competing bid distribution is continuous at $\beta_{1}$. 
        
        We will inductively apply the above argument to show there are $\numunits$ agents whose $\supportinf{i} > \alpha$. This is a contradiction to $(\alpha, \beta) \subset [\supportinforderstat{\numunits},\supportsuporderstat{\numunits}]$.
        
        \textit{Inductive hypotheses}: Suppose we have a set of agents $\mathcal{I}_{l} = \{i_{0},\dots,i_{l-1}\}$, where $|\mathcal{I}_{l}| < \numunits$. In addition, suppose this set satisfies the conditions stated below:
        \begin{enumerate}[(a)]
            \item There is a corresponding set of bids $\bidsinductive{l} :=  \{\beta_{j}\}_{j=0}^{l-1}$ where $\beta_{0} = \beta$ and $\beta_{j} \in (\alpha, \beta)$ for any $j \in [l-1]$. Moreover, the bids are monotonically decreasing, that is,
            \begin{align*}
                \beta_{j} > \beta_{j+1} \quad \text{ for all } j \in \{0,\dots,l-2\} \text{.}
            \end{align*}
            \item For every agent $i_j \in \mathcal{I}_{l}$, the competing bid distribution $\criticalbidCDF{i_{j}}$ has some bid in $\left(\alpha,\beta_{j}\right)$ in its support.
            \item For every agent $i_{j} \in \mathcal{I}_{l}$, the infimum of the support of the equilibrium bid distribution satisfies
            \begin{align*}
                \supportinf{i_{j}} \geq \sup \left \{ \left(\alpha,\beta_{j}\right) \cap \supp\left(\criticalbidCDF{i_{j}}\right) \right \} \text{.}
            \end{align*} 
            \item For any agent $i_{j}$, where $j \in \{0,\dots,l-2\}$,
            \begin{align*}
                \supportinf{i_{j}} \geq \beta_{j+1} \text{.}
            \end{align*}
            \item There is an agent $i_{l} \not \in \mathcal{I}_{l}$ and a bid $\beta_{l}$ in its support such that
            \begin{align*}
               \alpha < \beta_{l} < \beta_{l-1} \text{ and } \supportinf{i_{l-1}} \geq \beta_{l} \text{.}
            \end{align*}
        \end{enumerate}
    
        \textit{Base case}: Recall, we have shown that the conditions above are satisfied for $l=1$. The set in consideration is $\mathcal{I}_{l} = \{i_{0}\}$. Part (a) is satisfied vacuously with the set of bids $\{\beta\}$. Part (b) is satisfied because we are analyzing the scenario where $\criticalbidCDF{i_{0}}$ has probability density or probability mass in $(\alpha, \beta)$. Part (c) is satisfied by (\ref{eq:infsuplb}). Part (d) holds vacuously. Part (e) is satisfied by agent $i_{1}$: there is a bid $\beta_{1} \in \supp(\bidCDF{i_{1}})$ such that $\beta_{1} \in (\alpha, \beta)$ and $\supportinf{i_{0}} \geq \beta_{1}$.
        
        \textit{Inductive argument}: Suppose the conditions hold for some $l$ where $1 \leq l < \numunits$. We will analyze the competing bid distribution of agent $i_{l}$. Consider the case $\criticalbidCDF{i_{l}}$ is flat in $(\alpha,\beta_{l})$. By \Cref{lma:flat-implies-no-point-mass}, no agent has a point mass at $\beta_{l}$ and thus $\criticalbidCDF{i_{l}}$ is continuous at $\beta_{l}$. An immediate consequence of this is $i_{l}$ has a constant allocation rule in the interval $(\alpha,\beta_{l}]$. By an argument similar to that in \Cref{lma:neighborhood-deviation-argument}, however, this is a contradiction to the fact that $\beta_{l} \in \supp(\bidCDF{i_{l}})$. To see this, if $\beta_{l} \in \supp(\bidCDF{i_{l}})$, then by continuity of the competing bid distribution at $\beta_{l}$, a sufficiently small interval of bids around $\beta_{l}$ is strictly dominated by bidding $(\alpha + \beta_{l})/2$.  
        
        Thus, it must be the case that some point in $(\alpha,\beta_{l})$ is in the support of $\criticalbidCDF{i_{l}}$ (at the expense of being repetitive, by \Cref{lma:flat-implies-no-point-mass}, $\criticalbidCDF{i_{l}}$ cannot have a point mass in this interval). Let us define the set
        \begin{align*}
            \mathcal{I}_{l+1} := \mathcal{I}_{l} \uplus \{i_{l}\} \text{.}
        \end{align*} By Part (e), $i_{l} \not \in \mathcal{I}_{l}$, and thus you are adding a unique element to $\mathcal{I}_{l}$. Let us also consider the corresponding set of bids
        \begin{align*}
            \mathbb{B}_{l+1} := \{\beta_{j}\}_{j=0}^{l-1} \cup \{\beta_{l}\} \text{.}
        \end{align*}
        By Part (e), $\beta_{l} < \beta_{l-1}$ and thus, applying Part (a),
        \begin{align*}
            \beta_{j} > \beta_{j+1} \text{ for all } j \in \{0,\dots,l-1\} \text{.}
        \end{align*}
        Thus $\mathcal{I}_{l+1}$ and $\bidsinductive{l+1}$ satisfies Part (a) for $l+1$ agents. Part (b) is satisfied for $(l+1)$ agents as well because we are analyzing the case $\criticalbidCDF{i_{l}}$ has some bid in $(\alpha,\beta_{l})$ in its support. 
        
        We will now show that Part (c), (d) and (e) are also satisfied for $(l+1)$ agents. There is a bid in $(\alpha,\beta_{l})$ that is in the support of $\criticalbidCDF{i_{l}}$. However, the distribution of the bid of the $\numunits^{\text{th}}$ highest agent is flat in $(\alpha, \beta)$. Thus, it must be the case that
        \begin{align}\label{eq:infsuplbgen}
            \supportinf{i_{l}} \geq \sup \left \{\left(\alpha,\beta_{l}\right) \cap \supp\left(\criticalbidCDF{i_{l}}\right) \right \} \text{.}
        \end{align}
        
        This is because, if it were the case that $\supportinf{i_{l}} < \sup \left\{ (\alpha,\beta_{l}) \cap \supp\left(\criticalbidCDF{i_{l}}\right) \right\}$, then the bid of the $\numunits^{\text{th}}$ highest agent apart from $i_{l}$ becomes the bid of the $\numunits^{\text{th}}$ highest agent overall with strictly positive probability. Moreover, such bids would lie in $(\alpha, \beta)$ which is a contradiction to the assumption that the distribution of the $\numunits^{\text{th}}$ highest bid overall is flat in $(\alpha, \beta)$. Since, by the inductive hypotheses, this condition also holds for agents $j \in \{0,\dots,l-1\}$, Part (c) is satisfied for $(l+1)$ agents. Next, Part (d) is satisfied for $(l+1)$ agents because in Part (e) of the induction hypotheses, we had that $\supportinf{i_{l-1}} \geq \beta_{l}$.
        
        Now, there is a point in $(\alpha,\beta_{l})$ that is in the support of $\criticalbidCDF{i_{l}}$. Thus, there must exist an agent $i_{l+1}$, other than $i_{l}$, and a corresponding bid $\beta_{l+1} \in (\alpha,\beta_{l})$ such that $\beta_{l+1} \in \supp(\bidCDF{i_{l+1}})$. In addition to this, an implication of (\ref{eq:infsuplbgen}) is that $\supportinf{i_{l}} \geq \beta_{l+1}$. Finally, the induction hypotheses tells us that for any agent in $\mathcal{I}_{l}$, the infimum of the support of their bid distribution
        \begin{align*}
            \supportinf{i_{j}} \geq \beta_{j+1} \geq \beta_{l} \text{.}
        \end{align*} Thus it cannot be that agent $i_{l+1}$, who has probability density in $(\alpha,\beta_{l})$, lies in the set $\mathcal{I}_{l}$. This satisfies Part (e). 
        
        To summarize the conclusions, at each index $l \in [\numunits - 1]$, either we have a contradiction in the case $\criticalbidCDF{i_{l}}$ is flat in $(\alpha,\beta_{l})$ or the conditions of the inductive hypotheses are satisfied for index $(l+1)$. If the latter is satisfied at each index, we have a set of $\numunits$ agents $\mathcal{I}_{\numunits}$ such that
        \begin{align*}
            \supportinf{i_{j}} \geq \beta_{j+1} > \alpha \quad \text{ for all } j \in \{0,\dots,\numunits - 1\} \text{.}
        \end{align*}
        But, this is a contradiction to the infimum of bid distribution of the $\numunits^{\text{th}}$ highest agent $\supportinforderstat{\numunits}$ being strictly smaller than $\alpha$.
            
        \item \textit{Case} 2: Suppose $\khighbidCDF(\beta) = \khighbidCDF(\alpha) > 0$. First, we will show that no agent can have a point mass at $\beta$. To see this, consider the argument in the proof of \Cref{lma:flat-implies-no-point-mass}. Suppose there exists an agent, labeled $i$, who has a point mass at $\beta$. Define $\lowagents$ as
        \begin{align*}
            \lowagents := \{ l : \bidCDF{l}(\beta) = 0 \} \text{.}
        \end{align*}
        The agents in $\lowagents$ bid above $\beta$ almost surely. It must be the case that $|\lowagents| \leq \numunits - 1$ otherwise we would have $\supportinforderstat{\numunits} \geq \beta$ which is a contradiction to $\supportinforderstat{\numunits} < \alpha$. Next, define $\topagents$ as
        \begin{align*}
            \topagents := \{ l : \supportsup{l} = \supportsuporderstat{\numunits} \} \text{.}
        \end{align*}
        An application of \Cref{lma:support-upper-bound-equivalence} tells us that $|\topagents| \geq \numunits + 1$. Consider a set $\topagents'$ such that
        \begin{align*}
            \topagents' \subseteq \topagents \setminus \left(\{i\} \cup \lowagents\right) \quad \text{ and } \quad |\topagents'| = \numunits - 1 - |\lowagents| \text{.}
        \end{align*}
        Such a set exists because $|\topagents| \geq \numunits + 1$ and $|\lowagents| \leq \numunits - 1$. We now consider the set $\mathcal{S} = \lowagents \uplus \topagents'$. We will show that agents in $\mathcal{S}$ bid above $\beta$ with strictly positive probability.  By construction, the size of $\mathcal{S}$ is $\numunits - 1$. The agents in $\lowagents$ bid above $\beta$ almost surely. Any agent $l \in \topagents'$ has $\supportsup{l} > \beta$. Thus, he bids strictly above $\beta$ with positive probability. It follows that agents in $\mathcal{S}$ bid above $\beta$ with strictly positive probability.
        
        Now, any agent $l$ that does not belong to $\mathcal{S}$ has $\bidCDF{l}(\beta) > 0$. Thus, agents that do not belong to the set $\mathcal{S}$ bid strictly smaller than $\beta$ with positive probability. It follows that $\beta$ is the bid of the $\numunits^{\text{th}}$ highest agent with positive probability because agent $i$ bids $\beta$ with positive probability. This is a contradiction to the assumption that $\khighbidCDF(\alpha) = \khighbidCDF(\beta)$. Thus, no agent can have a point mass at $\beta$. 
        
        We will show that the induction argument presented in \textit{Case 1} extends to this case as well.
        
        Recall, the interval $(\alpha, \beta)$ is maximal in the following sense: for any $\delta>0$, $\khighbidCDF$ has a realization in $(\alpha-\delta,\alpha]$ and $[\beta,\beta+\delta)$ with positive probability. Thus, there must exist an agent $i_{0}$ who has $\beta$ in the support of his equilibrium bid distribution $\bidCDF{i_{0}}$.
        
        If it is the case that agent $i_{0}$'s competing bid distribution $\criticalbidCDF{i_{0}}$ has no bid in the open interval $(\alpha, \beta)$ in its support, then agent $i_{0}$'s interim allocation is constant in $(\alpha,\beta]$. This is because no agent has a point mass at $\beta$ and thus $\criticalbidCDF{i_{0}}$ is continuous at that point. An argument similar to that in the previous case shows that it is suboptimal for $i_{0}$ to have $\beta$ in its support: a sufficiently small interval of bids around $\beta$ is strictly dominated by bidding $(\alpha+\beta)/2$.
        
        Thus, it must be that some bid in $(\alpha, \beta)$ is in the support of $\criticalbidCDF{i_{0}}$. However, because the distribution of the $\numunits^{\text{th}}$ highest bid was flat in $(\alpha, \beta)$ it must be the case that
        \begin{align*}
            \supportinf{i_{0}} \geq \sup \left\{ (\alpha, \beta) \cap \supp\left(\criticalbidCDF{i_{0}}\right) \right\} \text{.}
        \end{align*}
        
        Moreover, there must be an agent $i_{1}$ (clearly not the same as $i_{0}$ because we are analyzing the competing bid distribution) and a corresponding bid $\beta_{1} \in (\alpha, \beta)$ such that $i_{1}$ has $\beta_{1}$ in its support and $\supportinf{i_{0}} \geq \beta_{1}$ (an application of \Cref{lma:flat-implies-no-point-mass} tells us that this agent will not have a probability mass at $\beta_{1}$). Moreover, since no agent has a point mass at $\beta_{1}$, as shown in \Cref{lma:flat-implies-no-point-mass}, agent $i_{1}$'s competing bid distribution is continuous at $\beta_{1}$.
        
        The arguments above show that the base case for the induction argument is satisfied. The conclusion of the argument is the same and we thus have a contradiction.
    \end{enumerate}
   
    Finally, suppose $\khighbidCDF$ has no bid in $(\alpha, \supportsuporderstat{\numunits})$ in its interval. It must be that $\khighbidCDF(\supportsuporderstat{\numunits}) > \khighbidCDF(\alpha) > 0$. If not, this is a contradiction to the definition of $\supportsuporderstat{\numunits}$. Thus, there must exist at least $\numunits$ agents with an atom at $\supportsuporderstat{\numunits}$. But this is a contradiction to Part (a) of \Cref{t:bnecontmono}.
    
    All of these cases imply that there does not exist an open interval in $[\supportinforderstat{\numunits}, \supportsuporderstat{\numunits}]$ such that $\khighbidCDF$ has no probability density or mass in it. Thus, the bid distribution $\khighbidCDF$ is strictly increasing on $[\supportinforderstat{\numunits}, \supportsuporderstat{\numunits}]$.
\end{proof}

The above result shows that the $\numunits$-th order statistic $\khighbidCDF$ has a strictly increasing distribution function on the interval $[\supportinforderstat{\numunits}, \supportsuporderstat{\numunits}]$. Below, we also establish that the competing bid distribution for any agent $i$ --- $\numunits$-th highest bid among agents other than $i$ --- has a strictly increasing distribution function in $[\supportinforderstat{\numunits}, \supportsuporderstat{\numunits}]$. We will use this result to prove Part (c) of \Cref{t:bnecontmono}.  

\begin{lemma}\label{thm:critical-bid-cdf-increasing}
    Consider any agent $i \in \agentset$. The competing bid distribution $\criticalbidCDF{i}$ has a strictly increasing CDF on the interval $[\supportinforderstat{\numunits}, \supportsuporderstat{\numunits}]$.
\end{lemma}

\begin{proof}
    Consider any agent $i \in \agentset$. Suppose the competing bid distribution $\criticalbidCDF{i}$ is flat in an open interval $(\alpha, \beta) \subset [\supportinforderstat{\numunits}, \supportsuporderstat{\numunits}]$. It is without loss for this interval to be maximal: for any $\delta > 0$, distribution $\criticalbidCDF{i}$ has probability density or mass in $(\alpha - \delta, \alpha]$ and $[\beta, \beta + \delta)$. By \Cref{t:bnecontmono} Part (b), however, the bid of the $\numunits^{\text{th}}$ highest agent is strictly increasing in $(\alpha, \beta)$. This implies there is a bid $b$ in the interval $(\alpha, \beta)$ such that $b$ is in the support of agent $i$'s equilibrium bid distribution.
    
    Since $b \in (\alpha, \beta)$, the competing bid distribution is flat in $(\alpha, b]$. This means that agent $i$ has a constant interim allocation $\interimalloc{i}(\cdot)$ in $(\alpha, b]$. By an argument similar to that in the proof of \Cref{lma:neighborhood-deviation-argument}, this is a contradiction to $b \in \supp(\bidCDF{i})$: a sufficiently small interval close to $b$ is strictly dominated by bidding $(\alpha + b)/2$. Thus, the competing bid distribution $\criticalbidCDF{i}$ is strictly increasing in $[\supportinforderstat{\numunits},\supportsuporderstat{\numunits}]$.
\end{proof}

Finally, we provide a proof for  \Cref{t:bnecontmono} Part (c)
which states the following: no agent's equilibrium distribution has an atom in $\left(\supportinforderstat{\numunits},\supportsuporderstat{\numunits}\right)$. 

\begin{proof}[Proof of \Cref{t:bnecontmono} Part (c)]
    
    Suppose agent $i$ has a point mass at $b^*$ in the support of his equilibrium bid distribution. By \Cref{lma:at-most-one-agent-with-point-mass}, only agent $i$ has a point mass at this bid. But since, by \Cref{thm:critical-bid-cdf-increasing}, his competing bid distribution $\criticalbidCDF{i}$ is strictly increasing at $b^*$, there is a different agent $i'$ such that $b^*$ is in the support of the equilibrium bid distribution $\bidCDF{i'}$. An argument similar to the one used in the proof of \Cref{lma:at-most-one-agent-with-point-mass} will imply that the interim allocation $\interimalloc{i'}(\cdot)$ is discontinuous at $b^*$. This means there exists a $\delta' > 0$ such that \begin{align}\label{eq:discontalloc}
        \interimalloc{i'}(b') \geq \interimalloc{i'}(b^*) + \delta' \quad \text{ for all } b' > b^* \text{.}
    \end{align}
    We will show that this is a contradiction to the following statement: there are bids smaller than $b^*$ and arbitrarily close to it that lie in the support of the equilibrium distribution of some agent.
    
    \begin{enumerate}[(i)]
        \item Suppose for every $\delta>0$, the bid distribution $\bidCDF{i'}$ has positive probability in $(b^*-\delta,b^*]$. Fix a $l \in \mathbb{N}$ and consider the interval $\left(b^*-\frac{1}{l},b^*\right)$. There must exist a bid-value pair $\left(b_{l},v_{l}\right)$ satisfying the following properties: $b_{l} \in \left(b^*-\frac{1}{l},b^*\right)$; bid $b_{l}$ is in the support of the equilibrium bid distribution of $v_{l}$; and $v_{l}$ is in the support of the value distribution of agent $i'$. Furthermore, by \Cref{lma:strategy-and-utility-monotonicity}, the sequence of values $(v_{l})_{l=1}^{\infty}$ and bids $b_{l}$ can be chosen to be weakly increasing. Note that the sequence $(b_{l})_{l=1}^{\infty}$ converges to $b^*$. The utility of agent $i'$ along this sequence is
        \begin{align*}
            (v_{l} - b_{l}) \interimalloc{i'}(b_{l}).
        \end{align*} We will establish that the limit infimum of the sequence of values must be strictly greater than $b^*$, that is, $
            \liminf_{l \to \infty} v_{l} > b^*.$ If this was not the case, then large enough values in the sequence obtain equilibrium utility that is arbitrarily close to $0$. This is a contradiction because large enough values in the sequence can guarantee a strictly positive utility by bidding at $\left(b^*+\supportinforderstat{\numunits}\right)/2$. Thus, it must be that $\liminf_{l \to \infty} v_{l} > b^*$. 
         
         Because of the discontinuity in the allocation rule at $b^*$, it follows by \Cref{lma:neighborhood-deviation-argument} that for values large enough in the sequence above, there exists a bid $b'$ strictly greater than $b^*$ which gives the agent a strictly better utility. This is a contradiction to bids being arbitrarily close to $b^*$ belonging to the support of $\bidCDF{i'}$.
    
        \item Suppose there exists a $\delta > 0$ such that $\bidCDF{i'}$ has $0$ probability in $(b^*-\delta,b^*]$. But because $b^* \in \supp(\bidCDF{i'})$ it must be that for any $\delta>0$, bid distribution $\bidCDF{i'}$ has positive probability in $[b^*,b^*+\delta)$. We will show there exists an agent $i^{\dag}$, not the same as $i$ and $i'$, such that for all $\delta_{i^{\dag}} > 0$, he has positive probability in $(b^*-\delta_{i^{\dag}},b^*]$. Suppose that apart from $i$ and $i'$, no agent satisfies the above condition. By assumption, there exists a $\delta>0$ such that $\bidCDF{i'}$ has $0$ probability in $(b^*-\delta,b^*]$. Thus, the competing bid distribution of agent $i$ is flat in $(b^*-\delta_{i},b^*]$ for some $\delta_{i}>0$. This is a contradiction to the result that $\criticalbidCDF{i}$ is strictly increasing in the interval $[\supportinforderstat{\numunits},\supportsuporderstat{\numunits}]$ as established in \Cref{thm:critical-bid-cdf-increasing}. 
    
        Thus, there exists an agent $i^{\dag}$, not the same as $i$ and $i'$, such that for all $\delta_{i^{\dag}} > 0$, he has positive probability in $\left(b^*-\delta_{i^{\dag}},b^*\right]$. A consequence of this is that $b^* \in \supp(\bidCDF{i^{\dag}})$. Moreover, this agent will have a discontinuous interim allocation rule at $b^*$. Applying the argument in Part (i), we get a contradiction to bids arbitrarily close to $b^*$ belonging to the support of $\bidCDF{i^{\dag}}$.  
    \end{enumerate}
    Since we get a contradiction in both cases, it cannot be the case that the equilibrium bid distribution function $\equilibrium_{i}^{(\numunits)}$ is discontinuous in $(\supportinforderstat{\numunits},\supportsuporderstat{\numunits})$.
\end{proof}

\section{Conclusion}

This paper performs a robust analysis of multi-unit first-price auctions with independent private values in a large supply regime. We employ a smoothness framework to perform our analysis. The main barrier is a worst-case distribution that establishes a PoA lower bound of $1-\frac{1}{e}$. Our analysis shows that such distributions do not arise in equilibrium and we use this characterization to establish novel PoA bounds. 

A key result governing our analysis concerns the structure of (approximate) equilibrium bid distributions. We show that for a welfare analysis, it is without loss to analyze bid distributions without atoms. Furthermore, we derive continuity and monotonicity results for bid distributions in a Bayes-Nash equilibrium.

An important feature of the smoothness paradigm is its applicability to complex auction environments. The concepts of individual efficiency and competitive efficiency establish non-trivial PoA bounds in settings where it is intractable to analytically characterize equilibrium. A key open question is to investigate how our improved individual efficiency bounds apply to such environments, some examples of which include simultaneous composition of auctions or those involving combinatorial feasibility constraints. We leave these questions for future work.  

{\footnotesize \linespread{1.0}
\bibliographystyle{ecta}
\bibliography{references.bib}
}

\appendix 

\newpage

\section{Omitted Proofs}
\label{a:detailedproofs}

\subsection{Chernoff Bounds} \label{ss:chernoffbounds}This section presents a Chernoff bound which we utilize extensively to establish bounds on the allocation probability. 

\begin{fact}\label{fact:chernoffbound}
    Let $X=\sum_{i=1}^{n}X_{i}$, where $X_{i}=1$ with probability $p_{i}$ and $X_{i}=0$ with probability $1-p_{i}$, and all $X_{i}$ are independent. Let $\mu = \expect(X) = \sum_{i=1}^{n}p_{i}$. Then \begin{enumerate}[(i)]
        \item \textbf{Upper Tail}: $\text{Pr}\left(X \geq (1+\lambda)\mu\right) \leq e^{-\frac{\lambda^2}{2+\lambda} \mu}$ for all $\lambda>0$;
        \item \textbf{Lower Tail}: $\text{Pr}\left(X \leq (1-\lambda)\mu\right) \leq e^{-\frac{\mu \lambda^2}{2}}$ for all $\lambda \in (0,1)$.
    \end{enumerate}
\end{fact}

\subsection{Allocation Rule and $\numunits$-th Order Statistic} This section presents a useful fact that relates the allocation probability of an agent with the distribution function of the $\numunits$-th order statistic.

\begin{fact}
    \label{fact:lballoc}
    Consider any profile of bid distributions $\equilibrium$. For any bid $b \in \mathbb{R}_{\geq 0}$ and any agent $i \in \agentset$, the allocation probability and $\numunits$-th order statistic satisfies \begin{align*}
        x_{i}(b+\Delta) \geq \khighbidCDF\left(b\right) \quad \text{ for all } \Delta > 0.
    \end{align*}

Furthermore, if the distribution $\equilibrium_{i}$ for every agent $i$ has no atoms, then $$x_{i}(b) \geq \khighbidCDF(b).$$
\end{fact}

\begin{proof}
    Consider any $b \geq 0$. Suppose the $\numunits^{\text{th}}$ highest bid is at most $b$. Then, the $\numunits^{\text{th}}$ highest bid among agents other than $i$ is strictly smaller than $b+\Delta$ for any $\Delta >0$. Thus, we have $$x_{i}(b+\Delta) \geq \khighbidCDF(b) \quad \text{ for all } \Delta > 0.$$ 

Next, if no distribution $\equilibrium_{i}$ has atoms, ties occur with zero probability. Thus, if the $\numunits$-th highest bid overall is strictly smaller than $b$, then the $\numunits$-th highest bid among agents other than $i$ is also strictly smaller than $b$. This implies $x_{i}(b) \geq \khighbidCDF(b)$.
\end{proof}

\section{Additional Properties of Equilibrium Bid Distributions}\label{ss:eqlbbidproperties}

This section establishes some standard properties of equilibrium bid distributions in the multi-unit setting. We introduce key definitions used throughout this section. We define $\supportinf{i}$ as the infimum and $\supportsup{i}$ as the supremum of the support of agent $i$'s equilibrium bid distribution: 
\begin{align*}
    \supportinf{i} := \inf(\supp (\bidCDF{i})) \text{ and } \supportsup{i} := \sup(\supp(\bidCDF{i})) \text{.}
\end{align*}
Additionally, we define $\supportinfset$ as the collection of $\supportinf{i}$ values, sorted in descending order:
\begin{align*}
    \supportinfset := \left \{\supportinf{i} \right \}_{i \in \agentset} \text{.}
\end{align*}
We denote the $i^{\text{th}}$ element in this ordered set by $\supportinfhighest{i}$. We define the infimum of the $\numunits$-th highest bid $\khighbidCDF$ as:
\begin{align*}
    \supportinforderstat{\numunits} := \inf(\supp(\khighbidCDF)) \text{.}
\end{align*} 

\begin{lemma}\label{lma:support-infimum-equivalence}
    The infimum of the support of the bid distribution of the $\numunits^{\text{th}}$-highest agent is the $\numunits^{\text{th}}$ element in the order $\supportinfset$, that is,
    \begin{align*}
        \supportinforderstat{\numunits} = \supportinfhighest{\numunits} \text{.}
    \end{align*}
\end{lemma}

\begin{proof}
    Consider any agent $i$ such that $\supportinf{i}$ is in the top $\numunits$ elements of the order $\supportinfset$. By definition of $\supportinf{i}$, this agent almost surely has a bid $b \geq \supportinf{i}$. Since this holds for any agent corresponding to a top $\numunits$ element in the order $\supportinfset$, the bid of the $\numunits^{\text{th}}$ highest agent is almost surely at least $\supportinfhighest{\numunits}$. Thus it must be that
    \begin{align*}
        \supportinforderstat{\numunits} \geq \supportinfhighest{\numunits} \text{.}
    \end{align*}
    We will now show that in fact $\supportinforderstat{\numunits} = \supportinfhighest{\numunits}$. For every $\delta>0$, consider the following event: bid profile $\bm{b} = \left(b_{1},\dots,b_{n}\right)$ is such that the bid $b_{i} \in \left[\supportinf{i},\supportinf{i}+\delta \right)$ for every agent $i \in \agentset$. By definition of $\supportinf{i}$, this event occurs with strictly positive probability. Thus for every $\delta>0$, the probability that the $\numunits$-th highest bid lies in the interval $\left[\supportinfhighest{\numunits},\supportinfhighest{\numunits}+\delta\right)$ is strictly positive. Thus $\supportinforderstat{\numunits} \leq \supportinfhighest{\numunits}$. The statement of the result follows. 
\end{proof}

The result above demonstrated an equivalence of the infimum of equilibrium bid distributions and the distribution of the $\numunits^{\text{th}}$ highest bid. In addition to this, we prove a crucial result which shows that the $\numunits + 1$ highest suprema of equilibrium bid distributions must all be the same.  

\begin{lemma}[Support upper bound equivalence]\label{lma:support-upper-bound-equivalence}
    Suppose $\supportsuphighest{i}$ is the $i^{\text{th}}$ highest value in the set $\{ \supportsup{j} \}_{j \in \agentset}$. Then,
    \begin{align*}
        \supportsuphighest{1} = ... = \supportsuphighest{\numunits + 1} \text{.}
    \end{align*}
\end{lemma}

\begin{proof}
    Suppose for a contradiction it is not true that $\supportsuphighest{1} = ... = \supportsuphighest{\numunits + 1}$. Then, since $\supportsuphighest{1} \geq ... \geq \supportsuphighest{\numunits + 1}$, we must have
    \begin{align*}
        \supportsuphighest{\numunits + 1} < \supportsuphighest{1} \text{.}
    \end{align*}
    This implies the following: $\supportsuphighest{\numunits + 1}$ is finite and the region $(\supportsuphighest{\numunits + 1}, \supportsuphighest{1}] \subseteq \mathbb{R}$ is non-empty. In addition to this, there must exist an agent $i$ with $\supportsup{i} = \supportsuphighest{1}$ who bids in $(\supportsuphighest{\numunits + 1}, \supportsuphighest{1}]$ with positive probability, and for all $z \in (\supportsuphighest{\numunits + 1}, \supportsuphighest{1}]$, this agent's interim allocation is $\interimalloc{i}(z) = 1$.

    This permits agent $i$ to adjust their bidding strategy such that, whenever they would have bid $b_{i} \in (\supportsuphighest{\numunits + 1}, \supportsuphighest{1}]$, they instead bid
    \begin{align*}
        b_{i}^{*} := \supportsuphighest{\numunits + 1} + (b_{i} - \supportsuphighest{\numunits + 1}) / 2 \text{.}
    \end{align*}
    This keeps their interim allocation $\interimalloc{i}(b_{i}^{*}) = \interimalloc{i}(b_{i}) = 1$ constant, but strictly decreases their bid, since $b_{i}^{*} < b_{i}$. Thus, this new strategy strictly increases agent $i$'s ex-ante utility, and so their original bidding strategy must not have been optimal, which is a contradiction.
\end{proof}

We now present some standard results that have been established in the single-unit setting (see for example \cite{JL-23}) and show their equivalent results in the multi-unit setting. The first result concerns an agent's interim allocation rule and shows several standard properties. Specifically, an agent's interim allocation weakly increases in their bid. Additionally, in equilibrium, a bid above $\supportinforderstat{\numunits}$ secures a unit with strictly positive probability, while a bid below $\supportinforderstat{\numunits}$ has a zero probability in receiving a unit.

\begin{lemma}\label{lma:interim-allocation-prop-1}
    The interim allocation rule $\interimalloc{i}(b)$ for any agent $i$ satisfies the following properties:
    \begin{enumerate}[(a)]
        \item $\interimalloc{i}(b)$ is weakly increasing in $b$.
        \item $\interimalloc{i}(b)$ is bounded as
        \begin{align*}
            \sup\{\criticalbidCDF{i}(t): t<b \}\leq \interimalloc{i}(b) \leq \criticalbidCDF{i}(b) \quad \text{ for all } b \text{.}
        \end{align*}
        \item Consider any bid $b$ in the support of the bid distribution $\bidCDF{i}$. It has interim allocation
        \begin{enumerate}[(i)]
            \item $\interimalloc{i}(b)=0$ when $b < \supportinforderstat{\numunits}$.
            \item $\interimalloc{i}(b) > 0$ when $b > \supportinforderstat{\numunits}$. 
        \end{enumerate}
    \end{enumerate}
\end{lemma}

\begin{proof}
    The allocation rule of agent $i$ for a bid $b$ is
    \begin{align}\label{eq:interimallocation}
        \interimalloc{i}(b) &= \prob[\criticalbidCDF{i} < b] + \prob[\criticalbidCDF{i} = b] \prob[i \text{ wins in a tie-break at bid } b] \text{.}
    \end{align}
    We will use this expression to prove the stated properties.

    \begin{enumerate}[(a)]
        \item \textit{Proof of Part (a)}: It is clear from (\ref{eq:interimallocation}) that the interim allocation is bounded as
        \begin{align*}
            \interimalloc{i}(b) \leq \prob[\criticalbidCDF{i} \leq b] \text{.}
        \end{align*}
        Consider an arbitrary $\epsilon>0$ and a bid $b'=b+\epsilon$. The interim allocation for bid $b'$ satisfies
        \begin{align*}
            \interimalloc{i}(b') &= \prob[\criticalbidCDF{i} < b'] + \prob[\criticalbidCDF{i} = b'] \prob[i \text{ wins in a tie-break at bid } b'] \text{,} \\
            & \geq \prob[\criticalbidCDF{i} < b + \epsilon] \text{,} \\
            & \geq \prob[\criticalbidCDF{i} \leq b] \text{,} \\
            & \geq \interimalloc{i}(b) \text{.}
        \end{align*}
        The equality follows from (\ref{eq:interimallocation}) and the first inequality follows from the fact that a probability is always non-negative. Since, we show that for any $\epsilon>0$, the interim allocation $\interimalloc{i}(b+\epsilon) \geq \interimalloc{i}(b)$, the statement in Part (a) follows.
    
        \item \textit{Proof of Part (b)}: As was shown in the proof of Part (a), an upper bound on the interim allocation is
        \begin{align*}
            \interimalloc{i}(b) \leq \criticalbidCDF{i}(b) \text{.}
        \end{align*}
        To show the lower bound, it holds that
        \begin{align*}
            \criticalbidCDF{i}(t) \leq \prob[\criticalbidCDF{i} < b] \quad \text{ for all } t < b \text{.}
        \end{align*}
        It is straightforward to see that
        \begin{align*}
            \sup\{\criticalbidCDF{i}(t): t<b \} & \leq \prob[\criticalbidCDF{i} < b] \text{,} \\
            & \leq \interimalloc{i}(b) \text{.}
        \end{align*}
        The statement of Part (b) follows.
    
        \item \textit{Proof of Part (c)}: Consider an agent $i$. We will now analyze its competing bid distribution $\criticalbidCDF{i}$. Let us denote the infimum of its support by $\supportinfcritical{i}$. Moreover, let us define the collection
        \begin{align*}
            \supportinfcriticalset{i} := \{\supportinf{j}\}_{j \neq i}
        \end{align*}
        An application of \Cref{lma:support-infimum-equivalence} tells us that
        \begin{align}\label{eq:competingbidinfsupport}
            \supportinfcritical{i} = \supportinfcriticalsetelement{i}{\numunits} \text{.}
        \end{align}
        
        Suppose agent $i$ is such that $\supportinf{i} < \supportinforderstat{\numunits}$. The infimum of the support of the competing bid distribution satisfies
        \begin{align*}
            \supportinfcritical{i} &= \supportinfcriticalsetelement{i}{\numunits} \text{,} \\
            &= \supportinfhighest{\numunits} \text{,} \\
            &= \supportinforderstat{\numunits} \text{.}
        \end{align*}
        The first equality follows from (\ref{eq:competingbidinfsupport}), the second follows from the assumption $\supportinf{i} < \supportinforderstat{\numunits}$ and the third follows from \Cref{lma:support-infimum-equivalence}. Consider any bid $b$ in the support of agent $i$'s bid distribution. In the case $b < \supportinforderstat{\numunits}$, the interim allocation $\interimalloc{i}(b) = 0$ because the bid lies strictly below the infimum of the support of the competing bid distribution. In the case $b > \supportinforderstat{\numunits}$, he is allocated with strictly positive probability because the bid lies strictly above the support of the competing bid distribution.
        
        Suppose agent $i$ is such that $\supportinf{i} \geq \supportinforderstat{\numunits}$. By definition of $\supportinf{i}$, the agent almost surely has a bid $b \geq \supportinf{i}$. Thus, Part (i) vacuously holds true. The infimum of the support of the competing bid distribution
        \begin{align*}
            \supportinfcritical{i} &= \supportinfcriticalsetelement{i}{\numunits} \text{,} \\
            &\leq \supportinfhighest{\numunits} \text{,} \\
            &= \supportinforderstat{\numunits} \text{.}
        \end{align*} The second equality follows from observing that removing an element from the order $\supportinfset$ only decreases the value of entry $\numunits$. It follows that any bid $b > \supportinforderstat{\numunits}$ is allocated with strictly positive probability because the infimum of the competing bid distribution is strictly below the bid.
    \end{enumerate}
    The $3$ parts of the result statement are proved.
\end{proof}

Next, we examine the equilibrium strategy of an agent. The agent obtains positive utility if and only their value is strictly greater than the infimum $\supportinforderstat{\numunits}$. Furthermore, they submit a bid at or above this threshold if their value exceeds it strictly. 

\begin{lemma}\label{lma:bidding-dichotomy}
    Consider an agent $i$ and value $v \in \supp(V_{i})$. The equilibrium strategy $s_{i}(\cdot)$ for any agent $i$ satisfies the following properties:
    \begin{enumerate}[(a)]
        \item If $v \leq \supportinforderstat{\numunits}$, the optimal utility $u_{i}(v)=0$ and the equilibrium bid $s_{i}(v) \leq \supportinforderstat{\numunits}$ almost surely over the random strategy $s_{i}(\cdot)$.
        \item If $v > \supportinforderstat{\numunits}$, the optimal utility $u_{i}(v)>0$ and the equilibrium bid $\supportinforderstat{\numunits} \leq s_{i}(v) < v$ almost surely over the random strategy $s_{i}(\cdot)$.
    \end{enumerate}
\end{lemma}

\begin{proof}
    We will consider two cases: one where $\supportinf{i} < \supportinforderstat{\numunits}$ and the other where $\supportinf{i} \geq \supportinforderstat{\numunits}$.
    
    Suppose $\supportinf{i} < \supportinforderstat{\numunits}$. As we have shown in the proof of \Cref{lma:interim-allocation-prop-1}, the infimum of the support of the competing bid distribution $\supportinfcritical{i}$ is equal to $\supportinforderstat{\numunits}$. 
    
    \begin{enumerate}[(i)]
        \item Proof of Part (a): Consider a value $v \leq \supportinforderstat{\numunits}$. If the agent were to bid $b > \supportinforderstat{\numunits}$, this would give them utility $u_{i}(v,b) < 0$ because it is allocated with strictly positive probability. If the agent bid $b < \supportinforderstat{\numunits}$, the agent is allocated with probability $0$ because the infimum of the support of the competing bid distribution is $\supportinforderstat{\numunits}$. It follows that they receive $u_{i}(v,b) = 0$ for such a bid. If the agent bid $b = \supportinforderstat{\numunits}$, the utility he gets for such a bid is at most $0$ (the utility would be $0$ in the case he is not allocated an item). Thus, the optimal utility for such a value is $u_{i}(v)=0$. Moreover, the bids which guarantee this utility must be at most $\supportinforderstat{\numunits}$. Thus, the equilibrium bid $s_{i}(v) \leq \supportinforderstat{\numunits}$ almost surely. 
        \item Proof of Part (b): Consider a value $v > \supportinforderstat{\numunits}$. We will show that the bid $b = \frac{v+\supportinforderstat{\numunits}}{2}$ gives the agent a strictly positive utility. Observe that the bid $b > \supportinforderstat{\numunits}$ and has allocation $\interimalloc{i}(b) > 0$ because the bid is strictly greater than the infimum of the support of the competing bid distribution $\supportinforderstat{\numunits}$. Since the bid $b < v$, the utility for this bid $u_{i}(v,b) > 0$. The optimal bid can only do better. Thus, the optimal utility $u_{i}(v) > 0$. We will now show that $\supportinforderstat{\numunits} \leq s_{i}(v) < v$ almost surely. If the agent bids $b < \supportinforderstat{\numunits}$, his allocation is $\interimalloc{i}(b) = 0$ (the bid is strictly smaller than the infimum of the support of the competing bid distribution). This would give the agent a utility of $0$ and thus cannot be optimal. It follows that $\supportinforderstat{\numunits} \leq s_{i}(v) < v$ almost surely (the bound $s_{i}(v) < v$ follows from the observation that a bid $b \geq v$ will be allocated with strictly positive probability but given a utility $u_{i}(v,b) \leq 0$, thus not being optimal).
    \end{enumerate}
    
    Suppose $\supportinf{i} \geq \supportinforderstat{\numunits}$. We will state two observations which will be helpful in proving both parts of the statement. First, the infimum of the support of $\bidCDF{i}$ is at least $\supportinforderstat{\numunits}$. Thus, the equilibrium bid of this agent is at least $\supportinforderstat{\numunits}$ almost surely. Second, as we have shown in the proof of \Cref{lma:interim-allocation-prop-1}, the infimum of the support of the competing bid distribution $\supportinfcritical{i}$ is at most $\supportinforderstat{\numunits}$. 
    \begin{enumerate}[(i)]
        \item Proof of Part (a): Consider a value $v \leq \supportinforderstat{\numunits}$. In this case, there is no value $v_{i}$ in the support of $\mathcal{V}_{i}$ that satisfies $v_{i} \in \left(\supportinforderstat{\numunits}_{-i},\supportinforderstat{\numunits}\right]$. This is because if $v_{i} \in \left(\supportinforderstat{\numunits}_{-i},\supportinforderstat{\numunits}\right]$, then it can obtain strictly positive utility by bidding $(v+\supportinforderstat{\numunits}_{-i})/2$. Furthermore, for a sufficiently small open interval around $v$, every value in the support of $\mathcal{V}_{i}$ and in that interval must bid strictly below $\supportinforderstat{\numunits}$ almost surely. This contradicts $\gamma_{i} \geq \supportinforderstat{\numunits}$ since for every interval around $v$, the value is realized with positive probability. For any value $v \leq \supportinforderstat{\numunits}_{-i}$, any bid $b < \supportinforderstat{\numunits}_{-i}$ gives zero allocation and any bid $b \geq v$ has non-positive utility. Thus $u_{i}(v) =0$ and $s_{i}(v) \leq \supportinforderstat{\numunits}$ almost surely because any bid above $\supportinforderstat{\numunits}$ guarantees allocation with positive probability.
        \item Proof of Part (b): Consider a value $v > \supportinforderstat{\numunits}$. We will show that the bid $b = \frac{v+\supportinforderstat{\numunits}}{2}$ gives the agent a strictly positive utility. Observe that the bid $b > \supportinforderstat{\numunits}$ and has allocation $\interimalloc{i}(b) > 0$ because the bid is strictly greater than $\supportinfcritical{i}$. To be precise, $\supportinfcritical{i} \leq \supportinforderstat{\numunits}$ for agent $i$. Since the bid $b < v$, the utility for this bid $u_{i}(v,b) > 0$. The optimal bid can only do better. Thus, the optimal utility $u_{i}(v) > 0$. Moreover, by definition of $\supportinf{i}$, the equilibrium bid satisfies $\supportinforderstat{\numunits} \leq s_{i}(v) < v$ almost surely. 
    \end{enumerate}
    The statement of the result holds.
\end{proof}

Furthermore, we reference a result from \cite{CH-13}, which establishes that an agent obtains strictly higher utility with a larger value, and, almost surely, submits a higher bid. For completeness, we include a proof below.

\begin{lemma}[\citealp{CH-13}]\label{lma:strategy-and-utility-monotonicity}
    Consider values $v, w$ in the support of value distribution $V_{i}$ such that $v > w > \supportinforderstat{\numunits}$. The following properties hold:
    \begin{enumerate}[(a)]
        \item The optimal utility at these values is ordered as
        \begin{align*}
            u_{i}(v) > u_{i}(w) > 0 \text{.}
        \end{align*}
        \item The equilibrium bids are ordered as $s_{i}(v) \geq s_{i}(w)$ almost surely over the random strategy $s_{i}(\cdot)$.
    \end{enumerate}
\end{lemma}

\begin{proof}
    \begin{enumerate}[(i)]
        \item Proof of Part (a): At value $v$, agent $i$ could bid according to random strategy $s_{i}(w)$. Because the allocation is exactly what the agent would get if the value were $w$, such a strategy would give a strictly higher utility than $u_{i}(w)$. The optimal strategy for value $v$ can only do weakly better and thus $u_{i}(v) > u_{i}(w)$. The fact that both of these are strictly greater than $0$ follows from \Cref{lma:bidding-dichotomy}.
        \item Proof of Part (b): The proof of this is exactly as that in the $1$-unit setting.
    \end{enumerate} The statement of the result holds.
\end{proof}

Finally, we present the neighborhood deviation argument from \citet{JL-23}.

\begin{lemma}[\citealp{JL-23}]\label{lma:neighborhood-deviation-argument} Consider any agent $i$ with interim allocation rule $x_{i}(\cdot)$. The following conditions are satisfied for any bid $b$: \begin{enumerate}[(a)]
    \item Suppose there is a sequence of bids $\left(b_{m}\right)_{m=1}^{\infty}$ such that $\lim_{m \to \infty}b_{m} = b$ and $b \in \left(\supportinforderstat{\numunits},\supportsuporderstat{\numunits}\right]$. Moreover, there is a $\delta>0$ for which $x_{i}(b_{m}) \geq x_{i}(b) + \delta$ for any $m$. Then $b$ is not an equilibrium bid.
    \item Suppose there is a sequence of bids $\left(b_{m}\right)_{m=1}^{\infty}$ and a $\delta>0$ such that $b-b_{m} \geq \delta$ for every $m$ and $b \in \left(\supportinforderstat{\numunits},\supportsuporderstat{\numunits}\right]$. Moreover, the allocation probability $\lim_{m \to \infty}x_{i}(b_{m}) = x_{i}(b)$. Then $b$ is not an equilibrium bid.
\end{enumerate}
\end{lemma}

\begin{proof}
     We first prove part (a). Consider any value $v \in \left(\supportinforderstat{\numunits},b\right]$. The utility it gets from bid $b$ is non-positive. However, this is a contradiction to \Cref{lma:bidding-dichotomy}. Consider any value $v > b$. Take a bid from the sequence $\left(b_{m}\right)_{m=1}^{\infty}$ that is arbitrarily close to $b$. This would give an agent strictly better utility than bidding $b$ because $x_{i}(b_{m}) \geq x_{i}(b) + \delta$ for $\delta>0$. Finally, for every $v \leq \supportinforderstat{\numunits}$, by \Cref{lma:bidding-dichotomy}, the equilibrium bid is at most $\supportinforderstat{\numunits}$ almost surely. 

     We next prove part (b).  Consider any value $v \in \left(\supportinforderstat{\numunits},b\right]$. The utility it gets from bid $b$ is non-positive. However, this is a contradiction to \Cref{lma:bidding-dichotomy}. Consider any value $v > b$. Take a bid from the sequence $\left(b_{m}\right)_{m=1}^{\infty}$ such that its allocation probability is arbitrarily close to $x_{i}(b)$. This would give an agent strictly better utility than bidding $b$ because $b-b_{m} \geq \delta$ and their allocation probabilities are arbitrarily close with $x_{i}(b) > 0$ (see the proof of \Cref{lma:bidding-dichotomy}). 
\end{proof}

\end{document}